\documentclass[twocolumn,showpacs,preprintnumbers,amsmath,amssymb,prb]{revtex4-1}
\usepackage{graphicx}% Include figure files
\usepackage{dcolumn}% Align table columns on decimal point
\usepackage{bm}% bold math
\usepackage{color}
\usepackage{physics}
\usepackage{multirow}

\renewcommand{\d}{{\bm{\delta}}}
\newcommand{\e}{{\bm e}}

\newcommand{\B}{{\bf B}}
\newcommand{\E}{{\bf E}}

\renewcommand{\a}{{\bf a}}

\renewcommand{\v}{{\bf v}}

\renewcommand{\j}{{\bf j}}
\renewcommand{\k}{{\bf{k}}}
\newcommand{\q}{{\bf{q}}}
\renewcommand{\r}{{\bf{r}}}

\newcommand{\rr}{{\bm{r}}}

\renewcommand{\d}{{\bf d}}

\usepackage[normalem]{ulem}

\def\lsim{\lower.35em\hbox{$\stackrel{\textstyle<}{\textstyle\sim}$}}
\def\gsim{\lower.35em\hbox{$\stackrel{\textstyle>}{\textstyle\sim}$}}

\begin{document}

\title{Ising superconductivity induced from spin-selective valley symmetry breaking in twisted trilayer graphene}

\author{J. Gonz\'alez$^{1*}$ and T. Stauber$^{2}$}
\email{j.gonzalez@csic.es}
\email{tobias.stauber@csic.es}

\affiliation{
$^{1}$Instituto de Estructura de la Materia, CSIC, E-28006 Madrid, Spain.\\ 
$^{2}$Instituto de Ciencia de Materiales de Madrid, CSIC, E-28049 Madrid, Spain.}
\date{\today}

\begin{abstract}
We show that the $e$-$e$ interaction induces a strong breakdown of valley symmetry for each spin channel in twisted trilayer graphene, leading to a ground state where the two spin projections have opposite sign of the valley symmetry breaking order parameter. This leads to a spin-valley locking in which the electrons of a Cooper pair are forced to live on different Fermi lines attached to opposite valleys. Furthermore, we find an effective intrinsic spin-orbit coupling explaining the protection of the superconductivity against in-plane magnetic fields. The effect of spin-selective valley symmetry breaking is validated as it reproduces the experimental observation of the reset of the Hall density at 2-hole doping. It also implies a breakdown of the symmetry of the bands from $C_6$ to $C_3$, with an enhancement of the anisotropy of the Fermi lines which is at the origin of a Kohn-Luttinger (pairing) instability. The isotropy of the bands is gradually recovered, however, when the Fermi level approaches the bottom of the second valence band, explaining why the superconductivity fades away in the doping range beyond 3 holes per moir\'e unit cell in twisted trilayer graphene.
\end{abstract}

\maketitle

\section*{Introduction} 
The discovery of superconductivity and its parent insulating phases at the magic angle of twisted bilayer graphene (TBG)\cite{Cao18a,Cao18b} has opened a new era in the investigation of strongly correlated phenomena in two-dimensional electron systems. There is an ongoing debate about the origin of the superconductivity in TBG\cite{Yankowitz19,Codecido19,Shen19,Lu19,Chen19,Xu18,Volovik18,Yuan18,Po18,Roy18,Guo18,Dodaro18,Baskaran18,Liu18,Slagle18,Peltonen18,Kennes18,Koshino18,Kang18,Isobe18,You18,Wu18b,Zhang18,Ochi18,Thomson18,Carr18,Guinea18,Zou18,Laksono18,Gonzalez19,Kang19,Pizarro19,Gonzalez19b,Choi19,Sharpe19,Moriyama19,Jiang19,Xie19,Kerelsky19}, which could also clarify whether a similar phenomenon can arise in other moir\'e van der Waals materials. In this regard, superconductivity has been already observed in twisted trilayer graphene (TTG),\cite{Park21,Zeyu21} showing unconventional features like reentrant behavior under large magnetic fields.\cite{Khalaf19,Carr20,Lopez-Bezanilla20,Cao21,Fischer21,LiuXiaoxue22,Christos22} Moreover, in the presence of spin-orbit coupling, a valley symmetry (VS) broken state can lead to a zero-field superconducting diode effect.\cite{Lin22,Scammell22}

%Apart from the superconductivity, a remarkable effect in TBG is the observation of symmetry breaking at integer fillings of the lowest valence and conduction bands.\cite{Zondiner20,Wong20,He21} This effect is most evident in the recurrent reset of the Fermi level to a sequence of Dirac points as the lowest valence (conduction) bands are emptied (filled), which points at the splitting of valley and spin degenerate bands. 

%\footnote{A more direct evidence of symmetry breaking has been also found at three-quarter filling of the lowest conduction band, where a clear signal of ferromagnetism has been observed.} 

TTG has also shown a striking phenomenon of reset of the Hall density at integer fillings of the highest valence and lowest conduction bands.\cite{Park21,Zeyu21} Specifically at 2-hole doping, it has been found that the Hall density jumps down to zero. This observation is particularly important, since the effect of reset precedes the development of the superconducting regime right below 2-hole doping as well as right above 2-electron doping in the conduction band. 
%This feature of Hall density reset may be used then as a source of valuable information about the effects of the electronic interaction.   
    
\begin{figure}[t]
\includegraphics[width=0.80\columnwidth]{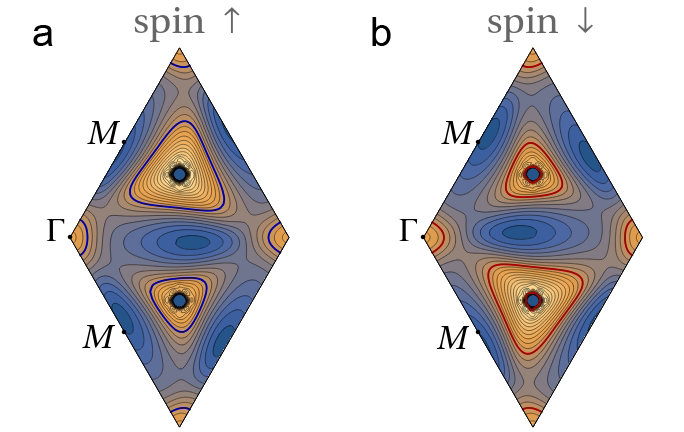}
\caption{{\bf Energy contour maps of the second valence band at filling fraction ${\bm \nu}$=-2.4.} {\bf a} Fermi lines for spin-up electrons. {\bf b} Fermi lines for spin-down electrons. Energy contours are shown on the moir\'e Brillouin zone of TTG with twist angle $\theta \approx 1.61^\circ$, for dielectric constant $\epsilon = 48$ and filling fraction of 2.4 holes per moir\'e unit cell. Contiguous contour lines differ by a constant step of 0.2 meV.}
\label{one}
\end{figure}

\begin{figure}[th]
\includegraphics[width=0.6\columnwidth]{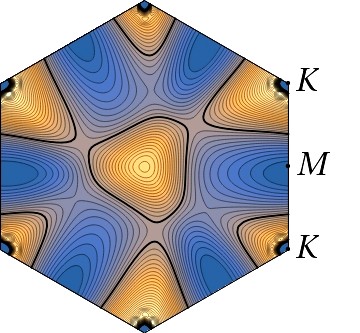}
\\
\caption{{\bf Energy contour map of the second valence band at filling fraction ${\bm \nu}$=-2.8.} Energy contour map of the second valence band (for spin-up projection) in the Brillouin zone of TTG at twist angle $\theta \approx 1.61^\circ$, computed in a self-consistent Hartree-Fock approximation with dielectric constant $\epsilon = 48$ and filling fraction of 2.8 holes per moir\'e unit cell. The thick contour stands for the Fermi line and contiguous contour lines differ by a constant step of 0.1 meV.}
\label{two}
\end{figure}

Here, we show within a self-consistent Hartree-Fock resolution in real space that the extended Coulomb interaction has a natural tendency to induce the breakdown of the VS of TTG. This lifts the degeneracy of the Dirac cones by moving them up and down in energy, respectively. The effect becomes strongest at 2-hole doping such that the Fermi level is pushed up to the vertices of the Dirac cones in the lower valley. At that filling, the Dirac nodes turn out to be unstable against time-reversal symmetry breaking with condensation of a Haldane mass, opening a gap at the Fermi level. As we show below, this is the mechanism responsible for the experimentally observed reset of the Hall density. We also show that the Fermi lines for spin-up and spin-down electrons are different but related by inversion symmetry, i.e., by the exchange of the two $K$ points in the Brillouin zone, as seen in Fig. \ref{one}. However, within one spin-channel, VS breaking leads to inversion breaking, as seen in Fig. \ref{two}. Ultimately, this can explain the violation of the Pauli limit by a factor of 2-3, observed in experiments.

%above a certain interaction strength, a gap opens up between the empty first valence band (VB) and the occupied second VB. The top of the second VB has quadratic dispersion, but the empty first VB develops an extremely flat shape, with the alignment of the van Hove singularity (vHS) to the bottom of the band. We show that this different behavior of the two bands is at the origin of the experimentally observed reset of the Hall density. 

\section*{Results}
\subsection{Spin-selective valley symmetry breaking} 
We deal with the setup of TTG usually realized in the experiments, in which the two outer layers are rotated by the same angle $\theta $ with respect to the central layer. We model this configuration by taking a twist angle $\theta \approx 1.61^\circ$ belonging to the set of commensurate superlattices realized by TBG. Then, the low-energy states are distributed into a Dirac-like band, with states odd under mirror symmetry with respect to the central plane, and two additional valence and conduction bands, with states even under the mirror symmetry (see the Supplemental Material (SM)\cite{SI}). The latter are the counterpart of the flat bands of TBG, and they become progressively flatter when approaching the magic angle of TTG, which is $\approx 1.6^\circ$.

In what follows, we apply an atomistic approach to TTG, based on a tight-binding model for the $\pi$ orbitals of the carbon atoms. The Hamiltonian $H$ is written as
\begin{align}
H = H_0 + H_{\rm int}\;,
\end{align} 
where $H_0$ stands for the non-interacting tight-binding Hamiltonian and $H_{\rm int}$ is the interaction part. This is expressed in terms of creation (annihilation) operators $a_{i\sigma}^+$ ($a_{i\sigma}$) for electrons at each carbon site $i$ with spin $\sigma$
\begin{align}  
H_{\rm int} = \frac{1}{2} \sum_{i,j,\sigma,\sigma'} a_{i\sigma}^{\dagger}a_{i\sigma} \: v_{\sigma \sigma'} (\r_i-\r_j) \: a_{j\sigma'}^{\dagger}a_{j\sigma'}\;,
\end{align}
For $\r_i \neq \r_j$, we take $v_{\sigma \sigma'} (\r_i-\r_j)=v(\r_i-\r_j)$, $v$ being the extended Coulomb potential with the long-range tail cut-off at a distance dictated by the screening length $\xi$, arising from the presence of nearby metallic gates, and with the strength further reduced by a dielectric constant $\epsilon$. For $\r_i = \r_j$, we have the Hubbard interaction $v_{\sigma \sigma'} = U \delta_{\sigma,-\sigma'}$, where we take $U = 0.5$ eV. The precise value of this rather small coupling is not relevant, as long as it is nonvanishing, but it plays a very important role to constrain the relative orientation of the spin projections in the two valleys of TTG (see the SM for all the details about the interaction). 

We resort to a self-consistent Hartree-Fock approximation in order to study the effects of the $e$-$e$ interaction. In this approach, the full electron propagator $G$ is represented in terms of a set of eigenvalues $\varepsilon_{a\sigma}$ and eigenvectors $\phi_{a\sigma} (\r_i)$ modified by the interaction, in such a way that in the static limit
\begin{align}
\left(  G  \right)_{i\sigma,j\sigma} = -\sum_a \frac{1}{\varepsilon_{a\sigma}}  \phi_{a\sigma} (\r_i)  \phi_{a\sigma}^* (\r_j)\;.
\label{hf}
\end{align}
We seek then the self-consistent resolution of the Dyson equation involving $G$, the noninteracting propagator $G_0$ and the self-energy $\Sigma $
\begin{align}
G^{-1} = G_0^{-1} - \Sigma \;.
\label{sd}
\end{align}
The self-consistent approach becomes feasible as the electron self-energy $\Sigma $ is expressed entirely in terms of the set of $\phi_{a\sigma} (\r_i)$. In the static limit, we have 
\begin{align}
\left( \Sigma  \right)_{i\sigma,j\sigma}  = &  \;  \mathbb{I}_{ij} \:  \sideset{}{'}\sum_a  \sum_{l, \sigma' } v_{\sigma \sigma'} (\r_i-\r_l)   \left|\phi_{a\sigma'} (\r_l)\right|^2      \notag    \\ 
    &  - v_{\sigma \sigma} (\r_i-\r_j)  \sideset{}{'}\sum_a \phi_{a\sigma} (\r_i)  \phi_{a\sigma}^* (\r_j)\;,
\label{selfe}
\end{align}
where the prime means that the sum is to be carried over the occupied levels \cite{Fetter71}.

%This is done in practice by applying an iterative procedure, which typically converges to a fixed-point solution for the eigenvectors $\phi_{a\sigma}$. 
The Fock contribution in Eq. (\ref{selfe}) becomes essential in order to account for the dynamical symmetry breaking. In TTG, we find that the dominant patterns correspond to the breakdown of time-reversal invariance\cite{SI}. This may be characterized by two different order parameters 
\begin{align}
P_{\pm}^{(\sigma )} = {\rm Im} \left( \sum_{i \in A}  \left(   h_{i_1 i_2}^{(\sigma )} h_{i_2 i_3}^{(\sigma )} h_{i_3 i_1}^{(\sigma )}   \right)^{\frac{1}{3}}
          \pm \sum_{i \in B}  \left(   h_{i_1 i_2}^{(\sigma )} h_{i_2 i_3}^{(\sigma )} h_{i_3 i_1}^{(\sigma )}   \right)^{\frac{1}{3}}  \right)
\label{ssigma}
\end{align}
where the sums run over the loops made of three nearest neighbors $i_1, i_2$ and $i_3$ of each atom $i$ in graphene sublattices $A$ and $B$, with matrix elements
\begin{align}
h_{ij}^{(\sigma )} =  \sideset{}{'}\sum_a \phi_{a\sigma} (\rr_i) \phi_{a\sigma}^* (\rr_j)\;,
\end{align}
which can be interpreted as an effective hopping between sites $i$ and $j$. One can check that $P_{-}^{(\sigma)}$ gives a measure of the mismatch in the energy shift of the bands in the two valleys of the electron system. On the other hand, a nonvanishing $P_{+}^{(\sigma)}$ is the hallmark of a Chern insulating phase, as described originally by Haldane\cite{Haldane88}. 

%In the continuum theory of Dirac fermions, the condensate (\ref{ssigma}) translates into the generation of a term proportional to the identity in pseudospin space. 
%This does not open a gap in the Dirac cones at the $K$ point, but instead it 
%This leads to a different shift in the energy of the bands in the two valleys of the electron system, with the consequent valley symmetry breaking.

\begin{figure}[th]
\includegraphics[width=0.9\columnwidth]{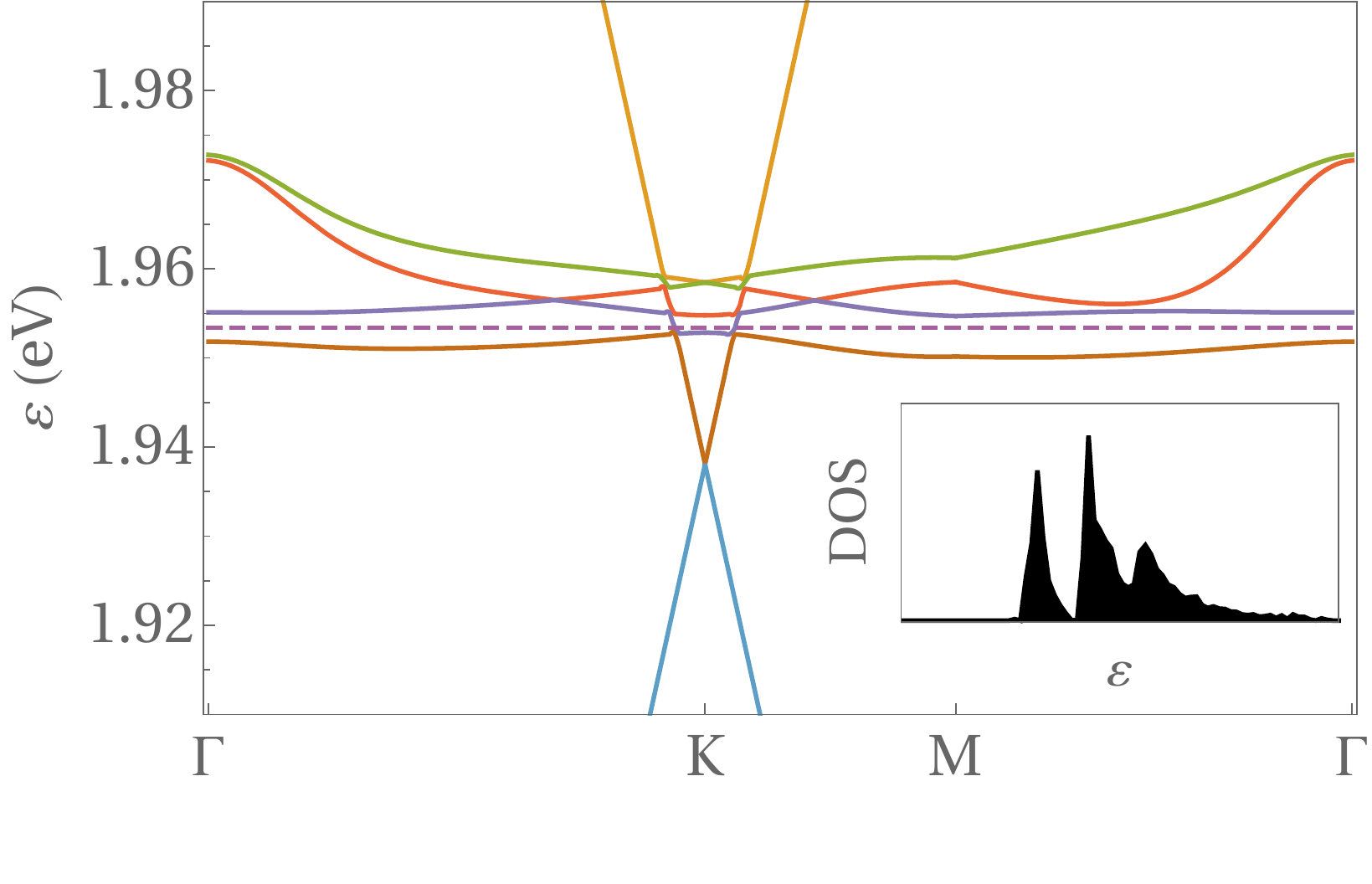}
\\
\caption{{\bf Self-consistent band structure along high-symmetry lines at filling fraction ${\bm \nu}$=-2.} Highest valence and lowest conduction bands of TTG at twist angle $\theta \approx 1.61^\circ$, computed in a self-consistent Hartree-Fock approximation with dielectric constant $\epsilon = 48$ and filling fraction of 2 holes per moir\'e unit cell (the dashed line stands for the Fermi level). The inset shows the density of states in the energy interval between 1.94 eV and 1.98 eV.}
\label{three}
\end{figure}

The analysis of internal screening in TTG reveals that the effective value of the dielectric constant must have in our model a magnitude of $\epsilon \sim 50$ (see SM\cite{SI}). The extended Coulomb interaction is then in a regime where the dominant order parameter is that of VS breaking, while $P_{+}^{(\sigma)}$ becomes also nonvanishing at filling fraction $\nu = -2$ \cite{SI}. This can be seen in Fig. \ref{three}, which shows the splitting at the $K$ point of the Dirac cones from the two valleys, as an effect of VS breaking. At 2-hole doping, the Fermi level should be then at the vertex of the Dirac cone of the lower valley. However, the interaction is strong enough to trigger the condensation of the Haldane mass, which leads to the gap seen in Fig. \ref{three} at the Fermi level. In this discussion, the effect of the ^^ ^^ third", lowest Dirac cone can be safely neglected as this band belongs to a different representation of the mirror symmetry.   

\subsection{Hall density reset}
From the resistivity tensor $\rho$ as function of the magnetic field $B$, the Hall density $n_H$ can be obtained which is usually directly related to the electronic density $n$: 
\begin{align}
n_{H}=-\left[e\frac{d\rho_{xy}}{d B}\right]^{-1}
\end{align}
Experimentally, a reset from a large value down to zero Hall density is observed in TTG at 2-hole doping (as well as at 2-electron doping in the conduction side). In our interacting model, we can explain such a discontinuity as a result of the jump of the Fermi level across the gap shown in Fig. \ref{three}, from the bottom of the first valence band (VB) to the top of the second VB.

As shown in the SM, in the semiclassical approximation, closed trajectories quite generally lead to a universal Hall density $n_H=n$, in terms of the electronic density $n$. Even extreme elliptic trajectories still fall under this universality class and anharmonic effects due to trigonal warping usually lead only to slight deviations. 
Thus, linear (universal) behavior $n_H=n$ is obtained starting from filling factor $\nu=0$.
%Also, for hole doping up to $n\approx-1.8$, no gap has developed yet, leading to $n_H=n$ in this regime as first approximation.

Non-universal behavior with $n_H\neq n$ only comes from open trajectories which are usually linked to van Hove singularities (vHSs).\cite{Guerci21} Around these points, the diverging Hall density is given by
\begin{align}
\label{vHFormula}
n_{H}&=\frac{n}{\pi}\ln\frac{\alpha\Lambda^2}{|\mu|+k_BT}\;,
\end{align} 
where $\alpha$ is related to the inverse reduced mass, $\Lambda$ is the phenomenological band-cutoff, and $\mu$ the relative chemical potential corresponding to the electronic density $n$. We also introduce the finite temperature $T$ that smears out the logarithmic divergence, which shall also include disorder effects. Details on the derivation of Eq. (\ref{vHFormula}) and the fitting procedure are given in the SM.

For a quantitative discussion of the Hall density in TTG, we consider the first and second VBs for $\nu=-2$ and $\nu=-2.8$, respectively, see SM. We expect deviations due to varying filling factors to only slightly shift the energy of the vHS corrections. Due to the pronounced gap between the first and the second VB, there is a reset of the Hall density at $\nu=-2$, which leads to $n_H=\nu+2$ for $\nu<-2$ due to the closed semi-classical orbits of the band structure near the band edge. As mentioned above, the linear (universal) behavior is also obtained around filling factors $\nu=0$ and $\nu=-4$ (neglecting the contribution of the Dirac cone that becomes relevant for $\nu \approx -4$). 

\begin{figure}[th]
\includegraphics[width=0.9\columnwidth]{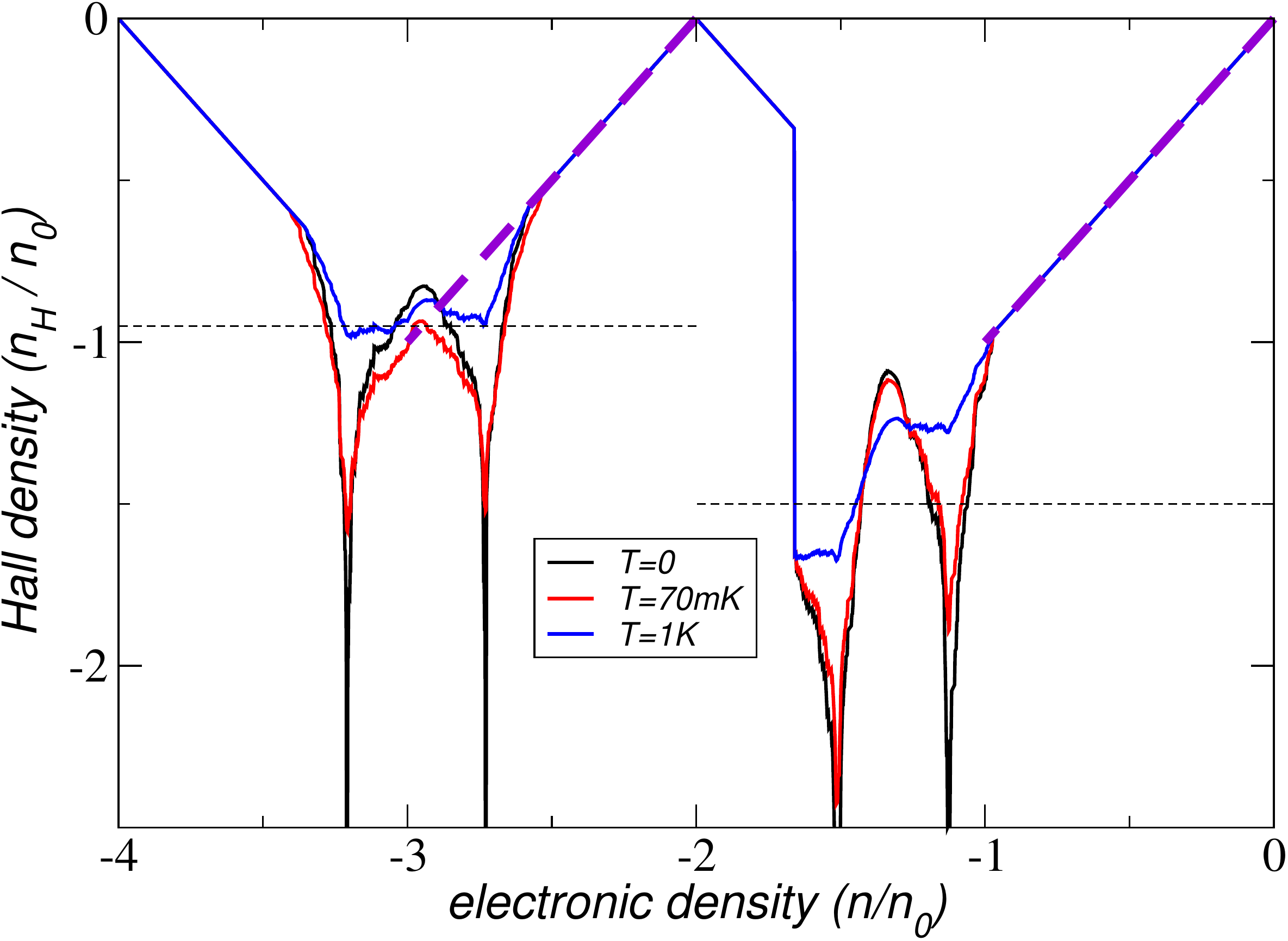}\\
\caption{{\bf Hall density for the two highest valence bands.} Hall density as function of the filling factor in units of the density $n_0$ of one electron per moir\'e supercell for three different temperatures $T=0,$ 70 mK, 1K. Also shown are the maximal values for each sub-band  of the Hall density measured in Ref. \onlinecite{Park21}, as well as the dashed purple lines indicating the universal behavior. The reset at 2-hole doping emerges due to the gap at the half-filled VB, see Fig. \ref{three}.}
\label{five}
\end{figure}

Fig. \ref{five} shows the Hall density $n_H$ as function of the filling factor for different temperatures $T=0$, 70 mK, 1 K. The energies and respective filling factors of the vHSs are indicated by the logarithmic divergences for $T=0$.
%that are expected to change slightly their position when the doping dependent band structure is taken into account.  
Also shown are the maximal values for each sub-band of the Hall density measured in Ref. \onlinecite{Park21}, as well as the dashed purple lines indicating the universal behavior. The curve for $T=1$ K agrees well with the experimental results performed at $T=70$ mK, which suggests a considerable amount of disorder in the unbiased sample. 

\subsection{Ising superconductivity} 
\begin{figure}[th]
\includegraphics[width=0.9\columnwidth]{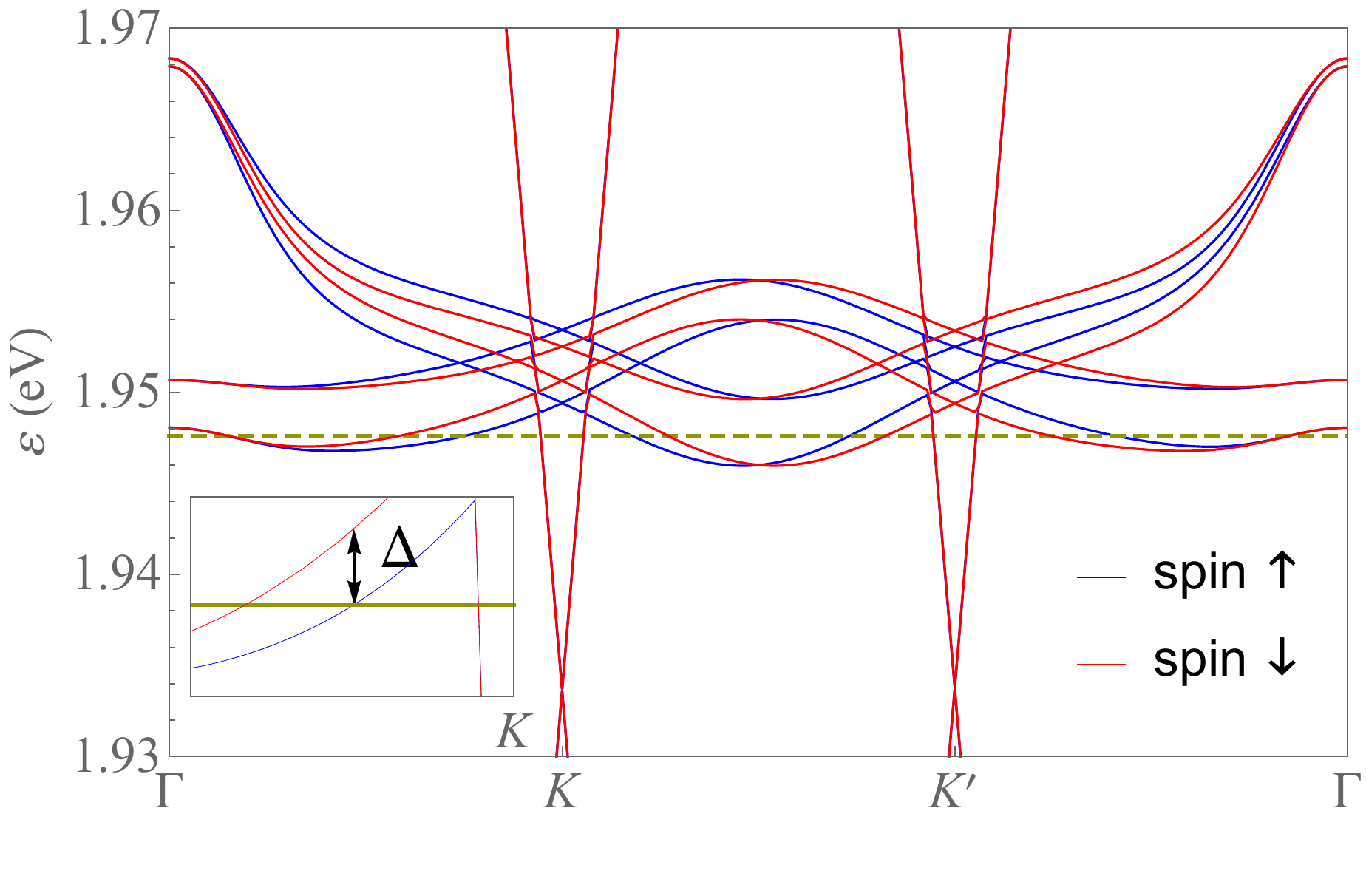}
\\
\caption{{\bf Self-consistent band structure for both spin projections along high-symmetry lines at filling fraction ${\bm \nu}$=-2.4.} Energy bands of TTG around charge neutrality (computed for dielectric function $\epsilon = 48$ and filling fraction $\nu = -2.4$) along a rectilinear path $\Gamma K K' \Gamma $, discerning the dispersion for spin-up and spin-down electrons.} %The inset shows a zoom view of the dispersion near the Fermi level (dashed line) with the definition of the spin splitting $\Delta $.}
\label{six}
\end{figure}

The strong spin-selective VS breaking leads to ground states where the inversion symmetry is broken for each spin projection, but in which this symmetry is recovered upon exchange of the two spin projections, as shown in Fig. \ref{six}. This opens the possibility of having Ising superconductivity, in which each spin projection in a Cooper pair is attached to a different Fermi line and the singlet is polarized in  out-of-plane direction.\cite{JMLu15,SaitoYu16,XiXiaoxiang16} This lends protection to the superconductivity against in-plane magnetic fields as no Zeeman term arises.

The actual pairing instability takes place as a result of the anisotropy of the $e$-$e$ scattering along the Fermi lines, which is strong enough to induce an effective attraction. This is characterized by the appearance of a negative coupling when projecting the Cooper pair vertex $V$ onto the different harmonics along the Fermi line.
The vertex $V$ is indeed a function of the angles $\phi $ and $\phi'$ of the respective momenta of the spin-up incoming and outgoing electrons on each contour line of energy $\varepsilon $. The scattering of Cooper pairs in the particle-particle channel leads to a reduction of the amplitude of the vertex, given by the equation
\begin{eqnarray}
&&V(\phi, \phi')= V_0 (\phi, \phi')-   \nonumber        \\ 
 && \frac{1}{(2\pi )^2} \int^{\Lambda_0} \frac{d \varepsilon }{\varepsilon } 
   \int_0^{2\pi } d \phi'' 
  \frac{\partial k_\perp }{\partial \varepsilon}  
      \frac{\partial k_\parallel }{\partial \phi''} 
  V_0 (\phi, \phi'')    
         V(\phi'', \phi')
\label{pp}
\end{eqnarray}
where $k_\parallel ,k_\perp $ are the longitudinal and transverse components of the momentum for each energy contour line while $V_0 (\phi, \phi') $ is the bare vertex at an energy cutoff $\Lambda_0$ (see SM\cite{SI}). By differentiating Eq. (\ref{pp}), we get 
\begin{equation}
\varepsilon \: \frac{\partial \widehat{V}(\phi, \phi' )}{\partial \varepsilon } 
 =   \frac{1}{2\pi }  \int_0^{2\pi } d \phi''  
 \widehat{V} (\phi, \phi'' )  \widehat{V}(\phi'', \phi' )
\label{scaling}
\end{equation}
with $\widehat{V} (\phi, \phi' ) = F(\phi ) F(\phi' ) V (\phi, \phi' )$ and $F(\phi ) = \sqrt{ (\partial k_\perp / \partial \varepsilon )
  (\partial k_\parallel / \partial \phi )/2\pi  }$. Then, when there is a negative eigenvalue in the expansion of $\widehat{V}$ in harmonics, Eq. (\ref{scaling}) leads to a divergent flow for that particular eigenvalue as $\varepsilon \rightarrow 0$, which is the signature of the pairing instability.

The crucial point is the determination of $V_0 (\phi, \phi')$ at the upper cutoff, for which one usually takes the interaction $v$ dressed at the scale $\Lambda_0$. The relevant electron-hole processes can be summed up to give (see SM\cite{SI})   
\begin{equation}
%\frac{\widehat{V}_0 (\theta, \theta')}{F(\theta )  F(\theta' )}=  U + 
V_0 (\phi, \phi') =  \frac{v_{\k-\k^\prime}}{1 + v_{\k-\k^\prime} \: \chi_{\k-\k^\prime }}
    + \frac{v_{\bm{Q}}^2 \: \widetilde{\chi}_{\k+\k^\prime }}{1 - v_{\bm{Q}} \: \widetilde{\chi}_{\k+\k^\prime }}   \;,
\label{init}
\end{equation} 
where $\k, \k^\prime$ are the respective momenta for angles $\phi, \phi'$ and $\chi_{\q}, \widetilde{\chi}_{\q}$ are particle-hole susceptibilities at momentum transfer $\q$, defined in the SM\cite{SI}.

It now remains to expand the vertex $V_0$ in the different harmonics $\cos (n\phi),\sin (n\phi)$. We illustrate here this analysis taking in particular the dispersion of the second VB represented in Fig. \ref{one}, for filling fraction $\nu = -2.4$. Similar analyses corresponding to $\nu = -2.8$ and $\nu = -3.6$ can be found in the SM\cite{SI}, showing the trend of decreasing pairing strength. 

The results of the expansion can be grouped in terms of irreducible representations of the symmetry group of the dispersion, as shown in Table \ref{table} for  $\nu = -2.4$. We observe that there are several negative eigenvalues corresponding to different harmonics (with angles measured from one of the corners of the triangle-like Fermi lines in Fig. \ref{one}). From the resolution of Eq. (\ref{scaling}), the dominant negative eigenvalue $\lambda $ leads to a pole at a critical energy scale (see SM\cite{SI})
\begin{equation}
\varepsilon_c = \Lambda_0 \: e^{-1/|\lambda |}
\end{equation}
This can be translated into the critical temperature $T_c$ of the pairing instability. At $\nu = -2.4$, the Fermi level is near the middle of the second VB shown in Fig. \ref{one}, so we can take $\Lambda_0$ as half the bandwidth ($\approx 1.5$ meV). Then, we estimate $T_c \sim 1$ K, which is consistent with the order of magnitude found in the experiments.

%When evaluating $\widehat{V}_0 (\phi, \phi')$ for a band like that in Fig. \ref{one}, the vertex has a natural expansion in terms of irreducible representations of the approximate symmetry $C_{3v}$. These can be characterized in terms of Fourier components, so that we have $A_1\rightarrow\{\cos (3n\phi)\}$, $A_2\rightarrow\{\sin (3n\phi)\}$ for integer $n$, and $E\rightarrow\{\cos (m\phi),\sin (m\phi)\}$ for $m\neq3n$. We find that there is in general a dominant pairing instability characterized by a negative coupling for the degenerate terms $\cos(\phi ), \sin(\phi )$ in the expansion of $\widehat{V}_0 (\phi, \phi')$. Besides, another negative coupling appears but with smaller magnitude for the harmonic $\cos(3\phi )$. The couplings with largest magnitude in the expansion of the vertex can be seen in Table \ref{table}, for the particular case of Hubbard repulsion $U/a_M^2 = 2.7$ meV ($a_M$ being the moir\'e lattice constant of TTG) and filling fraction of 2.4 holes per moir\'e unit cell.   

\begin{table}[t]
\begin{tabular}{c|c|c}
Eigenvalue $\lambda$  &   harmonics  &   Irr. Rep.\\
\hline \hline
2.66  &     $     1     $         &                  \\
\hline
1.80   &    \multirow{2}{*}{ $\{\cos (\phi),\sin (\phi)\}$ }    & \multirow{2}{*}{E}  \\
1.80   &                                  &                                      \\
\hline
0.65  &     $     \cos (3\phi)     $         &       $A_1$            \\
\hline
0.42   &    \multirow{2}{*}{  $\{\cos (4\phi),\sin (4\phi)\}$  }     &  \multirow{2}{*}{E}   \\
0.42   &                             &                                               \\
\hline
$-0.37$  &     \multirow{2}{*}{ $\{\cos (4\phi),\sin (4\phi)\}$ }    &  \multirow{2}{*}{E}  \\
$-0.37$   &                                 &                                         \\
\hline
$-0.37$   &     $\sin (3\phi)  $       &     $A_2$               \\                
\hline
0.22   &    \multirow{2}{*}{  $\{\cos (5\phi),\sin (5\phi)\}$  }     &  \multirow{2}{*}{E}   \\
0.22   &                                  &                         \\
\hline
0.18   &     $\sin (6\phi)  $       &     $A_2$                     
\end{tabular}
\caption{{\bf Dominant eigenvalues of the Cooper-pair vertex.} Eigenvalues of the Cooper-pair vertex with largest magnitude and dominant harmonics, grouped according to the irreducible representations of the approximate $C_{3v}$ symmetry, for the Fermi line shown in Fig. \ref{one}. The modes $\{\cos (4\phi),\sin (4\phi)\}$ appear twice in the list, as they only denote the dominant harmonic, but they actually represent different eigenvectors.} 
  \label{table}
\end{table} 

A detailed inspection shows that the nesting between parallel segments of the triangular Fermi lines for opposite spin projections (as seen in Fig. \ref{one}) is the effect behind the large magnitude of the negative couplings in Table \ref{table}. Once the Fermi line crosses to the other side of the vHS shown in Fig. \ref{two} at $\nu \approx -2.8$, the triangular patches are replaced by elliptical Fermi lines. This comes with a decrease in the magnitude of the negative couplings, leading to a substantial drop of the critical temperature (see SM\cite{SI}) which may explain why the superconductivity is suppressed in the experiments in that doping range. 

Finally, we can estimate the critical magnetic field that is needed to break up the Cooper pairs. For an in-plane field, orbital effects can be neglected and the Zeeman term will usually shift the energy of the spin up and spin down dispersions by $\pm\mu_B B$, respectively. This energy can be related to the pairing energy, giving rise to the Clogston-Chandrasekhar or Pauli limit $B_p=1.86 \: T_c$ (in Tesla for $T_c$ in Kelvin).\cite{Clogston62,Chandrasekhar62} However, due to the emergence of an imaginary hopping element between next-nearest in-plane neighbours, a Haldane flux arises which is opposite for the two spin-projections. There is thus a renormalized intrinsic spin-orbit coupling just as in the Kane-Mele model, leading to Cooper pair singlets which are polarized in out-of-plane direction. As a consequence, there is no Zeeman coupling arising from an in-plane magnetic field unless the field energy is larger than the characteristic effective spin-orbit gap $\Delta\sim1$meV, see SM. The critical field can then be estimated as $B_c=\Delta/2\mu_B \sim 8$ T, assuming the electron $g$-factor equal to 2. For $T_c \approx 2$ K, we thus find a violation of the Pauli limit by a factor 2-3, consistent with the experimental findings of Ref. \onlinecite{Cao21}. 
%Let us finally note that the effective Zeeman splitting seen in the dispersion may arise from a chirality induced effective spin-orbit coupling.\cite{Yananose21,Bahamon20}

\section*{Discussion} 
We have shown that the $e$-$e$ interaction induces a strong breakdown of spin-selective VS in TTG, with the two spin projections having opposite sign of the VS breaking order parameter.
The two spin projections are preferentially attached to opposite $K$ points, leading to an effect of spin-valley locking. In these conditions, the electrons with opposite momenta of a Cooper pair are forced to live on different Fermi lines attached to opposite valleys, giving rise to Ising superconductivity. We stress that in a conventional Ising superconductor such as NeSb$_2$, the bare spin-orbit coupling leads to spin projections perpendicular to the plane,\cite{JMLu15,SaitoYu16,XiXiaoxiang16} whereas here, a renormalized spin-orbit coupling emerges as discussed by Kane and Mele,\cite{Kane05} leading to the same effect. Thus, a weak in-plane magnetic field cannot couple to the singlet of the Cooper pair which explains the violation of the Pauli limit, as observed experimentally.   

The breakdown of VS in each spin channel leads also to a reduction of the symmetry of the bands from $C_6$ to $C_3$, as the latter is the symmetry enforced in a single valley. This enhanced anisotropy induces a strong modulation of the $e$-$e$ scattering, which is able to trigger a Kohn-Luttinger (pairing) instability, driven solely by electron interactions\cite{Kohn65,Baranov92}. The instability is here amplified by the strong nesting between the very regular triangular Fermi lines shown in Fig. \ref{one}, leading in particular to an attractive interaction in two channels corresponding to the $\sin (3\phi )$ harmonic and to the two-dimensional representation with $\{\cos (4\phi),\sin (4\phi)\}$. This mechanism of attraction is progressively weakened, however, for filling fraction $\nu < -3$ as the topology of the Fermi line changes into elliptic form (as seen around the $M$ points in the plot of Fig. \ref{two}), explaining why there is a limited range of superconductivity in the hole-doped regime of TTG.

VS breaking seems to be a ubiquitous feature in many moir\'e systems, and it is plausible that its role in the development of superconductivity may be also important in other derivatives of graphene. In this regard, it is remarkable that superconductivity has been recently found in rhombohedral trilayer graphene\cite{Zhou21,Chou21,Chatterjee21,Ghazaryan21,Dong21,Cea21,Szabo21,You21}, which is another system close to an isospin instability. It would be pertinent then to reexamine the superconductivity of such systems in the light of spin-selective VS breaking, including TBG, to confirm the connection between the enhanced anisotropy and the Kohn-Luttinger pairing instability established in this paper. 
Moreover, it should be interesting to confront preliminary results on twisted quadrilayer graphene, which make us expect an odd-even effect where the superconducting instability should be most protected in the central layer present for odd multilayers. 

\section*{Methods}
There are several Hartree-Fock studies using the continuum model for twisted bilayer \cite{Bultinck20,YiZhang20,Cea20,BiaoLian21,Bernevig21,FangXie21a} 
or trilayer\cite{FangXie21b} graphene. Here, however, we apply a self-consistent Hartree-Fock resolution in real space,\cite{Rademaker19,Seo19,Gonzalez21} which allows us to include microscopic details such as the correct Coulomb interaction between the layers or the out-of-plane interaction. For details, see the Supplemental Information.
\section*{Data availability}
The datasets generated and analyzed during the current study are available from the corresponding author on reasonable request.
\section*{Code availability}
The computer code used for the analysis and simulations in the current study are available from the corresponding author on reasonable request.

\section*{Acknowledgements}
This work has been supported by MINECO (Spain) under Grant No. FIS2017-82260-P, MICINN (Spain) under Grant No. PID2020-113164GB-I00, as well as by the CSIC Research Platform on Quantum Technologies PTI-001. The access to computational resources of CESGA (Centro de Supercomputaci\'on de Galicia) is also gratefully acknowledged.\\
\section*{Author Contribution} J.G. and T.S. jointly identified the problem, performed the analysis and wrote the paper.\\
\section*{Competing interests}
The authors declare no competing interests.

\onecolumngrid

\vspace{5cm}

\begin{center}
{\bf SUPPLEMENTAL MATERIAL}\\
{\bf Ising superconductivity induced from spin-selective valley symmetry breaking in twisted trilayer graphene}
\end{center}

\section*{Supplementary Note I.\\ Tight-binding approach for twisted trilayer graphene}

We model twisted trilayer graphene in a tight-binding approach, taking as starting point the non-interacting Hamiltonian:
\begin{align}
H_0 = -\sum_{\langle i,j\rangle} t_{\parallel} ({\bm r}_i-{\bm r}_j) \; (a_{i\sigma}^{\dagger}a_{j\sigma}+h.c.) 
   - \sum_{(i,j)} t_{\perp}({\bm r}_i-{\bm r}_j) \; (a_{i\sigma}^{\dagger}a_{j\sigma}+h.c.)\;,
\label{tbh}
\end{align}
The sum over the brackets $\langle...\rangle$ runs over pairs of atoms in the same layer, whereas the sum over the curved brackets $(...)$ runs over pairs with atoms belonging to different layers (1 to 3). $t_{\parallel} ({\bm r})$ and $t_{\perp} ({\bm r})$ are hopping matrix elements which have an exponential decay with the distance $|{\bm r}|$ between carbon atoms. A common parametrization is based on the Slater-Koster formula for the transfer integral\cite{Moon13} 
\begin{align}
-t(\d)=V_{pp\pi}(d)\left[1-\left(\frac{\d\cdot\e_z}{d}\right)^2\right]+V_{pp\sigma}(d)\left(\frac{\d\cdot\e_z}{d}\right)^2
\end{align}
with
\begin{align}
V_{pp\pi}(d)=V_{pp\pi}^0\exp\left(-\frac{d-a_0}{r_0}\right)\;,
V_{pp\sigma}(d)=V_{pp\sigma}^0\exp\left(-\frac{d-d_0}{r_0}\right)\;,
\end{align}
where $\d $ is the vector connecting the two sites, $\e_z$ is the unit vector in the $z$-direction, $a_0 $ is the C-C distance and $d_0$ is the distance between layers. A typical choice of parameters is given by $V_{pp\pi}^0=-2.7$ eV, $V_{pp\sigma}^0=0.48$ eV and $r_0=0.319 a_0$ \cite{Moon13}. 

\begin{figure*}
\includegraphics[width=0.4\columnwidth]{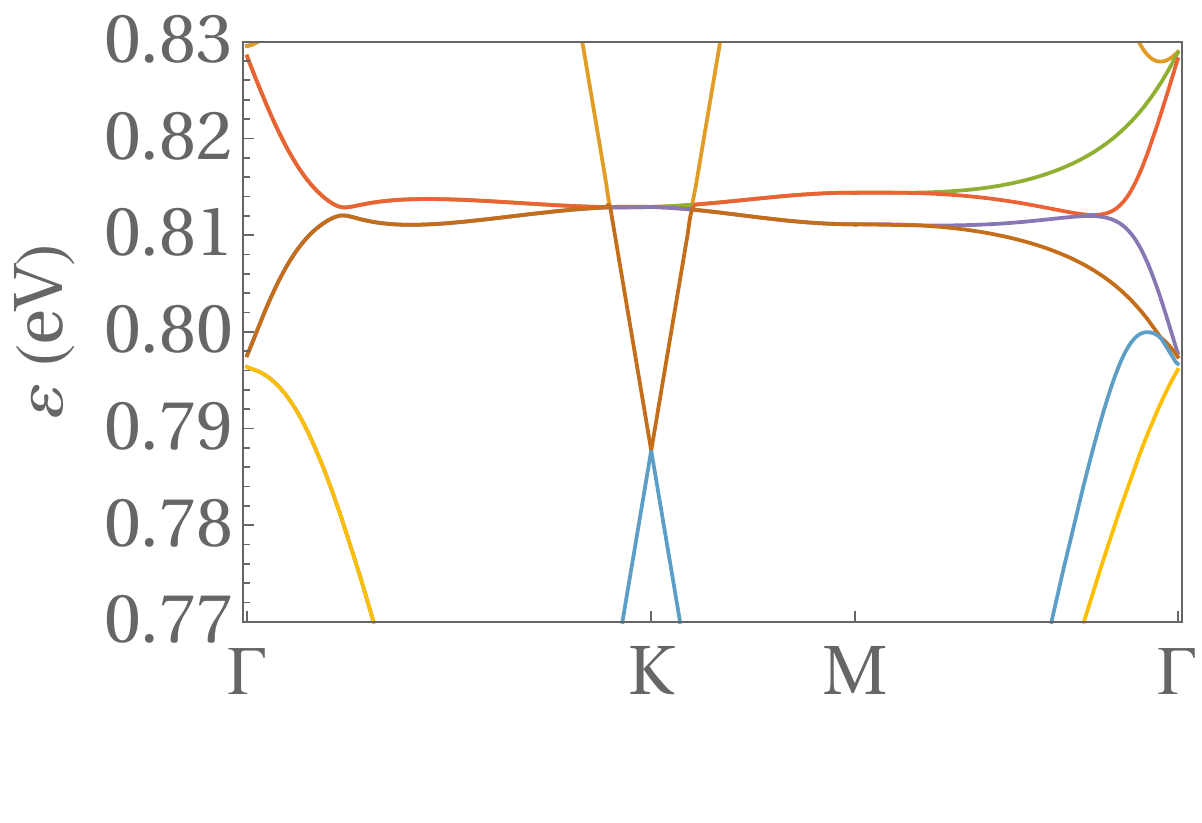}
\caption{Dispersion of the first valence and conduction bands about the charge neutrality point of twisted trilayer graphene with twist angle $\theta \approx 1.61^\circ$, obtained in a tight-binding approach with no out-of-plane corrugation.}

\label{norlx}
\end{figure*}

In practice, we have taken the above values to carry out the analysis reported in the main text. We have chosen a configuration of twisted trilayer graphene belonging to the set of commensurate superlattices also realized by twisted bilayer graphene, with a twist angle $\theta \approx 1.61^\circ$ (7566 atoms in the moir\'e unit cell) very close to the magic angle condition. At a first stage without out-of-plane relaxation, the tight-binding approach applied to this model leads to the low-energy bands shown in Fig. \ref{norlx} about the charge neutrality point.

At the twist angles considered in the paper, the in-plane lattice relaxation of twisted trilayer graphene does not have the important role that it plays at the magic angle of the twisted bilayer. However, the out-of-plane corrugation of the trilayer is a relevant effect, which arises from the dependence of the interlayer interaction on the stacking of the graphene layers. Thus, the lattice structure tends to relax in the out-of-plane direction, reaching a minimum interlayer distance in the regions of $AB$ stacking, and a maximum value in the regions of $AA$ stacking. To describe the interlayer interaction we have used a Kolmogorov-Crespi potential\cite{Kolmogorov00,Kolmogorov05} 
\begin{align}
U(z) = -A \left( \frac{z_0}{z} \right)^6 + C e^{-\lambda (z-z_0)}
\label{kc}
\end{align}    
where the first term stands for the van der Waals attraction and the second term accounts for an exponentially decaying repulsion due to the interlayer wave-function overlap. The effect of the registry of the carbon atoms is included in the second term, and we have adjusted it to interpolate between the different interaction energies in the regions of $AB$ and $AA$ stacking. In the relaxed structure we have left the central layer intact, so that the separation of the outer layers about the center becomes modulated across the superlattice according to the potential (\ref{kc}), reaching a minimum interlayer distance of 0.334 nm for $AB$ stacking and a maximum distance of 0.356 nm for $AA$ stacking. 

\begin{figure*}
\includegraphics[width=0.4\columnwidth]{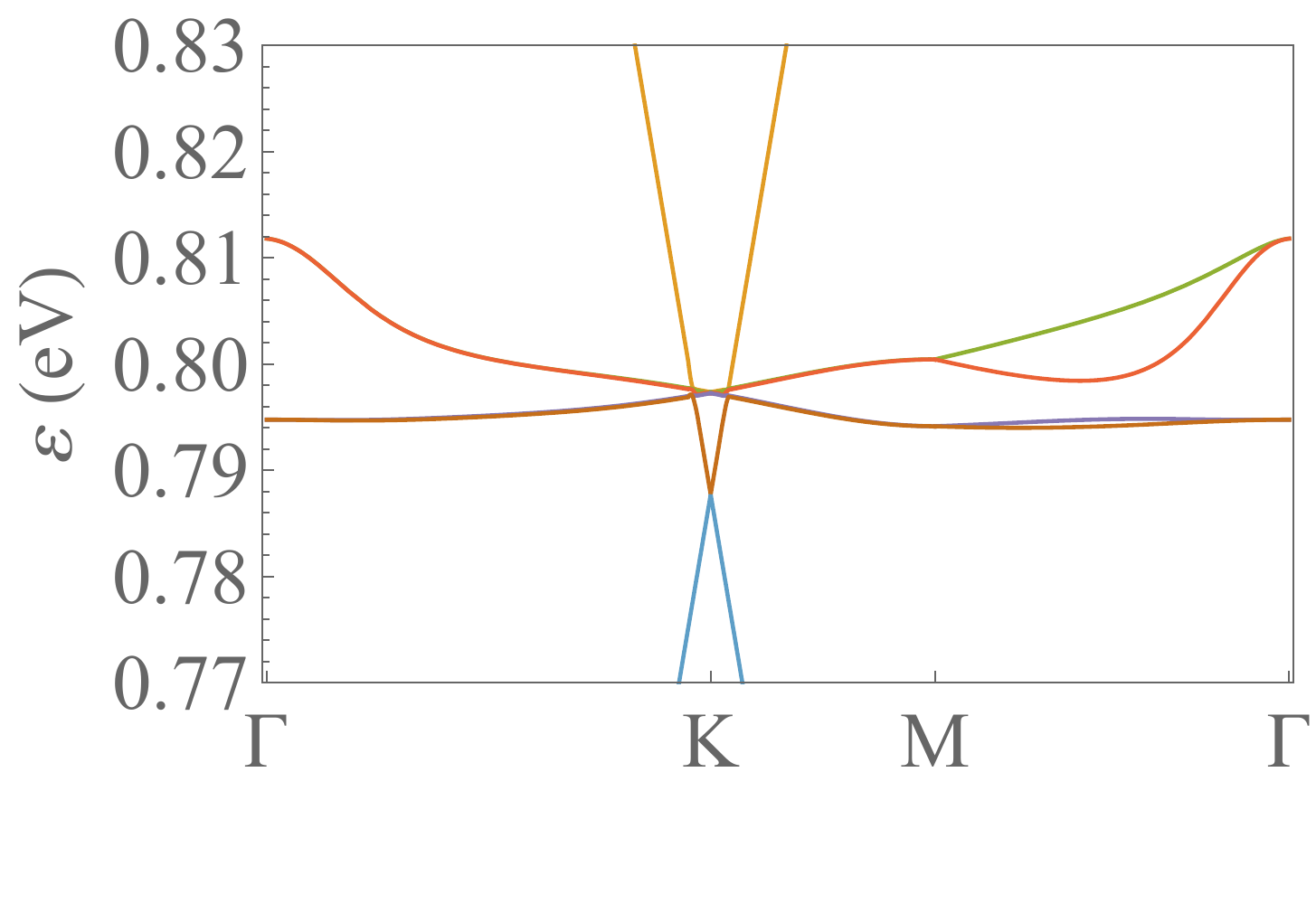}
\hspace{1cm}
\includegraphics[width=0.4\columnwidth]{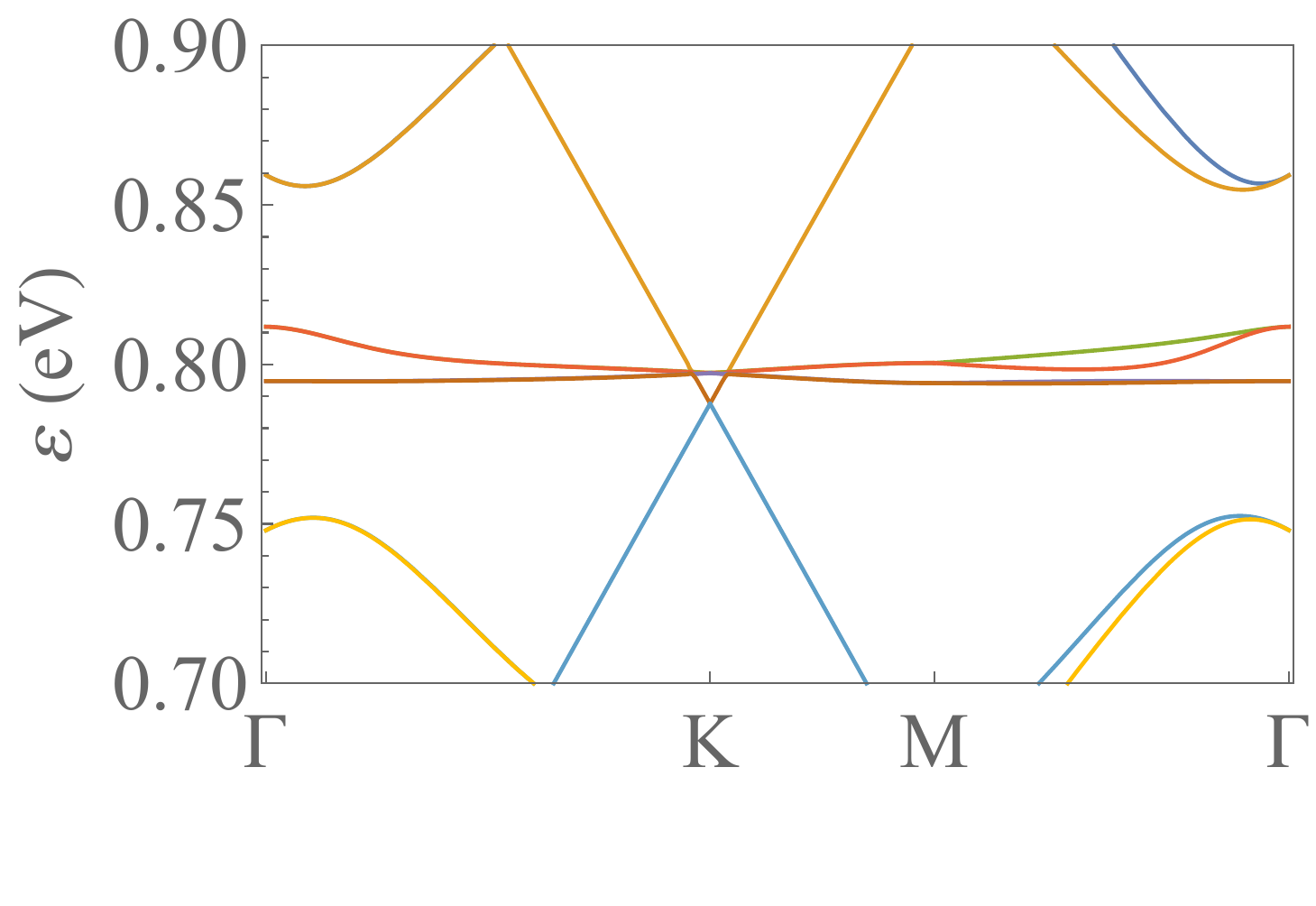}
\\
 \hspace*{1.0cm} (a) \hspace{7.7cm} (b)
\caption{Dispersion of the first valence and conduction bands (zoomed out in (b)) about the charge neutrality point of twisted trilayer graphene with twist angle $\theta \approx 1.61^\circ$, computed in a tight-binding approach with parameters given in the text and accounting for out-of-plane relaxation.}
\label{vcb}
\end{figure*}

Overall, including out-of-plane relaxation, our tight-binding approach leads to sensible results for the commensurate lattice studied in the main text with twist angle $\theta \approx 1.61^\circ$, whose first valence and conduction bands are shown in Fig. \ref{vcb}.

\section*{Supplementary Note II.\\ Hartree-Fock approximation}

In our microscopic approach, we may consider two different sources of electronic interaction, corresponding to the extended Coulomb interaction and the on-site (Hubbard) repulsion of electrons at the same carbon site. The first of them gives rise to a contribution to the interaction Hamiltonian which can be written in terms of creation (annihilation) operators $a_{i\sigma}^+$ ($a_{i\sigma}$) for electrons at each carbon site $i$ with spin $\sigma$
\begin{align}  
H_{\rm C} = \frac{1}{2} \sum_{i,j,\sigma,\sigma'} a_{i\sigma}^{\dagger}a_{i\sigma} \: v (\rr_i-\rr_j) \: a_{j\sigma'}^{\dagger}a_{j\sigma'}\;,
\label{coul}
\end{align}
We consider a form of the Coulomb potential $v$ which is adapted to the case where twisted trilayer graphene is surrounded by top and bottom metallic gates. The starting point is the unscreened Coulomb potential $v_0 (\r ) =  e^2/4\pi \epsilon r$, $\epsilon $ being the dielectric constant. 
In the presence of a gate at distance $z = \xi /2$, the electrostatic energy of two electrons lying in a plane parallel to the electric gate and being separated by a distance $r$ is given by
\begin{align}
v(\r ) =  \frac{e^2}{4\pi \epsilon} \left( \frac{1}{r} - \frac{1}{\sqrt{r^2 + \xi^2}}  \right)\;.
\end{align}
In the presence of an additional opposite gate also at distance $z = \xi /2$, and again using the  image-charge technique, one obtains for the electrostatic energy\cite{Throckmorton12}
\begin{align}
v(\r ) 
     & =  \frac{e^2}{4\pi \epsilon} \sum_{n=-\infty}^{\infty} \frac{(-1)^n}{\sqrt{r^2 + n^2 \xi^2}}    \to    \frac{e^2}{4\pi \epsilon}   \frac{2\sqrt{2} \: e^{-\pi r/\xi }}{\xi \sqrt{r/\xi }}\;.
\label{vscr}
\end{align}
In the main text, we have used the approximate expression in Eq. (\ref{vscr}), which is very accurate for $r/\xi\gtrsim0.2$. We have addressed the particular case of a setup with $\xi = 10$ nm. For this screening length, the use of the expression in Eq. (\ref{vscr}) does not modify the shape of the interacting flat bands, while determining correctly the phases and the position of the critical point for symmetry breaking and gap opening at 2-hole doping.

%In the main text, we have used the approximate formula valid for $r/\xi\gtrsim0.2$ and addressed the particular case of a setup with $\xi = 10$ nm. For this screening length, there are hardly differences for the band structure as well as for the phase diagram between the exact and approximate formula.

Moreover, we take also into account the Hubbard interaction, which can be seen as a regularization of the interaction in Eq. (\ref{coul}) when $\rr_i = \rr_j$. This leads to a contribution to the interaction Hamiltonian
\begin{align}  
H_{\rm U} =  U \: \sum_{i} a_{i\uparrow}^{\dagger}a_{i\uparrow}  \: a_{i\downarrow}^{\dagger}a_{i\downarrow}\;.
\end{align}
This on-site repulsion is actually the spin-dependent part of the interaction and, in this respect, it plays an important role as it helps to stabilize the iterative resolution of the self-consistent Hartree-Fock equations. For that purpose, we have taken a not too large value of the Hubbard repulsion, $U = 0.5$ eV, which is also a way of compensating the strong tendency of the Hartree-Fock approximation to overestimate the ferromagnetic instabilities arising from the spin-dependent interaction.

The self-consistent Hartree-Fock equations take then the form
\begin{align}
\sum_a \varepsilon_{a\sigma} \: \phi_{a\sigma} (\r_i)  \phi_{a\sigma}^* (\r_j) = & \sum_a \varepsilon_{a\sigma}^0  \: \phi_{a\sigma}^0 (\r_i)  \phi_{a\sigma}^0 (\r_j)    
    + \mathbb{I}_{ij} \:  U  \sideset{}{'}\sum_a  \left|\phi_{a \: -\sigma} (\r_i)\right|^2          \notag    \\
   & +  \mathbb{I}_{ij} \:  \sideset{}{'}\sum_a  \sum_{l, \sigma' } v (\r_i-\r_l)   \left|\phi_{a\sigma'} (\r_l)\right|^2  
    - v (\r_i-\r_j)  \sideset{}{'}\sum_a \phi_{a\sigma} (\r_i)  \phi_{a\sigma}^* (\r_j)
\label{schf}
\end{align}
where $\varepsilon_{a\sigma} \: (\varepsilon_{a\sigma}^0)$ and $\phi_{a\sigma} (\r_i) \: (\phi_{a\sigma}^0 (\r_i))$ represent respectively the eigenvalues and eigenvectors building the interacting (free) electron propagator, and the prime means that the sum is to be carried over the occupied levels \cite{Fetter71}. In our notation, $-\sigma$ represents the spin projection with opposite orientation to the spin $\sigma$. 

In Eq. (\ref{schf}) we already see that, if the set $\{ \phi_{a\sigma} (\r_i)  \}$ is a self-consistent solution for a given spin projection, the set $\{ \phi_{a\sigma'}^* (\r_i)  \}$ is an equally good solution. If the on-site repulsion $U$ were not present in the equations, then we could assemble a solution with $\{ \phi_{a\sigma} (\r_i)  \}$ and $\{ \phi_{a\sigma'}^* (\r_i)  \}$ corresponding to two different (arbitrary) spin orientations. The operation of complex conjugation implies the exchange of the two valleys of the twisted trilayer, so this possibility of choosing arbitrary spin orientations would mean the freedom to make independent spin rotations in each valley. 

In any event, the on-site Hubbard repulsion is not vanishing, and this forces possible solutions of Eq. (\ref{schf}) to have opposite spin projections. When there is spin-selective valley symmetry breaking, a combined solution of the form $\{ \phi_{a\sigma} (\r_i) , \phi_{a \: -\sigma}^* (\r_i) \}$ corresponds then to having a nonvanishing value of the valley polarization order parameter for one of the spins, and the opposite value for the opposite spin projection. This is the basis of the spin-valley locking mechanism discussed in the main text.

Turning to technical questions, the construction of the self-energy in Eq. (\ref{schf}) demands the knowledge of a relevant set of eigenvectors of the Hamiltonian. That self-energy is defined as the sum over all the occupied states in the electronic bands, but in practice one has to impose some kind of truncation when carrying out the calculation. In this respect, we have retained the first 51 valence bands in the self-consistent resolution. 
	
Moreover, we have adopted a mixed representation of the electronic states by performing a Fourier transform passing to momenta ${\bf k}$ in the superlattice of the twisted trilayer. That is, we build the electron operators as
\begin{eqnarray}
a_{n,i,\sigma} = \frac{1}{\sqrt{N_c}} \sum_{{\bf k} \in BZ} a({\bf k})_{i,\sigma} e^{i{\bf k} \cdot ({\bf r}_i + {\bf R}_n) }  
\end{eqnarray}
where ${\bf r}_i$ are the coordinates of the carbon atoms in the supercell, ${\bf R}_n$ are lattice vectors in the superlattice of the twisted trilayer, and the sum is over momenta in the Brillouin zone of the superlattice ($N_c$ is the number of unit cells). In practice, we compute the self-energy taking a grid with 192 momenta (plus the Gamma point) covering the Brillouin zone. We have checked that such a content of states is safe to capture the relevant symmetry-breaking patterns of twisted trilayer graphene, as well as to obtain a sensible description of its low-energy bands.

\section*{Supplementary Note III.\\ Internal screening and dielectric constant}

An important question in the discussion of the electronic properties is the determination of the dielectric constant $\epsilon $ to be used for twisted trilayer graphene. The magnitude of that quantity depends mainly on the internal screening of the Coulomb interaction, which becomes rather intense as a consequence of the reduced bandwidth of the lowest-energy valence and conduction bands. A good estimate of the dielectric constant can be obtained from the dielectric function $\epsilon ({\bf q}, \omega)$, which can be computed in the RPA for the two-dimensional Coulomb interaction as
\begin{align}
\epsilon ({\bf q}, \omega) = 1 + \frac{e^2}{2 \epsilon_0 |{\bf q}|} \chi({\bf q}, \omega)
\label{eps}
\end{align} 
where $\chi({\bf q}, \omega)$ stands for the particle-hole susceptibility. We are going to be interested in the effects of internal screening at length scales of the order of the size of the supercell of the twisted trilayer, as the interaction is already screened at long distances by the presence of metallic gates in our model. Then, we can estimate the magnitude of the dielectric constant from the values of the dielectric function $\epsilon ({\bf q}, 0)$ at momenta ${\bf q}$ of the order of the inverse of the lattice constant of the superlattice.

\begin{figure*}
\includegraphics[width=0.3\columnwidth]{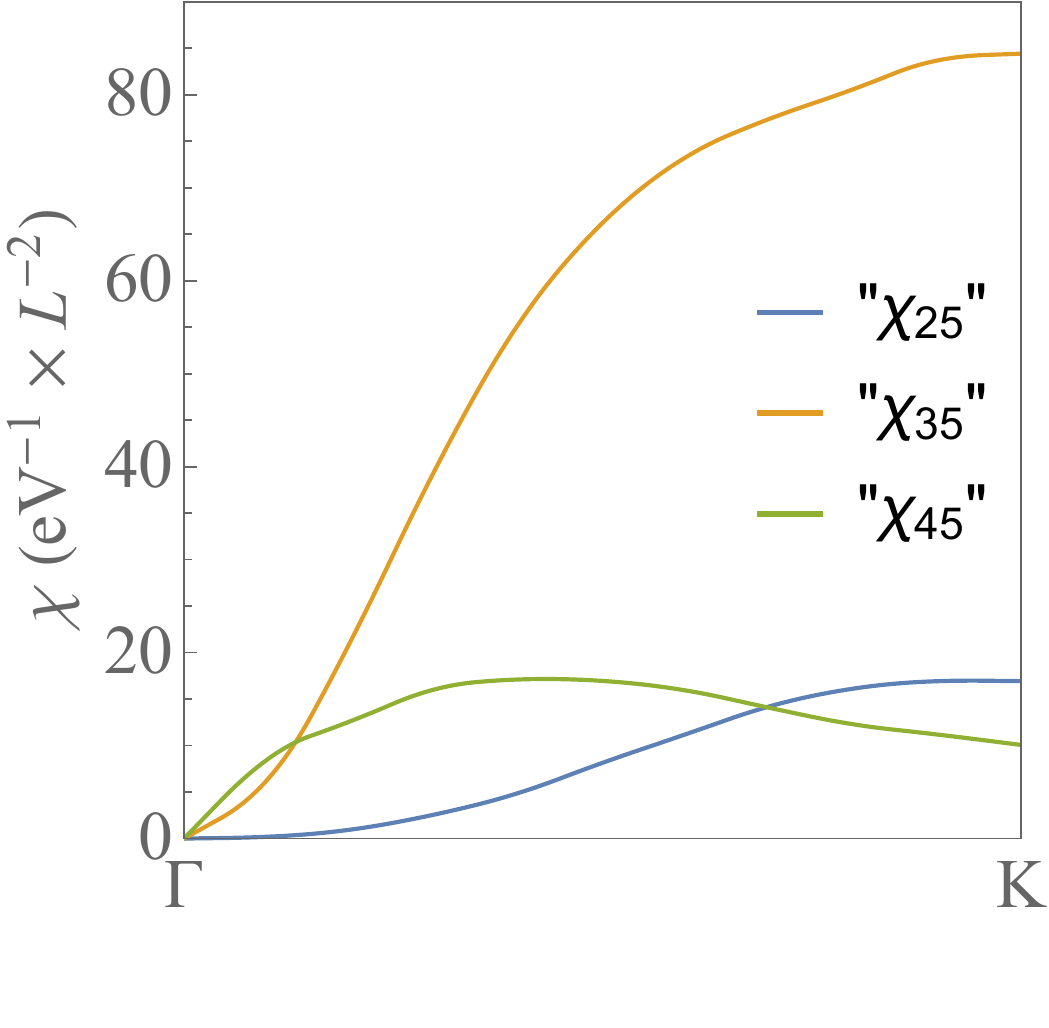}
\hspace{1cm}
\includegraphics[width=0.3\columnwidth]{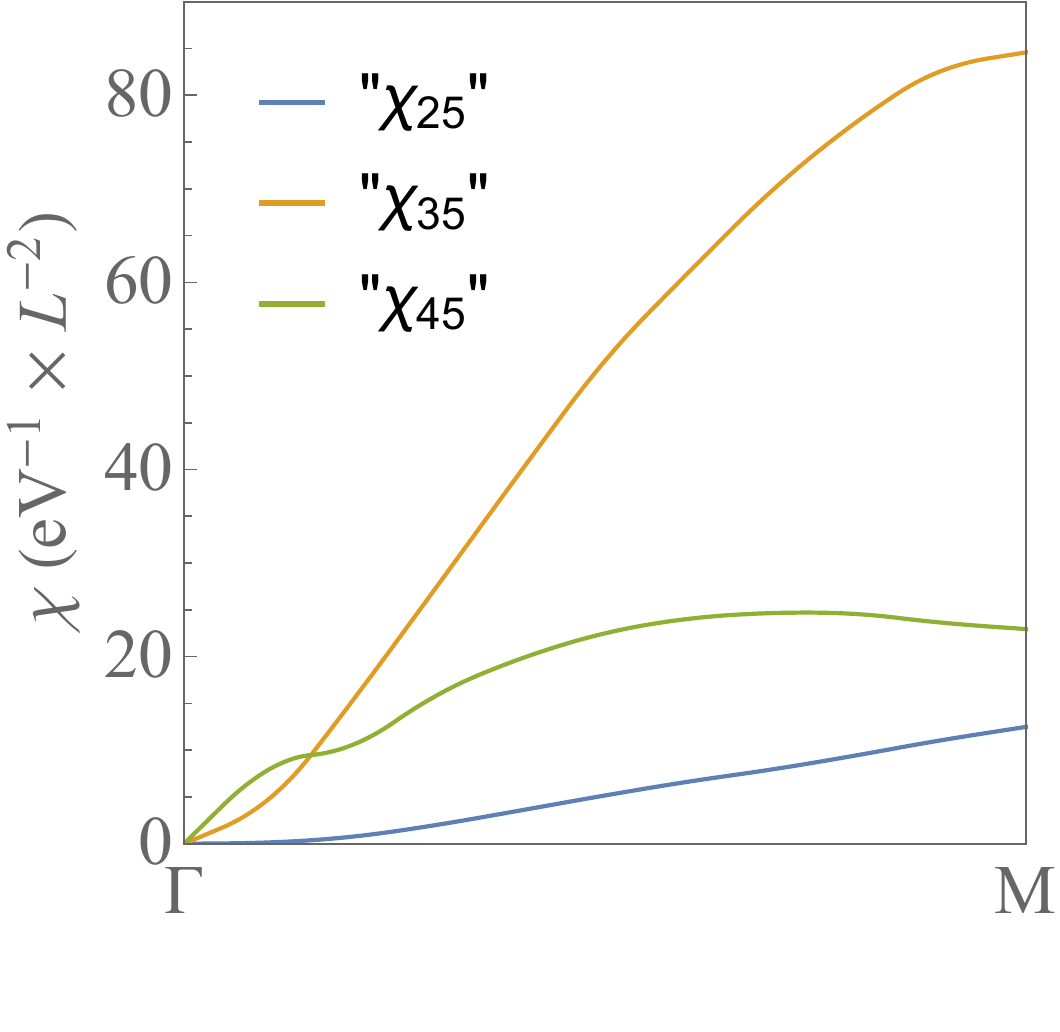}
\\
 \hspace*{1.0cm} (a) \hspace{7.7cm} (b)
\caption{Evolution along the $\Gamma K$ (a) and the $\Gamma M$ line (b) of the particle-hole susceptibility computed with the bands of the interacting theory represented in Fig. 3 of the main text. The plot shows the partial contributions $\chi_{ij}$ between the three flat conduction bands ($i=2,3,4$) and the flat valence band ($j=5$). The susceptibility is measured in units of eV$^{-1} \times L^{-2}$, where $L$ is the lattice constant of the moir\'e superlattice.}
\label{suscep}
\end{figure*}

In order to make a reliable estimate of the internal screening, we compute the dielectric function with the bands of the interacting theory represented in Fig. 3 of the main text. This is a suitable situation, as there is a gap separating the flat valence and conduction bands around the Fermi level. Then, we will be able to check the consistency of our estimates by comparing them with the actual value of the dielectric constant used to obtain the bands in the mentioned figure. The largest contributions to $\epsilon ({\bf q}, 0)$ come indeed from the four flat bands around the Fermi level. We can approximate the susceptibility by considering particle-hole excitations between the three flat conduction bands (numbered as $i=2,3, 4$) and the flat valence band below the Fermi level (numbered as $j=5$). These lead to the partial contributions shown as $\chi_{ij}$ in Fig. \ref{suscep} along the directions $\Gamma K$ and $\Gamma M$.

We have for the dielectric function computed for instance at the large momentum ${\bf Q}_K$ of the $K$ point in the moir\'e Brillouin zone
\begin{align}
\epsilon ({\bf Q}_K, 0)& = 1 + \frac{e^2}{2 \epsilon_0 |{\bf Q}_K|} \chi({\bf Q}_K, 0)   \\
       & = 1 + \frac{3}{8 \pi \epsilon_0} e^2  \:  L  \: \chi({\bf Q}_K, 0)   
\label{epskp}       
\end{align}
where $L$ stands for the lattice constant of the moir\'e superlattice. We take $e^2/\epsilon_0 \approx 17.7$ eV nm and the length $L \approx 8.46$ nm for a twisted trilayer belonging to the sequence of commensurate superlattices with $\theta \approx 1.61^\circ$. The particle-hole susceptibility can be obtained by adding the different contributions from Fig. \ref{suscep}. Taking into account the spin degeneracy, we get the estimate
\begin{align}
\epsilon ({\bf Q}_K, 0)  \approx  55
\label{est}
\end{align}       
Interestingly, we obtain a very similar magnitude if we carry out the estimate at the $M$ point of the moir\'e Brillouin zone, taking again the values for the particle-hole susceptibility from the curves shown in Fig. \ref{suscep}.

\begin{figure*}[h!]
\includegraphics[width=0.3\columnwidth]{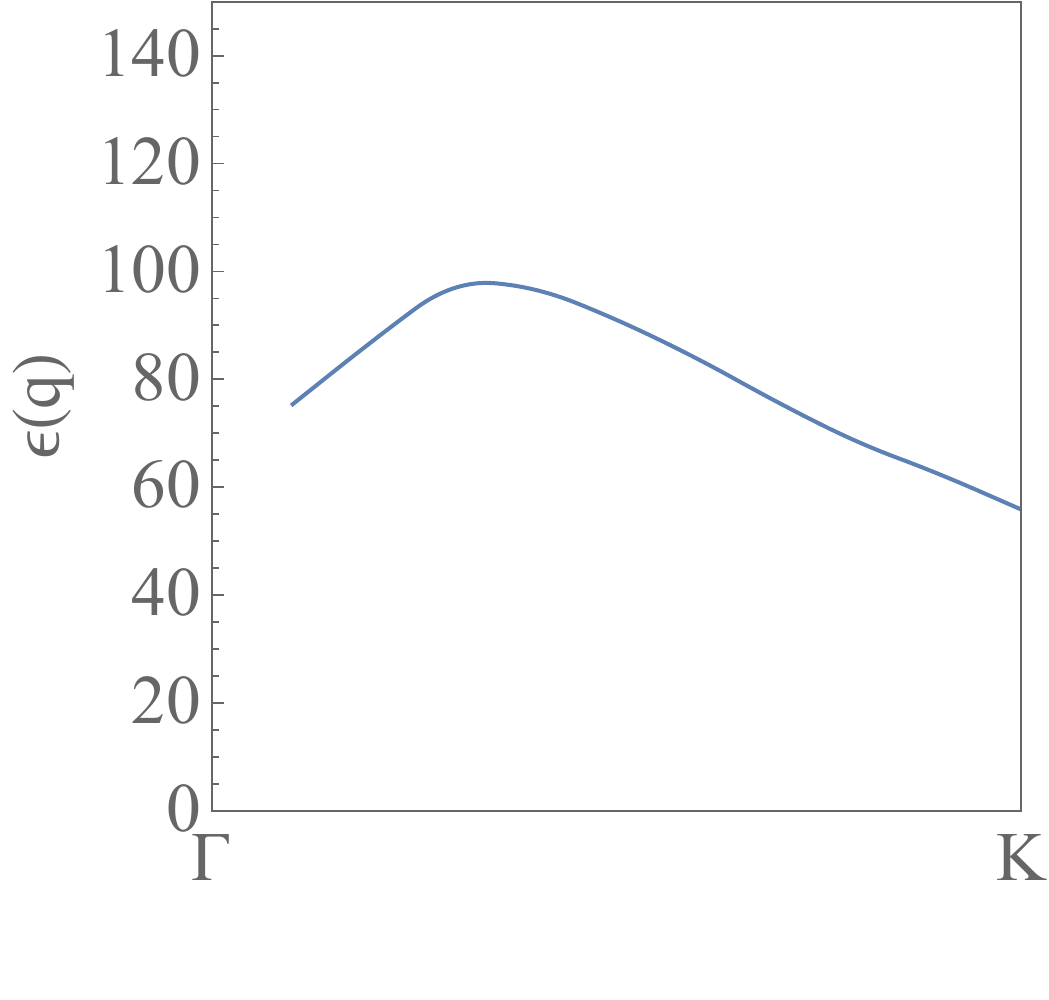}
\hspace{1cm}
\includegraphics[width=0.3\columnwidth]{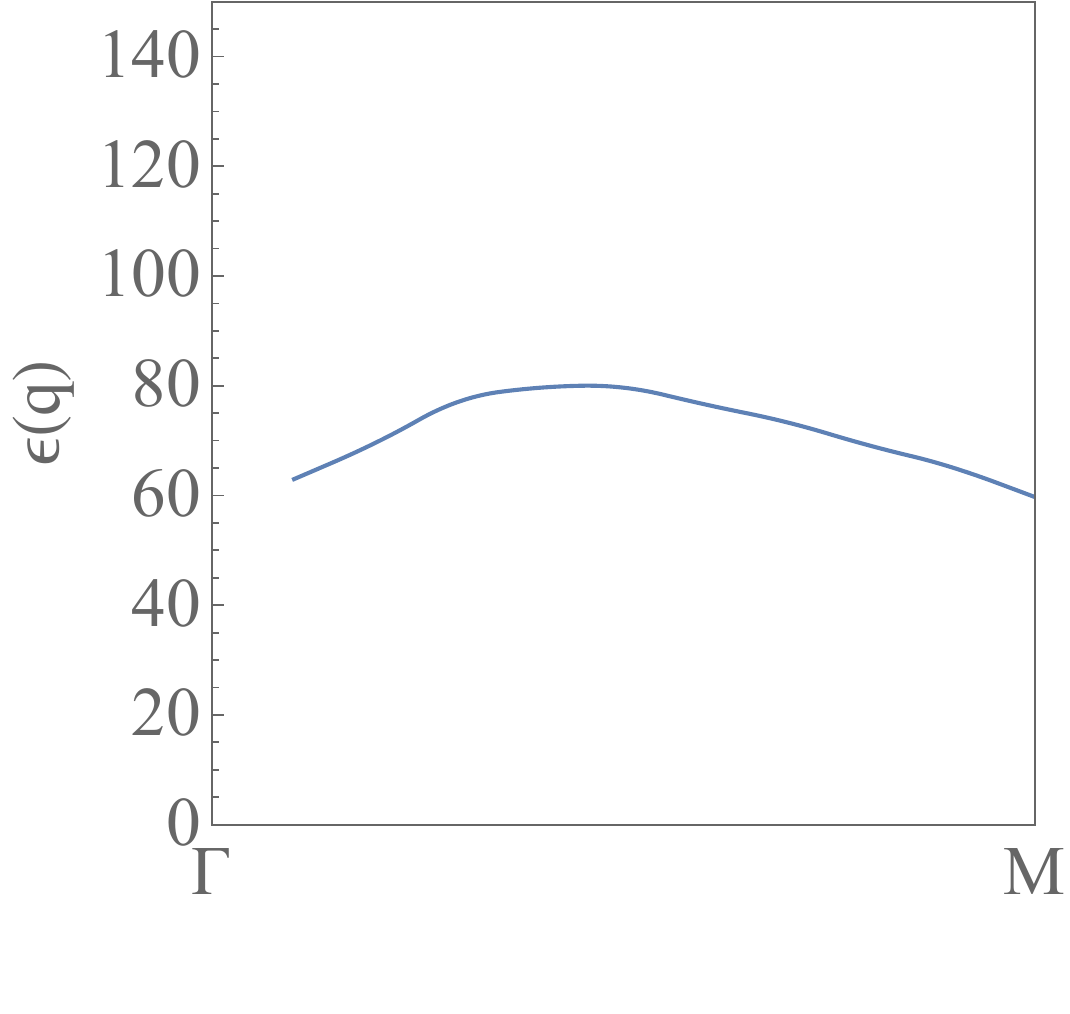}
\\
 \hspace*{1.0cm} (a) \hspace{7.7cm} (b)
\caption{Evolution along the $\Gamma K$ (a) and the $\Gamma M$ line (b) of the dielectric function computed with the bands of the interacting theory represented in Fig. 3 of the main text.}
\label{diel}
\end{figure*}

The full shape of the dielectric function, computed with the procedure just outlined, is represented along the $\Gamma K$ and $\Gamma M$ directions in Fig. \ref{diel}. As already mentioned, we take the Coulomb interaction in our calculations with a screening length $\xi = 10$ nm from the presence of metallic gates. This means that we can encode the internal screening in the form of an effective dielectric constant by looking at momenta not much smaller than the inverse of the lattice constant of the superlattice. As seen in Figs. \ref{diel}(a)-(b), the behavior of the dielectric function is rather smooth towards the boundary of the moir\'e Brillouin zone. The values of $\epsilon ({\bf q}, 0)$ at ${\bf Q}_K$ and ${\bf Q}_M$ are actually lower bounds, so it makes sense to take them as conservative estimates of the dielectric constant. Moreover, they turn out to be only slightly larger than the value used to obtain the bands in Fig. 3 of the main text, which shows the consistency of our determination of the internal screening in the model.

We remark that the effect of screening from the dielectric environment can be included in the above computation, but making almost no difference in the final result, as long as the value of the dielectric constant in (\ref{est}) is much larger than any typical dielectric constant $\epsilon_{env}$ of the substrate. That is, one can introduce $\epsilon_{env}$ instead of $\epsilon_0$ in the above derivation, but this would reduce correspondingly the effect of the particle-hole susceptibility, leading to the cancellation of $\epsilon_{env}$ in the final expression for the effective Coulomb potential  $e^2/2 \epsilon |{\bf q}|$, with the value of $\epsilon $ already given by (\ref{est}).

\section*{Supplementary Note IV.\\ Order parameters and phase diagram of twisted trilayer graphene}

In the self-consistent Hartree-Fock resolution, an important role is played by the matrix elements
\begin{align}
h_{ij}^{(\sigma )} =  \sideset{}{'}\sum_a \phi_{a\sigma} (\r_i) \phi_{a\sigma}^* (\r_j)\;.
\end{align}
where $\phi_{a\sigma} (\r_i)$ stand for the eigenvectors of the different states labeled by $a$ and the spin $\sigma $, and depending on the atomic positions $\r_i$. The prime means that the sum is to be carried over the occupied levels. These matrix elements become also very useful in the definition of the order parameters for broken-symmetry phases. This is due to the fact that they coincide with the values of the equal-time propagator for the electron operators $a_{i\sigma}$. It can be actually shown that
\begin{align}
\langle a_{j\sigma}^+ (t) a_{i\sigma} (t) \rangle = \sideset{}{'}\sum_a \phi_{a\sigma} (\r_i) \phi_{a\sigma}^* (\r_j)
\end{align} 
This means that different charge densities as well as hopping amplitudes can be written in terms of $h_{ij}^{(\sigma )}$.

\begin{figure}[t]
\includegraphics[width=0.5\columnwidth]{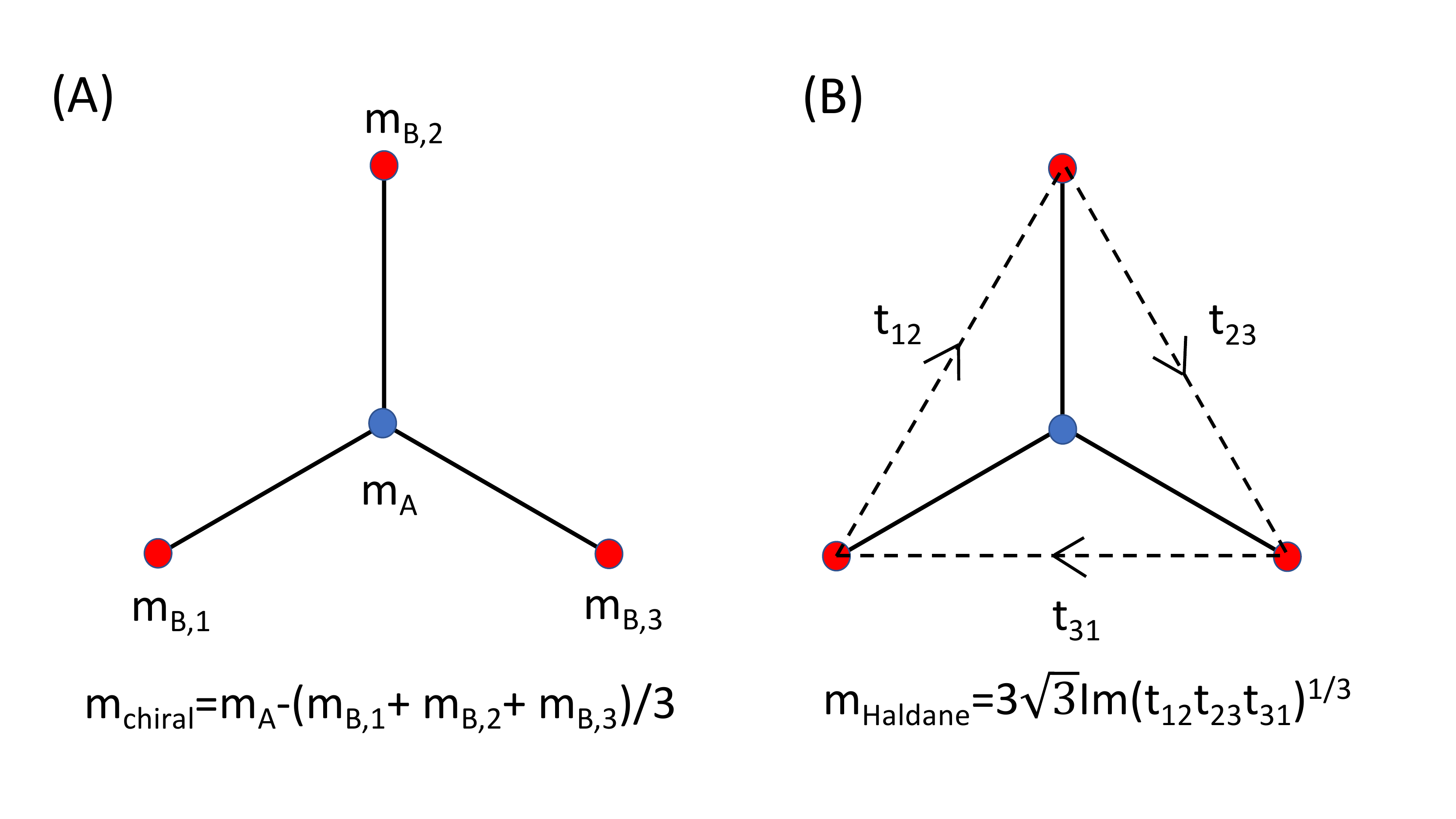}
\caption{Schematic definition of the two main symmetry breaking patterns opening a gap in the honeycomb lattice: (A) chiral symmetry breaking leading to a Dirac mass (where we have replaced $h_{ii}$ defined in the text by $m$), (B) time-reversal symmetry breaking leading to the Haldane mass (where we have replaced $h_{ij}$ defined in the text by $t_{ij}$).
\label{order}}
\end{figure}

The main charge instability corresponds indeed to a mismatch in the charge densities for the two different sublattices $A$ and $B$ in each graphene layer. This leads to chiral symmetry breaking, with the opening of a gap between the low-energy Dirac cones at the charge neutrality point. Locally, the order parameter is given by the charge asymmetry between each carbon atom and its nearest neighbors, as represented in Fig. \ref{order}. Globally, the order parameter is defined by the quantity
\begin{align}
C^{(\sigma )} = & \sum_{i \in A} h_{ii}^{(\sigma )}  - \sum_{i \in B} h_{ii}^{(\sigma )} 
\label{chiral}
\end{align}

The other way to open a gap between the low-energy Dirac cones consists in producing an effective magnetic flux at each atomic site, which can be assessed by adding the phases of the hopping matrix elements between nearest neighbors $i_1, i_2$ and $i_3$ of each atom $i$, as represented in Fig. \ref{order}. The effective flux leads to time-reversal and parity symmetry breaking, conferring a so-called Haldane mass to the low-energy Dirac fermions. The order parameter for this broken-symmetry phase is given by   
\begin{align}
P_+^{(\sigma)} &= {\rm Im} \left( \sum_{i \in A}  \left(   h_{i_1 i_2}^{(\sigma )} h_{i_2 i_3}^{(\sigma )} h_{i_3 i_1}^{(\sigma )}   \right)^{\frac{1}{3}}
          + \sum_{i \in B}  \left(   h_{i_1 i_2}^{(\sigma )} h_{i_2 i_3}^{(\sigma )} h_{i_3 i_1}^{(\sigma )}   \right)^{\frac{1}{3}}   \right)
\label{hald}
\end{align}
where the nearest neighbors $i_1, i_2$ and $i_3$ are always taken with a definite orientation.

Furthermore, there is also the possibility of having an effective magnetic flux but preserving parity, which is realized by attaching opposite fluxes at atoms belonging to different sublattices $A$ and $B$. The order parameter characterizing this phase is 
\begin{align}
P_-^{(\sigma)} &= {\rm Im} \left( \sum_{i \in A}  \left(   h_{i_1 i_2}^{(\sigma )} h_{i_2 i_3}^{(\sigma )} h_{i_3 i_1}^{(\sigma )}   \right)^{\frac{1}{3}}
          - \sum_{i \in B}  \left(   h_{i_1 i_2}^{(\sigma )} h_{i_2 i_3}^{(\sigma )} h_{i_3 i_1}^{(\sigma )}   \right)^{\frac{1}{3}}   \right)
\label{valley}
\end{align}
In the continuum theory of Dirac fermions, it can be shown that this breakdown of symmetry translates into the generation of a term proportional to the identity in pseudospin space. This does not open a gap in the Dirac cones at the $K$ point, but the shift in the energy of the cones becomes different in the two valleys of the electron system. The main effect corresponds therefore to spin-selective valley symmetry breaking, which is indeed a ubiquitous feature in graphene multilayers away from the charge neutrality point.

\begin{figure*}[h!]
\includegraphics[width=0.45\columnwidth]{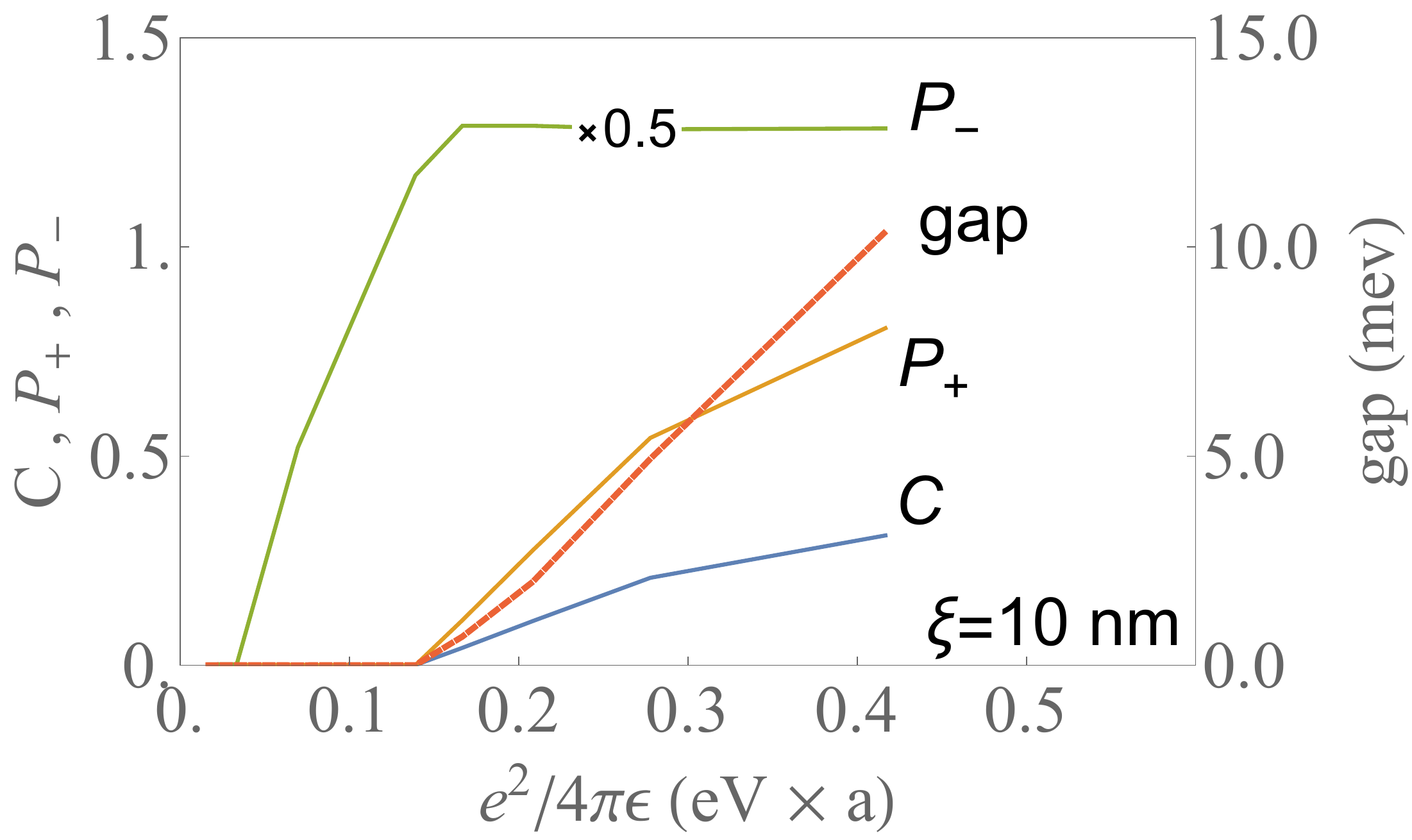}
\caption{Phase diagram showing the different order parameters of symmetry breaking at a filling fraction of 2 holes per moir\'e unit cell in twisted trilayer graphene, obtained by means of the self-consistent Hartree-Fock approximation with screening length $\xi = 10$ nm for the extended Coulomb potential. The interaction strength is measured in units of eV times the C-C distance $a$.}
\label{phd}
\end{figure*}

The evolution of the order parameters (\ref{chiral})-(\ref{valley}) can be studied as the strength of the extended Coulomb interaction is varied. The most interesting instance corresponds to a filling fraction of 2 holes per moir\'e unit cell. Then the dominant pattern of symmetry breaking corresponds to $P_-^{(\sigma)}$, while $P_+^{(\sigma)}$ and $C^{(\sigma )}$ open up beyond a certain interaction strength. This is illustrated in the phase diagram shown in Fig. \ref{phd}, which is the result of applying the self-consistent Hartree-Fock approximation for the screening length $\xi=10$nm. With our estimate of the dielectric constant (\ref{est}), we have $e^2/4 \pi \epsilon  \approx  0.22$ eV$\times  a$ (where $a$ is the C-C distance). This places the interaction in a regime where, apart from spin-selective valley symmetry breaking, there is also a breakdown of time-reversal symmetry leading to a Chern insulator phase. The origin of this phase lies in the fact that, at the filling fraction of 2-hole doping, valley symmetry breaking for each spin channel sets the Fermi level at the vertices of the Dirac cones of the lower valley. The Dirac nodes may then be destabilized for a sufficiently strong interaction, mainly due to the appearance of the Haldane mass. This explains the opening of the gap at 2-hole doping, which is the relevant instance discussed in the main text.

We finally comment on the possibility of having order parameters which reflect in the form of microscope structure in the twisted trilayer, at the level of the hexagonal lattices of the layers. In this respect, we have checked that the fluxes that make up $P_+^{(\sigma)}$ and $P_-^{(\sigma)}$ do not show any microscopic pattern in the graphene lattices, for all interactions strengths ranging from small to large values of the dielectric constant $\epsilon$. A particular instance is illustrated at the end of this Supplemental Material, where one can see that the fluxes in the microscopic triangular loops have a smooth envelop across the unit cell of the moir\'e superlattice.

However, it is more interesting the case of the order parameter for the so-called K-intervalley coherence, which has been discussed at length in Ref. \cite{Bultinck20} for magic-angle twisted bilayer graphene. This order parameter takes the form of a magnetization density wave at the wave vector $K$ of graphene, with circulating currents along the hexagonal rings combining into a typical kekul\'e pattern which triples the graphene unit cell. In our microscopic approach, we can characterize such an order parameter by measuring the flux enclosed in the six-fold rings made of consecutive nearest-neighbors sites $i_1$ to $i_6$ in the graphene lattice (with a fixed orientation). We then define the quantity
\begin{align}
P_{\rm KIVC}^{(\sigma)} ({\bm r}_i) &=   h_{i_1 i_2}^{(\sigma )} h_{i_2 i_3}^{(\sigma )} h_{i_3 i_4}^{(\sigma )}  
      h_{i_4 i_5}^{(\sigma )} h_{i_5 i_6}^{(\sigma )} h_{i_6 i_1}^{(\sigma )} 
\end{align}
This allows us to capture the signature of K-intervalley coherence by looking for microscopic structure in the angle $\theta_{\rm KIVC}$ given by 
\begin{align}
P_{\rm KIVC}^{(\sigma)} &=  | P_{\rm KIVC}^{(\sigma)} |  e^{i \theta_{\rm KIVC}}  
\end{align}

\begin{figure*}[h!]
	\includegraphics[width=0.22\columnwidth]{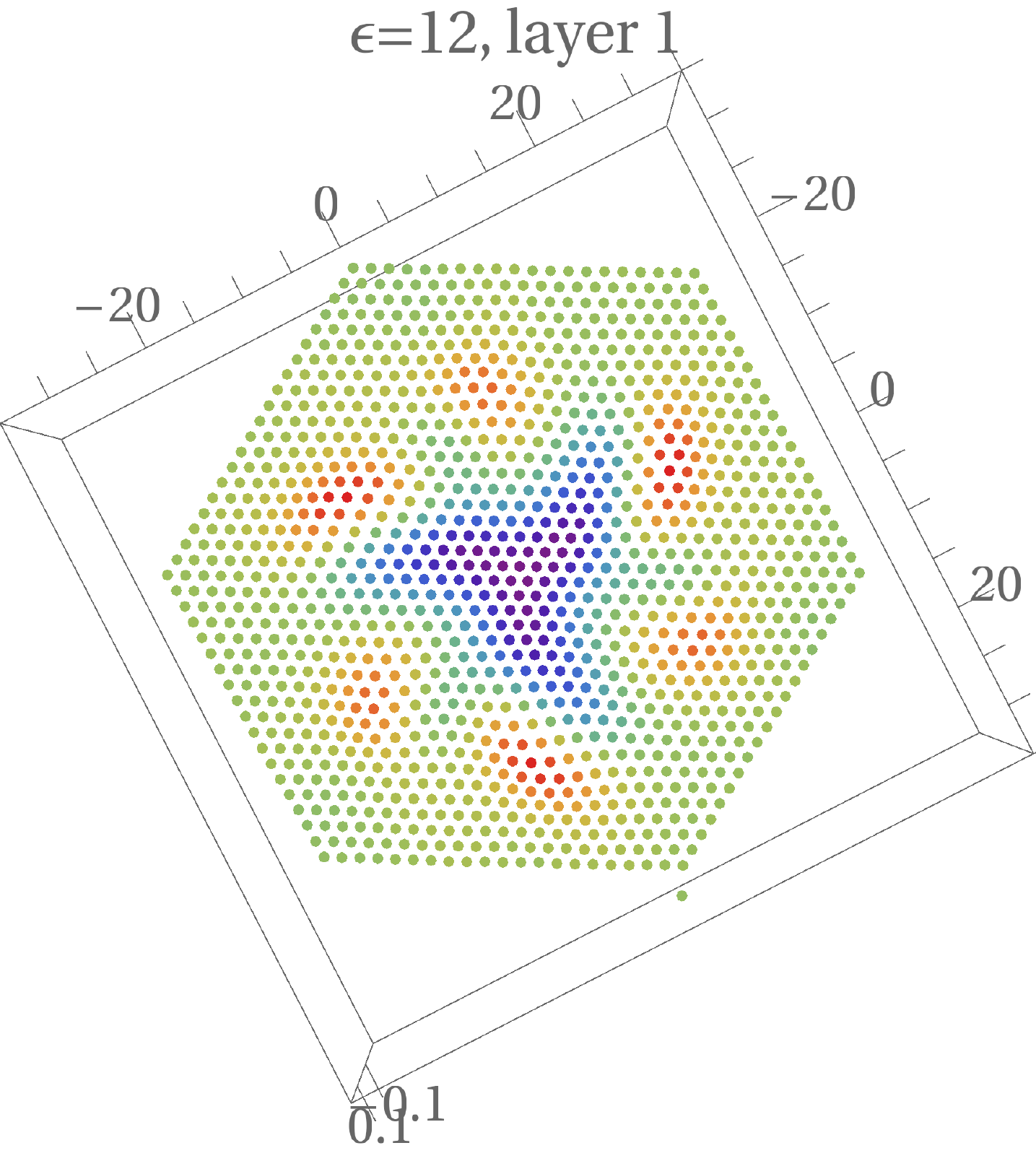}
	\includegraphics[width=0.24\columnwidth]{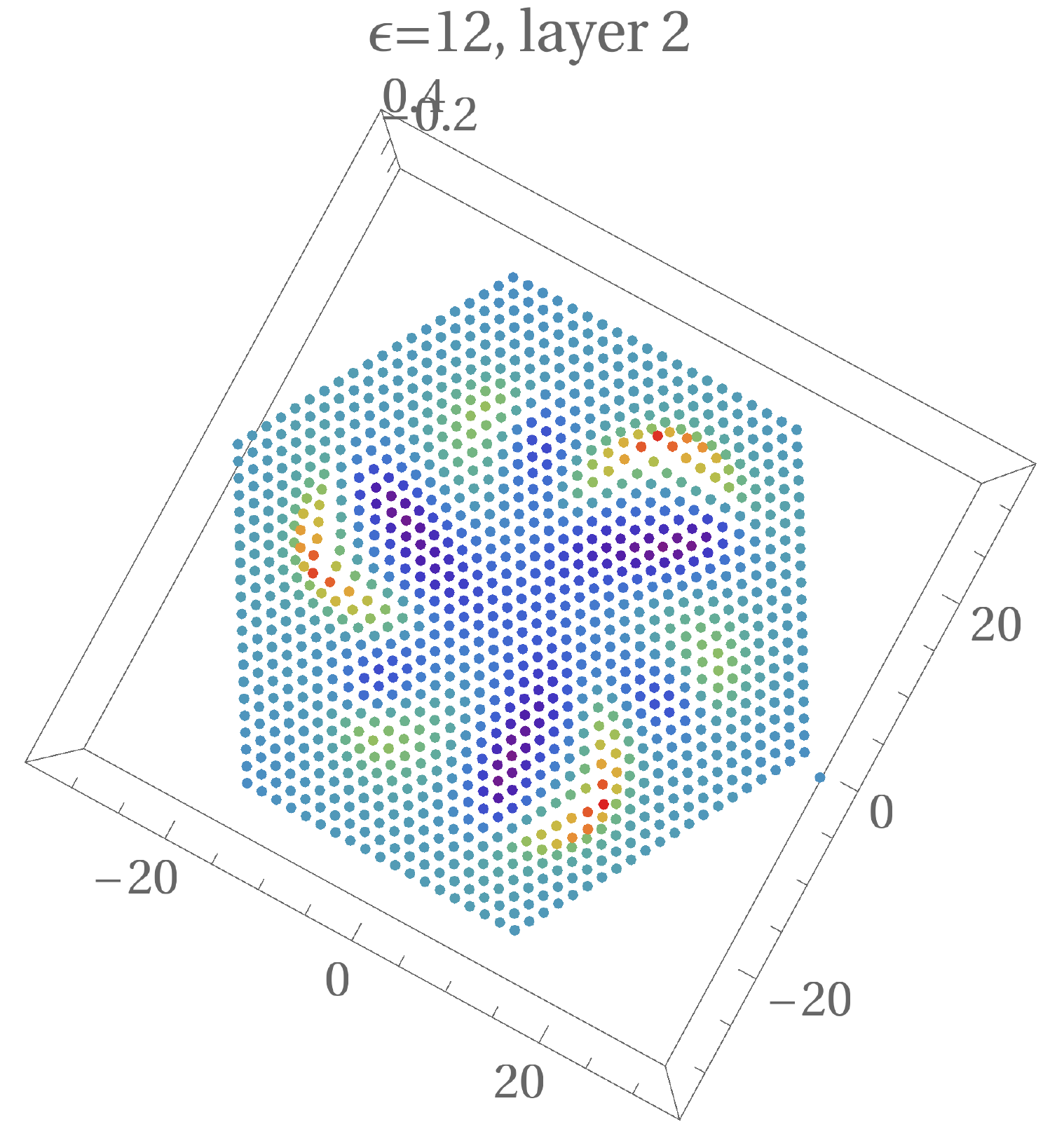}
	\includegraphics[width=0.22\columnwidth]{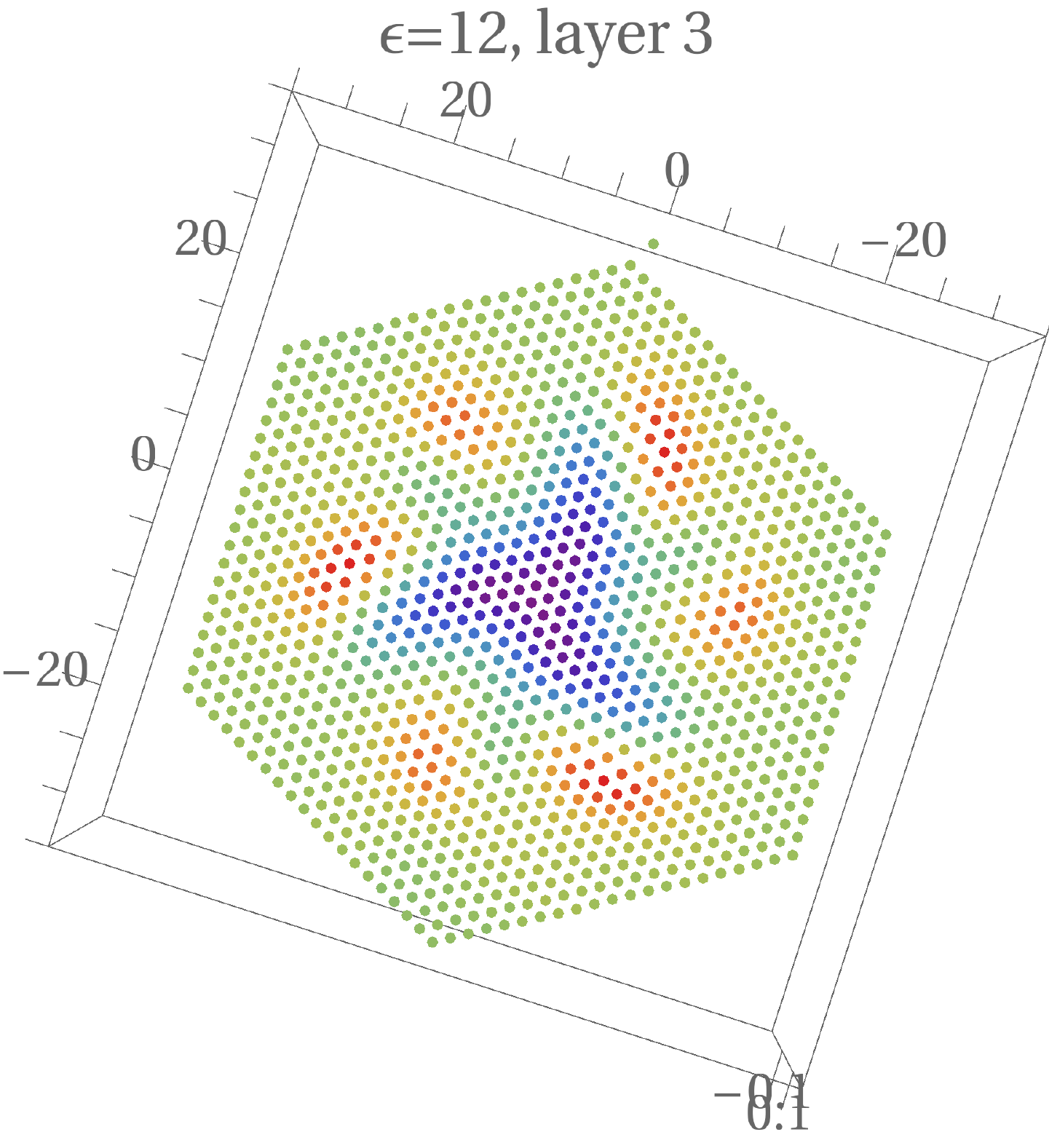}\\
	\vspace{0.5cm}
	\includegraphics[width=0.23\columnwidth]{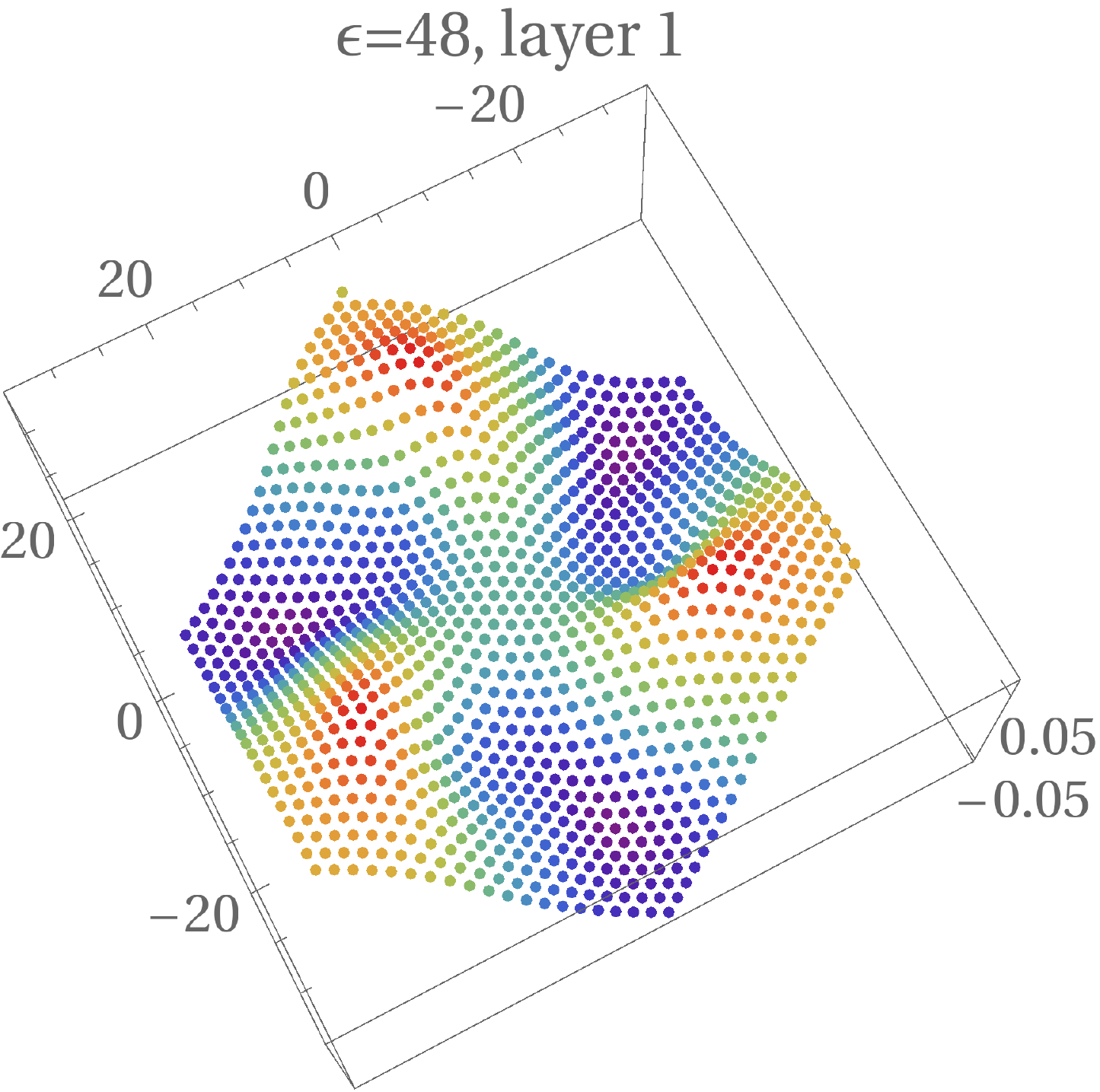} 
	\includegraphics[width=0.21\columnwidth]{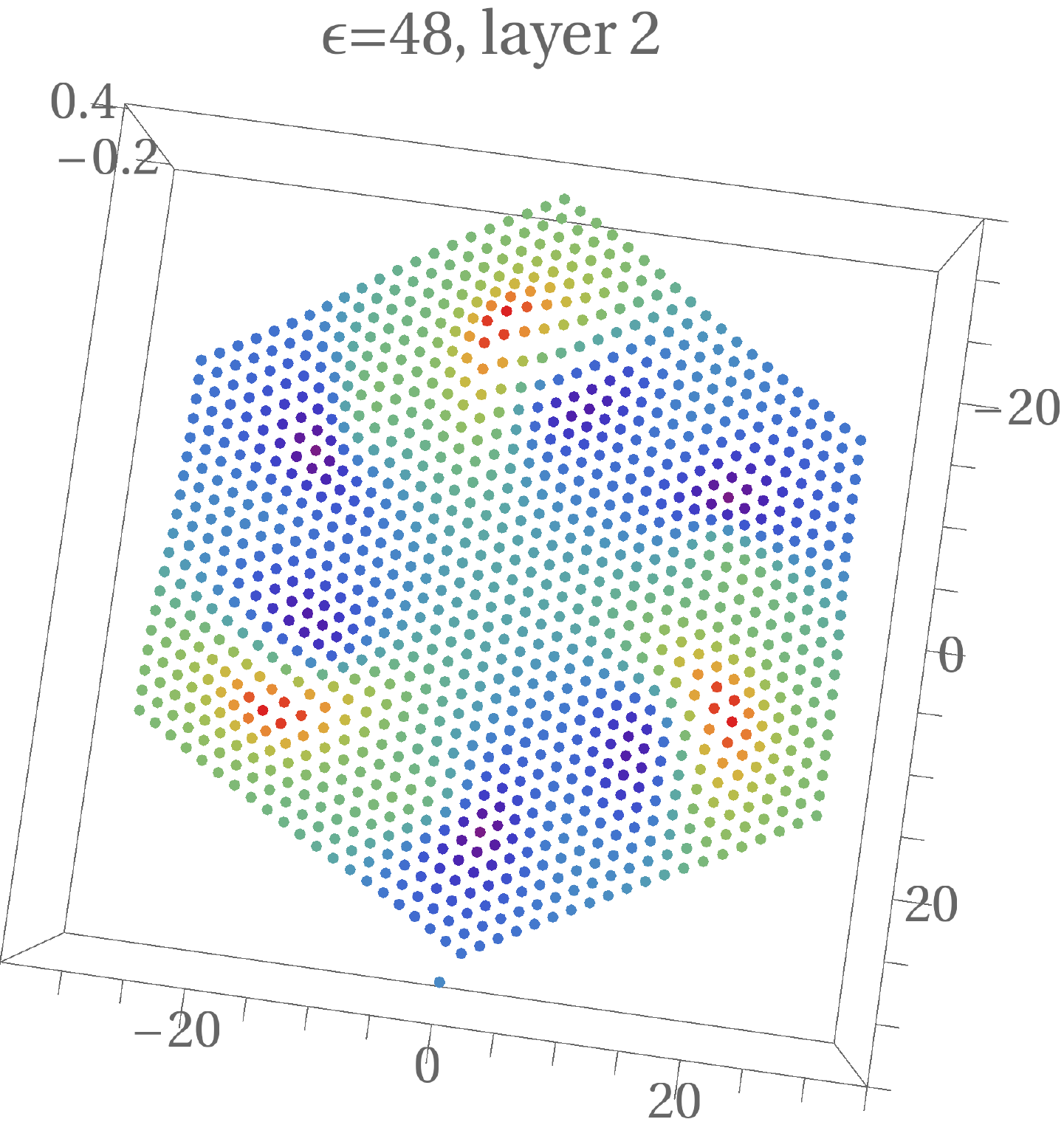}
	\includegraphics[width=0.21\columnwidth]{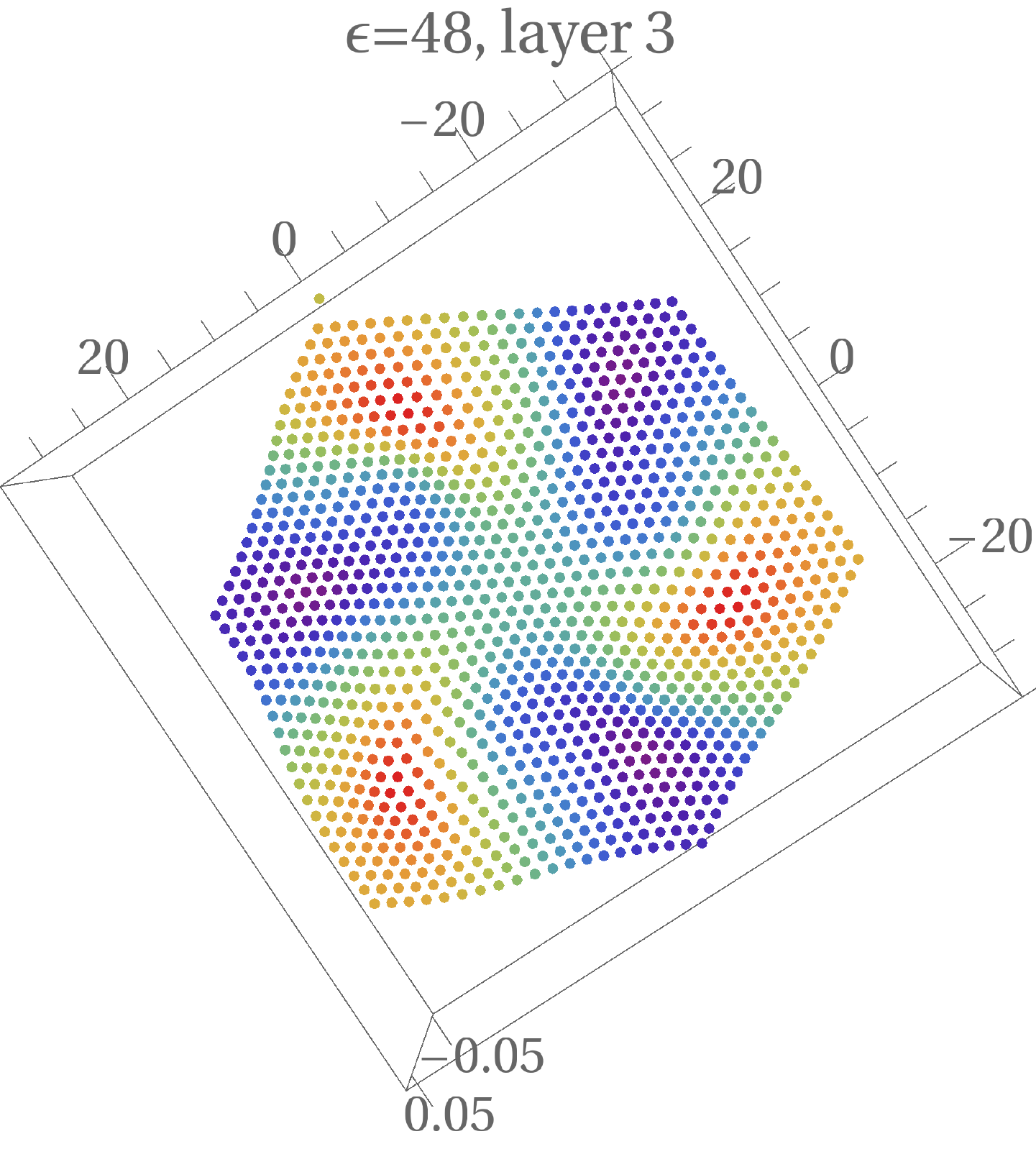}
	\caption{Plot of the fluxes enclosed in the hexagonal rings of each layer of twisted trilayer graphene, for two different values of the dielectric constant $\epsilon = 12$ (upper row) and 48 (lower row). Darker red (blue) color corresponds to higher positive (lower negative) values of the flux.}
	\label{kivc}
\end{figure*}

We have computed $\theta_{\rm KIVC}$ across the supercell of the twisted trilayer, looking for a definite pattern at the microscopic scale. However, we have only found negative evidence in that respect, for strong coupling of the Coulomb interaction as well as in the regime of the twisted trilayer considered in the paper. This can be seen in the plots shown in Fig. \ref{kivc}, which represent the values of $\theta_{\rm KIVC}$ at the hexagonal rings of each layer for two different values of the dielectric constant $\epsilon = 12$ and 48. The envelop of the angles in the supercell gives rise to a smooth surface in all cases, showing the absence of K-intervalley coherence in the twisted trilayer.

The above negative result points at a marked difference between the behavior of twisted bilayer and twisted trilayer graphene at the magic angle, as we have checked that a similar microscopic Hartree-Fock approach applied to the twisted bilayer (with in-plane relaxation) leads indeed to signatures of K-intervalley coherence for values of the dielectric constant as large as $\epsilon \sim 40$. This deviation between the two systems may come from the different type of relevant relaxation (in-plane versus out-of-plane) which one needs to consider in each case. This prevents from assuming a simple decoupling of twisted trilayer graphene as a system of twisted bilayer plus a single graphene layer. As shown in the first section of this Supplemental Material, the out-of-plane corrugation leads to important modifications in the shape of the flat bands of twisted trilayer graphene, inducing a significant particle-hole asymmetry which has a large impact in the symmetry breaking properties of the system.

\section*{Supplementary Note V.\\ Semiclassical theory of the Hall density} 
In this Section, we will analyze the Hall density, accessible in typical transport experiments \cite{Park21}. Within a semiclassical theory, electrons move on trajectories of constant energy. For small magnetic fields, these trajectories are not altered and we will thus use the energy contours of the flat-band dispersion obtained from the self-consistent Hartree-Fock calculations. For the first valence band we use the filling factor $\nu=-2$, for the second valence band we use the filling factor $\nu=-2.8$. In both cases, the dielectric constant is set to $\epsilon=48$. 

The contours can be divided into closed and open trajectories. Closed trajectories can be approximated by circular, elliptic or trigonal warped curves. Open trajectories shall be characterized by van Hove singularities. In the following, we will obtain analytical formulas for all these situations. 

\subsection*{Supplementary Note V.A.: Closed trajectories}
\noindent
For closed trajectories, we calculate the conductivity via the Chambers' formula
\begin{align}
\sigma_{ij}=\frac{g_sg_ve^2}{(2\pi)^2}\int d^2kv_i(\k)\int_{-\infty}^{0}dt'v_j(\k(t'))e^{t'/\tau}\left(-\frac{\partial f(E)}{\partial E}\right)\;.
\end{align}
By virtue of the Lorentz force rule, this can be transformed into the following expression\cite{Maharaj17}:
\begin{align}
\sigma_{ij}=\frac{g_sg_v}{(2\pi)^2}\frac{e^3B}{\hbar^2}\int_0^Tdtv_i(t)\int_{-\infty}^{t}dt'v_j(t')e^{(t'-t)/\tau}
\end{align}
The velocities $v_i(t)$ are obtained from the semiclassical equations,
\begin{align}
\dot\r&=\v(\k)=\frac{1}{\hbar}\partial_\k \epsilon_\k\;,\\
\dot\k&=-\frac{e}{\hbar}\left[\E+\v(\k)\times\B\right]\;.
\end{align}
We assume the magnetic field perpendicular to the plane and the electric field in $x$-direction, i.e., $\B=B\e_z$ and $\E=E\e_x$. The equations of motion in the plane can be integrated and yield 
\begin{align}
\label{EoM}
\r(t)=-\frac{\hbar}{eB}\e_z\times\k(t)-\frac{E}{B}\e_y t\;.
\end{align}
The drift velocity $\v_D=-\frac{E}{B}\e_y$ is in fact the velocity of the frame of reference in which the electric field vanishes.\cite{AshcroftMermin} With $\widetilde \epsilon_\k=\epsilon_\k-\hbar \k\cdot\v_D$, we can thus combine the equations of motion to the following compact form where only the magnetic field enters:
\begin{align}
\dot\k=-\frac{e}{\hbar^2}\partial_\k \widetilde \epsilon_\k\times\B\;,
\end{align}
In the following, we assume a small electric field and neglect the drift term.

Closed orbits are periodic in $T$ and can thus be expanded into a Fourier series. We can therefore set 
\begin{align}
k_i(t)=\sum_\nu\kappa_\nu^ie^{i\nu\omega_c t}\;,
\end{align}
where we introduced the two components $i=x,y$ and the cyclotron frequency $\omega_c=2\pi/T$.
We now obtain the velocities $v_i(t)$ by differentiating Eq. (\ref{EoM}) and with Chambers' formula, this finally yields
\begin{align}
\sigma&=\frac{g_sg_v}{2\pi}\frac{e^2\tau}{m_c}\sum_{\nu>0}\frac{\nu^2}{1+(\nu\omega_c\tau)^2}\left(\begin{array}{cc}
|\kappa_\nu^{y}|^2&-\Re\left[\kappa_{\nu}^x\kappa_{-\nu}^y(1+i\nu\omega_c\tau)\right]\\
-\Re\left[\kappa_{\nu}^x\kappa_{-\nu}^y(1-i\nu\omega_c\tau)\right]&|\kappa_\nu^{x}|^2
\end{array}\right)\;.
\end{align}
Since $\kappa_{\nu}^x$ is independent of $B$, this generally proves the Onsager relation $\sigma_{xy}(B)=\sigma_{yx}(-B)$. To be more explicit, we will now discuss isotropic and elliptic models. 

\subsubsection*{Supplementary Note V.A.1.: Circular and elliptic curves} 
For the general isotropic dispersion $\epsilon_\k=\alpha|k|^\xi$, the Fermi wave number is given by $k_F=(\mu/\alpha)^{1/\xi}$ with $\mu$ the chemical potential. This yields the circular trajectories $k_x(t)=k_F\cos(\omega_c t)$ and $k_y(t)=k_F\sin(\omega_ct)$ with the cyclotron frequency $\omega_c=\frac{eB}{m_c}$ and cyclotron mass $m_c=\frac{\hbar^2}{\xi\alpha}k_F^{2-\xi}$. Consistently, this is the same result as obtained from the general definition $m_c=\frac{\hbar^2}{2\pi}\frac{\partial A}{\partial \mu}$ with $A=\pi k_F^2$ denoting the area that is enclosed by the cyclotron orbit. With $(g_sg_v/4)k_F^2=\pi n$, where $g_s$ and $g_v$ denote spin and valley (or other) degeneracies, this then gives the final result
\begin{align}
\sigma=\frac{\sigma_{xx}}{1+\a^2}\left(
\begin{array}{cc}
1&\a\\
-\a&1
\end{array}\right)\;,
\end{align}
where the longitudinal conductivity is given by $\sigma_{xx}=e^2n\tau/m_c$ and $\a=\omega_c\tau$. With the resistivity tensor $\rho=\sigma^{-1}$, we get for all isotropic dispersion relations the universal Hall density
\begin{align}
\label{HallDensityDef}
n_{H}=-\left[e\frac{d\rho_{xy}}{d B}\right]^{-1}&=n\;.
\end{align}

The universal result also holds for an elliptic dispersion with $\epsilon_\k=\alpha_xk_x^2+\alpha_yk_y^2$ and $k_{F,i}=\sqrt{\mu/\alpha_i}$. With $k_x(t)=k_{F,x}\cos(\omega_ct)$,   $k_y(t)=k_{F,y}\sin(\omega_ct)$, $A=\pi k_{F,x}k_{F,y}$ and $\omega_c=\frac{eB}{m_c}$, we have $m_c=\frac{\hbar^2}{2\sqrt{\alpha_x\alpha_y}}$ and 
\begin{align}
\sigma=\frac{\sigma_{xx}}{1+\a^2}\left(
\begin{array}{cc}
\sqrt{\alpha_x/\alpha_y}&\a\\
-\a&\sqrt{\alpha_y/\alpha_x}
\end{array}\right)\;,
\end{align}
where $\sigma_{xx}=\frac{e^2n\tau}{m_c}$. With Eq. (\ref{HallDensityDef}), we again obtain the universal expression $n_H=n$ for the Hall density independent of the band parameters.

\subsubsection*{Supplementary Note V.A.2.: Trigonal warped trajectories}
\noindent
For trigonal warped Fermi-surfaces, there are deviations from the universal result. However, in a perturbative treatment the first non-vanishing term is quadratic in the expansion parameter $\epsilon\sim k_Fa$. This suggests that general closed orbits will lead to a Hall density close to the universal result, i.e., $n=n_H$. This shall be discussed below. 

To proceed analytically, let us discuss single-layer graphene in the trigonal warped regime. Graphene is characterized by the energy dispersion $\epsilon_\k=\pm t|\Phi_\k|$ where $\Phi_\k=\sum_\delta e^{i\k\cdot\delta}$ with the three nearest-neighbor vectors $\delta_1=a(1,0)$, $\delta_2=a(-1/2,\sqrt{3}/2)$, and $\delta_3=a(-1/2,-\sqrt{3}/2)$ as well as $t=-2.7$ the hopping matrix element. With the Jacobi-Anger expansion, the structure factor can also be written as $\Phi_\k=3\sum_nJ_{-1+3n}(ka)e^{i(-1+3n)\phi}$ \cite{Santos11}. To lowest order in the lattice effect, we then have the following expression for the Fermi surface in polar coordinates:
\begin{align}
k_F'(\phi)=k_F\left(1+\epsilon\cos(3\phi)+\frac{\epsilon^2}{4}\left[11+5\cos(6\phi)\right]\right)
\end{align}
The enclosed area is still given by $A=\pi k_F^2$ with $E_F=\hbar v_Fk_F$ and $v_F=\frac{3}{2}at$. We further introduced the trigonal warping parameter $\epsilon=\frac{k_Fa}{4}$.

With $\v=-\frac{\hbar}{eB}\e_z\times\dot\k$, we have $\partial_\k \epsilon_\k=\frac{\hbar^2}{eB}\left[k\dot\phi\e_k-\dot k\e_\phi\right]$. With the dimensionless parameter $\bar k=k/k_F$, this gives the following set of differential equations:
\begin{align}
\bar k\dot\phi&=\omega_c\left[1-2\bar k\epsilon\cos(3\phi)-\frac{3}{4} (\bar k\epsilon)^2\left[7+\cos(6\phi)\right]\right]\\
\dot {\bar k}&=-3\omega_c\bar k\epsilon\sin(3\phi)-\frac{3}{2}\omega_c(\bar k\epsilon)^2\sin(6\phi)
\end{align}
To second order, the solution thus reads $\bar k(t)=1+\epsilon\cos(3\omega_ct)+\frac{\epsilon^2}{4}(17-\cos(6\omega_ct))$ and $\phi(t)=\omega_ct-\epsilon\sin(3\omega_ct)+\epsilon^2(-12\omega_ct+\frac{1}{2}\sin(6\omega_ct))$. For the cartesian coordinates, we then have $k_x(t)=k_F'(\phi(t))\cos(\phi(t))$ and $k_y(t)=k_F'(\phi(t))\sin(\phi(t))$. Note that for $k_F'(\phi(t))$ only the expansion of $\phi$ up to first order is needed in order to be consistent. 

We can now again discuss the response in the presence of a magnetic field via the Chambers' formula. For the explicit solution of the trigonal warped graphene regime, we get
\begin{align}
\sigma&=\frac{e^2n\tau}{m_c}\frac{1}{(1+(\omega_c\tau)^2)(1+4(\omega_c\tau)^2)}\left(\begin{array}{cc}
1+12\epsilon^2+4(1+9\epsilon^2)(\omega_c\tau)^2&\omega_c\tau\left[1+4(1+6\epsilon^2)(\omega_c\tau)^2\right]\\-\omega_c\tau\left[1+4(1+6\epsilon^2)(\omega_c\tau)^2\right]&1+12\epsilon^2+4(1+9\epsilon^2)(\omega_c\tau)^2
\end{array}\right)\;.
\end{align}
The Hall number is usually defined by
\begin{align}
\frac{1}{n_H}=\frac{e}{B}\frac{\sigma_{xy}}{\sigma_{xx}\sigma_{yy}+\sigma_{xy}^2}\;.
\end{align}
This gives the final result
\begin{align}
\frac{n_H}{n}=1+\left[6+\frac{18}{1+4(\omega_c\tau)^2}\right]\epsilon^2\;.
\end{align}
In the clean limit $\tau\to\infty$, this simplifies to $n_H=n(1+6\epsilon^2)$ and in the low-field limit $\omega_c\to0$, we have $n_H=n(1+24\epsilon^2)$. In both cases, there is thus a slight increase of the Hall number due to the acceleration around the corners of the deformed Brillouin zone.

The Hall number is also sometimes defined by
\begin{align}
\frac{1}{n_H}=e\frac{d}{dB}\frac{\sigma_{xy}}{\sigma_{xx}\sigma_{yy}+\sigma_{xy}^2}\;.
\end{align}
This gives for the Hall density
\begin{align}
\frac{n_H}{n}=1+\frac{24\left[1-(\omega_c\tau)^2+4(\omega_c\tau)^4\right]}{(1+4(\omega_c\tau)^2)^2}\epsilon^2\;,
\end{align}
which is a slightly different expression than above. However, in the clean limit, this again simplifies to $n_H=n(1+6\epsilon^2)$ and we also have $n_H=n(1+24\epsilon^2)$ for the low-field limit as before.

To conclude, there is no linear correction in $\epsilon$ to the Hall density. The deviations from the universal result $n_H=n$ should thus be small and negligible. We shall, therefore, approximate $n_H=n$ for all closed trajectories.

\subsection*{Supplementary Note V.B.: Trajectories close to van Hove singularities}
\noindent
To discuss the semiclassical motion of electrons close to van Hove singularities, a well-defined regularization procedure is needed since   the orbits are unbounded for a continuum theory. Therefore, we will not use the Chambers' formula, but start from the macroscopic equations of motion for the current density.  The general response theory in the presence of an in-plane electric field $\E$ and a perpendicular magnetic field $\B$ then reads
\begin{align}
\label{SemiclassicalMotion}
\partial_t\j=\chi \E+\frac{e}{m_c}\bar\j\times\B-\j/\tau\;.
\end{align}
Above, we introduced the current-current response function $\chi$ in the dc-limit  and the ^^ ^^ average" current density $\bar\j$ which will both be discussed below. We also introduced the inverse relaxation time $\eta=\tau^{-1}$ and the cyclotron mass is defined by\cite{AshcroftMermin} 
\begin{align}
m_c=\frac{\hbar^2}{2\pi}\frac{\partial A}{\partial \mu}\;,
\end{align}
where $A$ denotes the area that is enclosed by the cyclotron orbit. Within this formalism, the above results for the isotropic and elliptic models can be obtained. Here, we will outline the specific case of a hyperbolic model. 
\subsubsection*{Supplementary Note V.B.1.: Drude response around a saddle-point}
\noindent
The van Hove singularity shall be described by the saddle-point dispersion $\epsilon_\k=-\alpha_- k_x^2+\alpha_+ k_y^2$. The so-called Drude response can entirely be obtained from the band structure and for $T=0$ at the chemical potential $\mu$, it is defined by
\begin{align}
\label{DrudeWeight}
\chi_{ij}=\frac{g_sg_ve^2}{(2\pi\hbar)^2}\int d^2k(\nabla \epsilon_\k)_i(\nabla \epsilon_\k)_j\delta(\mu-\epsilon_\k)\;.
\end{align}
As we have assumed the principle axes to be along the $x$- and $y$-direction, $\chi_{ij}\propto\delta_{ij}$. 

The above integral can be performed by first eliminating the $\delta$-function via the polar integration. For the radial integration, the following integrals are needed:
\begin{align}
\mathcal{I}_\pm(\Lambda,\gamma)=\int_{1}^{\Lambda^2} dx\sqrt{\frac{x-1}{\gamma x+1}}^{\pm1}
\end{align} 
This gives for $\mu=\pm|\mu|$ the final expression
\begin{align}
\chi_\pm&=\frac{g_sg_v}{(2\pi)^2}\frac{e^2}{\hbar^2}\frac{4\tilde\mu_\pm}{(\alpha_++\alpha_-)}\left(
\begin{array}{cc}
\alpha_-^2\mathcal{I}_\pm(\tilde\Lambda_\pm,\gamma_\pm)&0\\
0&\alpha_+^2\mathcal{I}_\mp(\tilde\Lambda_\pm,\gamma_\pm)
\end{array}\right)\;,
\end{align}
with $\gamma_\pm=\alpha_\mp/\alpha_\pm$, $\tilde\mu_\pm=|\mu|/\alpha_\pm$, and $\tilde\Lambda_\pm=\Lambda/\sqrt{\tilde\mu_\pm}$ where $\Lambda$ denotes the wavenumber cutoff. In the following, we will only discuss the response due to electron doping with $\mu>0$ and set $\gamma=\gamma_+$.

At the neutrality point, the susceptibility is proportional to $\Lambda^2$ and we will discuss the difference $\delta\chi=\chi_+-\chi_{\mu=0}$. To leading order, we have
\begin{align}
\label{responseHyper}
\delta\chi=\frac{g_sg_v}{(2\pi)^2}\frac{e^2}{\hbar^2}2\mu\left(
\begin{array}{cc}
-\sqrt{\gamma}\ln\frac{\alpha\Lambda^2}{\mu}&0\\
0&\sqrt{\gamma}^{-1}\ln\frac{\alpha\Lambda^2}{\mu}
\end{array}\right)\;,
\end{align}
where $\alpha=\frac{2\alpha_+\alpha_-}{\alpha_++\alpha_-}$. The area relative to the one of $\mu=0$ is given by $A=4\frac{\mu}{\sqrt{\alpha_+\alpha_-}}\ln\frac{4\alpha_-\Lambda^2}{\mu}$. Therefore, we get to leading order in $\Lambda$ the cyclotron mass $m_c=4\frac{\hbar^2}{2\pi\sqrt{\alpha_+\alpha_-}}\ln\frac{4\alpha_-\Lambda^2}{\mu}$. With $n=\frac{g_sg_v}{(2\pi)^2}A$, this yields
\begin{align}
\frac{m_c}{e^2}\delta\chi=\frac{n}{\pi}\left(
\begin{array}{cc}
-\sqrt{\gamma}\ln\frac{\alpha\Lambda^2}{\mu}&0\\
0&\sqrt{\gamma}^{-1}\ln\frac{\alpha\Lambda^2}{\mu}
\end{array}\right)\;.
\end{align}
\subsubsection*{Supplementary Note V.B.2.: Magnetic response around a saddle-point}
\noindent
Let us now include the magnetic field. A magnetic field does not break rotational invariance and for an anisotropic system, the field couples to the average velocity $v^2=v_xv_y$. For an elliptic dispersion, this yields the universal results $n_H=n$ as mentioned above. 

In the case of a saddle-point, however, we also have to keep track of the negative sign and we have to couple to the positive mean velocity $v^2=-v_xv_y$. From the cartesian velocities $v_i=\hbar^{-1}\partial_{k_i}\epsilon_\k$ and $\j=-en\v$, we thus set $\bar\j=(j_x/\sqrt{\gamma},-\sqrt{\gamma}j_y)$. The hyperbolic response with respect to $\mu=0$ gives then rise to the  following conductivity tensor:
\begin{align}
\label{ConductivityFull}
\sigma=\frac{\tau}{1-\a^2}\left(
\begin{array}{cc}
\delta\chi_{1}&-\a\sqrt{\gamma}\delta\chi_{2}\\
-\a\delta\chi_{1}/\sqrt{\gamma}&\delta\chi_{2}
\end{array}\right)\;,
\end{align}
where we used $\delta\chi=\text{diag}(\delta\chi_{1},\delta\chi_{2})$ of Eq. (\ref{responseHyper}). The resistivity tensor thus reads
\begin{align}
\label{ResistivityFull}
\rho=\frac{\eta}{\delta\chi_{1}\delta\chi_{2}}\left(
\begin{array}{cc}
\delta\chi_{2}&\a\sqrt{\gamma}\delta\chi_{2}\\
\a\delta\chi_{1}/\sqrt{\gamma}&\delta\chi_{1}
\end{array}\right)\;.
\end{align}

Therefore, we get for the Hall density the final result
\begin{align}
\label{SIvHFormula}
n_{H}&=-\left[e\frac{d\rho_{xy}}{d B}\right]^{-1} =\frac{n}{\pi}\ln\frac{\alpha\Lambda^2}{\mu}\;.
\end{align}
There is a logarithmic divergence for $\mu\to0$ which has been discussed also in the context of a tight-binding model.\cite{Maharaj17} However, for extended van Hove singularities there is also a possible divergence in the limit $\alpha\to0$ which is independent of $\mu$. 

\section*{Supplementary Note VI.\\ Numerical discussion of the Hall density}
\noindent
We will now numerically discuss the Hall density starting with the first valence band. For hole doping up to $\nu\approx-1.8$, no gap has developed yet and the transport is dominated by hole-doping. Close to the neutrality point, all semiclassical trajectories are closed and we have $n_H\sim n$. This ^^ ^^ universal behavior" is, however, modified by the presence of two van Hove singularities which shall be modeled by Eq. (\ref{SIvHFormula}). At the filling factor $\nu\sim -1.8$, we observe a merging of the three van Hove singularities at the $\Gamma$-point to form a so-called higher-order van Hove singularity.\cite{Gon22} Beyond that point, a gap is formed due to time-reversal symmetry breaking leading to electronic transport with universal behavior. This is consistent with the experimentally observed Hall reset at $\nu=-2$. 

To discuss the Hall density of the second valance band, we start from the symmetric transport model, i.e., half of the band is dominated by electron transport and the other half by hole transport. This is justified by noting that close to the band edges, the trajectories are all closed. Again, this ^^ ^^ universal behavior" is modified by the presence of two van Hove singularities which is also modeled by Eq. (\ref{SIvHFormula}).

\begin{figure*}
\includegraphics[width=0.4\columnwidth]{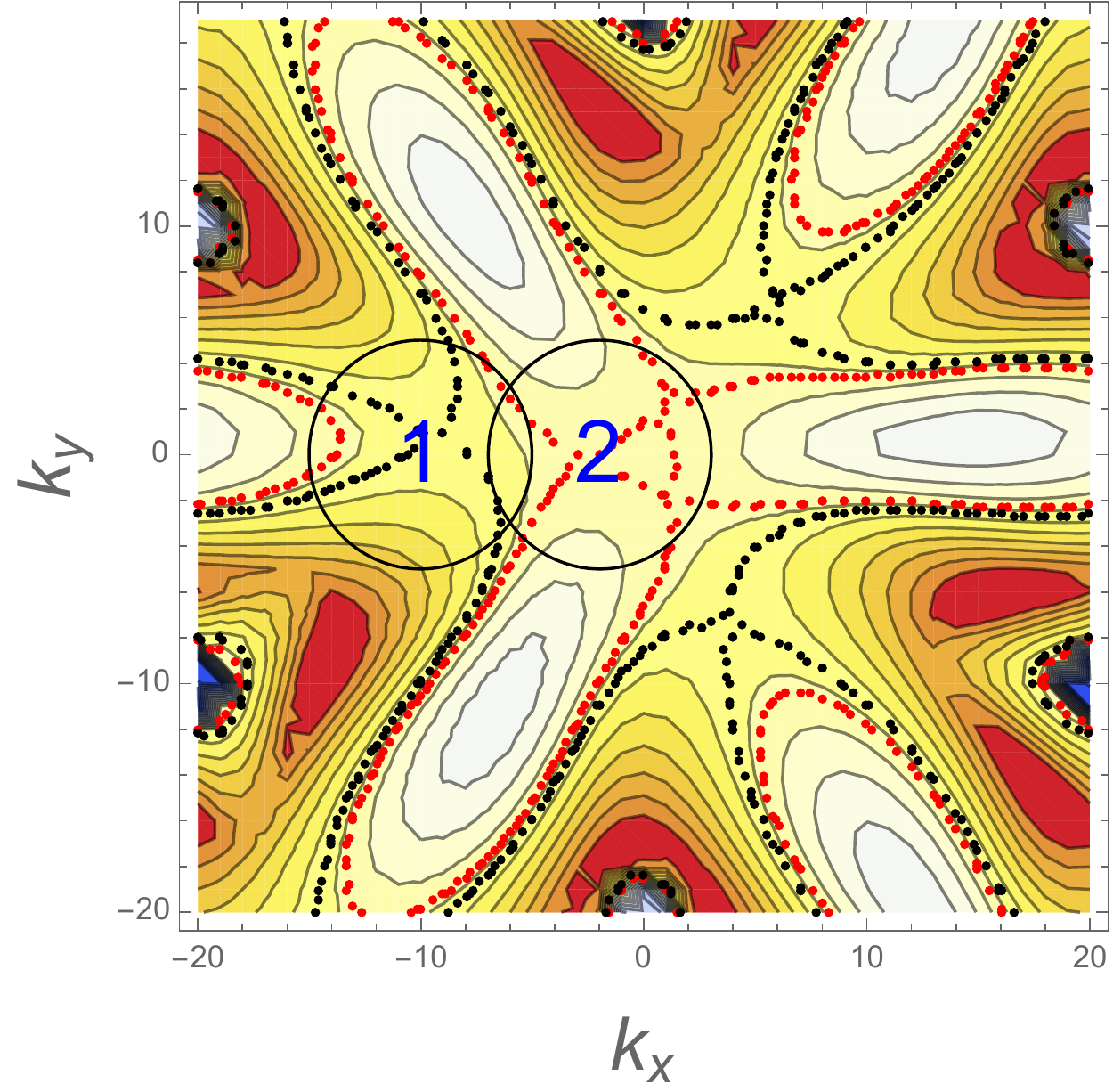}
\includegraphics[width=0.4\columnwidth]{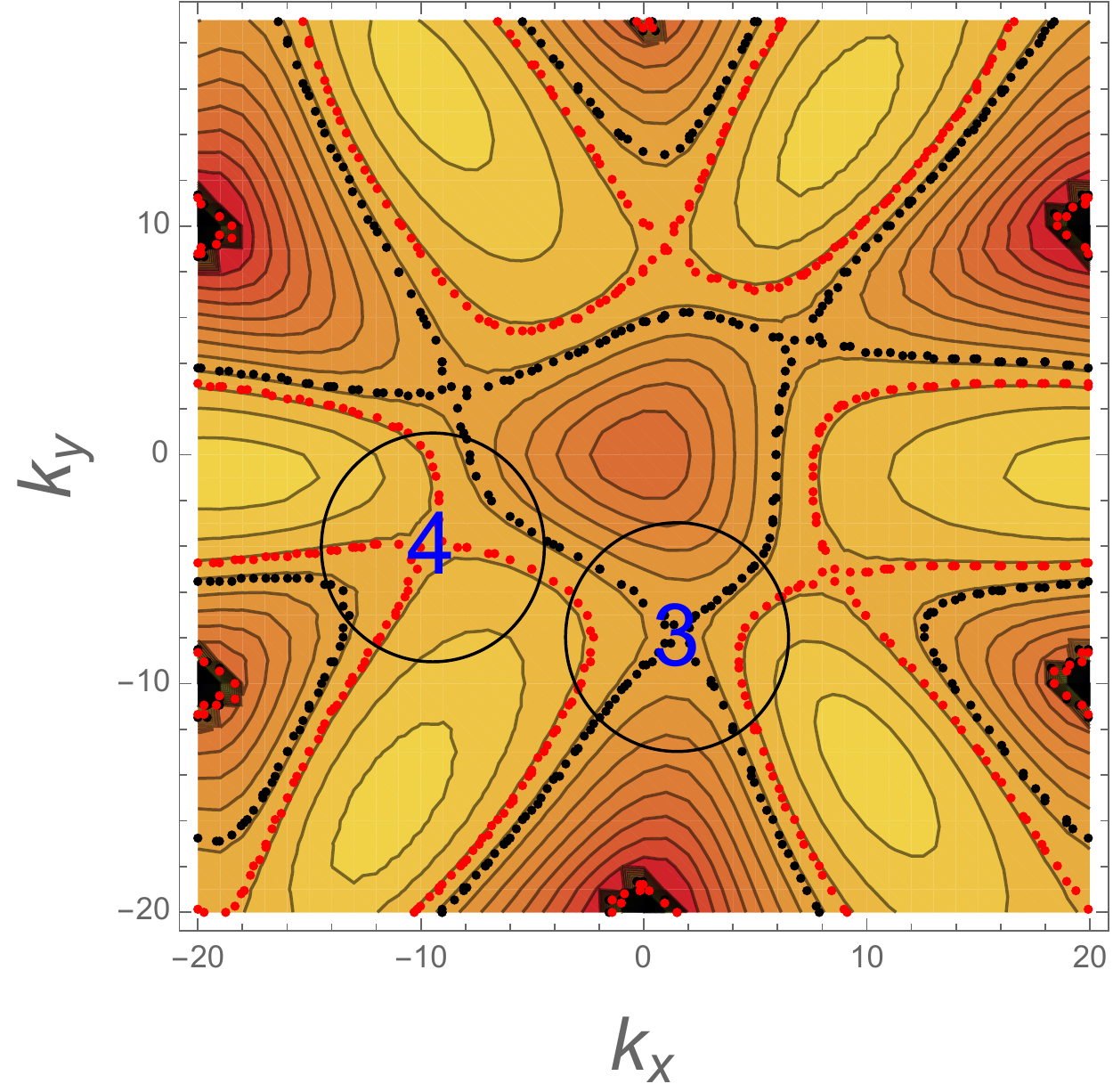}
\caption{Self-consistent energy dispersion of the first (left) and second (right) valence band for $\nu=-2$ and $\nu=-2.8$, respectively. The black and red dotted lines indicate the energy contour at the van Hove energies. The wave numbers are in units of $\frac{1}{20}\frac{4\pi}{3a_M}$ with $a_M$ the moir\'e lattice constant.}
\label{VanHoveContour}
\end{figure*}

Before we outline the fitting procedure, let us recall that we find a prominent valley symmetry breaking for each spin channel which reduces the inherent $C_6$-symmetry to a $C_3$-symmetry. This symmetry is also reflected in the van Hove singularities which are always composed of saddle points that come in triplets. The positions of the van Hove singularities usually lie on the six $\Gamma M$ directions. However, for $\nu\lesssim-1.8$, the two triplets lie on the same three $\Gamma M$ directions which enforces the valley-symmetry broken state and induces a gap. In Fig. \ref{VanHoveContour}, the contour plots of the two valence bands are shown, highlighting the energy contours at the two van Hove energies, respectively. The initial discretization of the Brillouin zone was given by 20 $k$-points between the two $K$ points. With the moir\'e supercell lattice constant $a_M$, the wave numbers $k_x$ and $k_y$ are thus in units of $\frac{1}{20}\frac{4\pi}{3a_M}$. 

The expression for the Hall density around a van Hove singularity depends on the parameters $\alpha$ and $\Lambda$ which shall now be determined. Due to numerical errors, the $C_3$-symmetry regarding the three-fold saddle points is not exact even though the appearance in the contour plot suggests this approximate symmetry. We thus choose to fit the saddle points along the principle axis by the general dispersion $\epsilon_\k=-\alpha_-k_-^2+\alpha_+k_+^2$ which is closest to a parabola with positive and negative mass. In Fig. \ref{VanHoveContour}, we indicate and numerate the saddle points that were used in the fitting process. 
\begin{figure*}
\includegraphics[width=0.24\columnwidth]{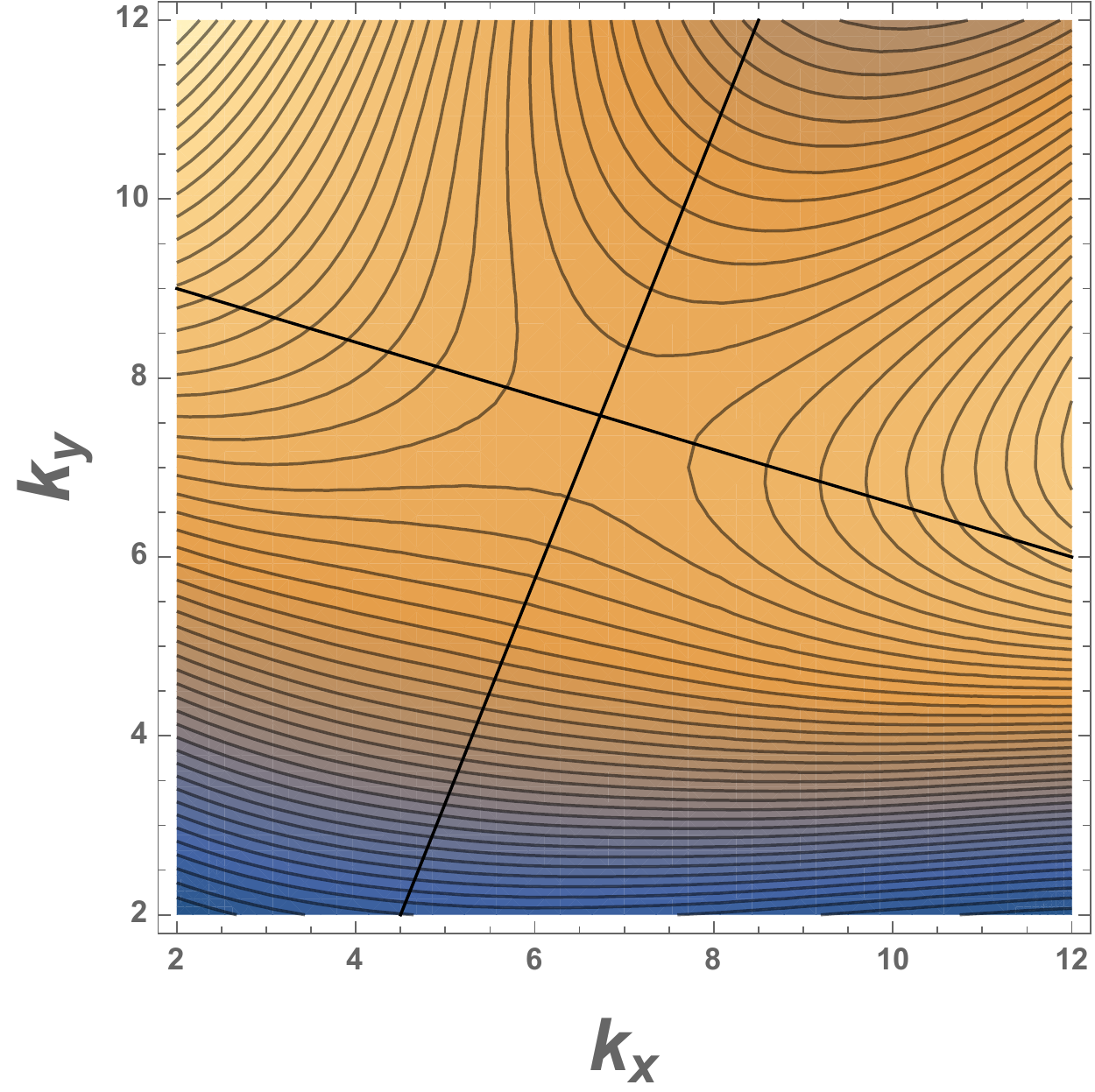}
\includegraphics[width=0.237\columnwidth]{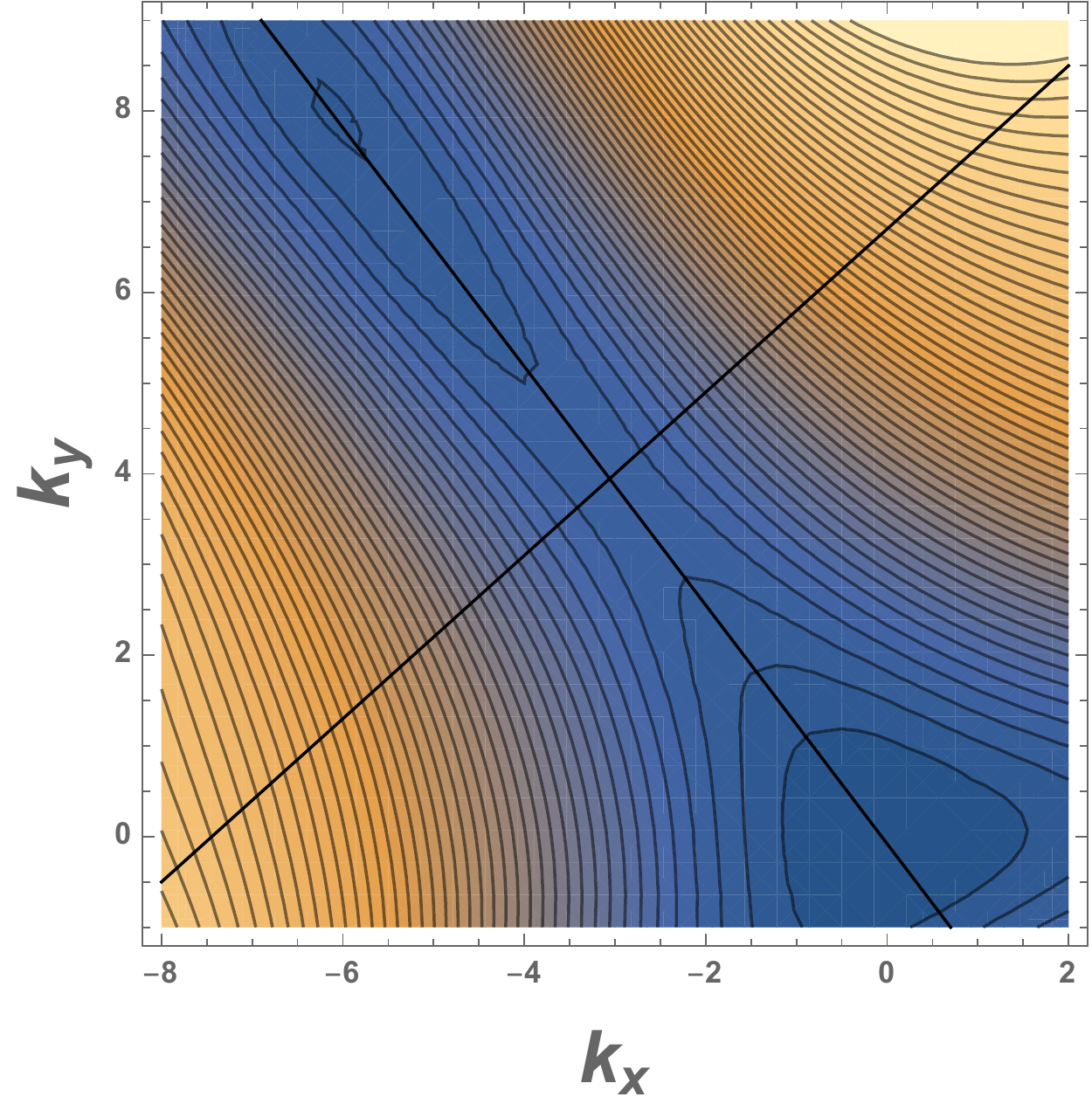}
\includegraphics[width=0.246\columnwidth]{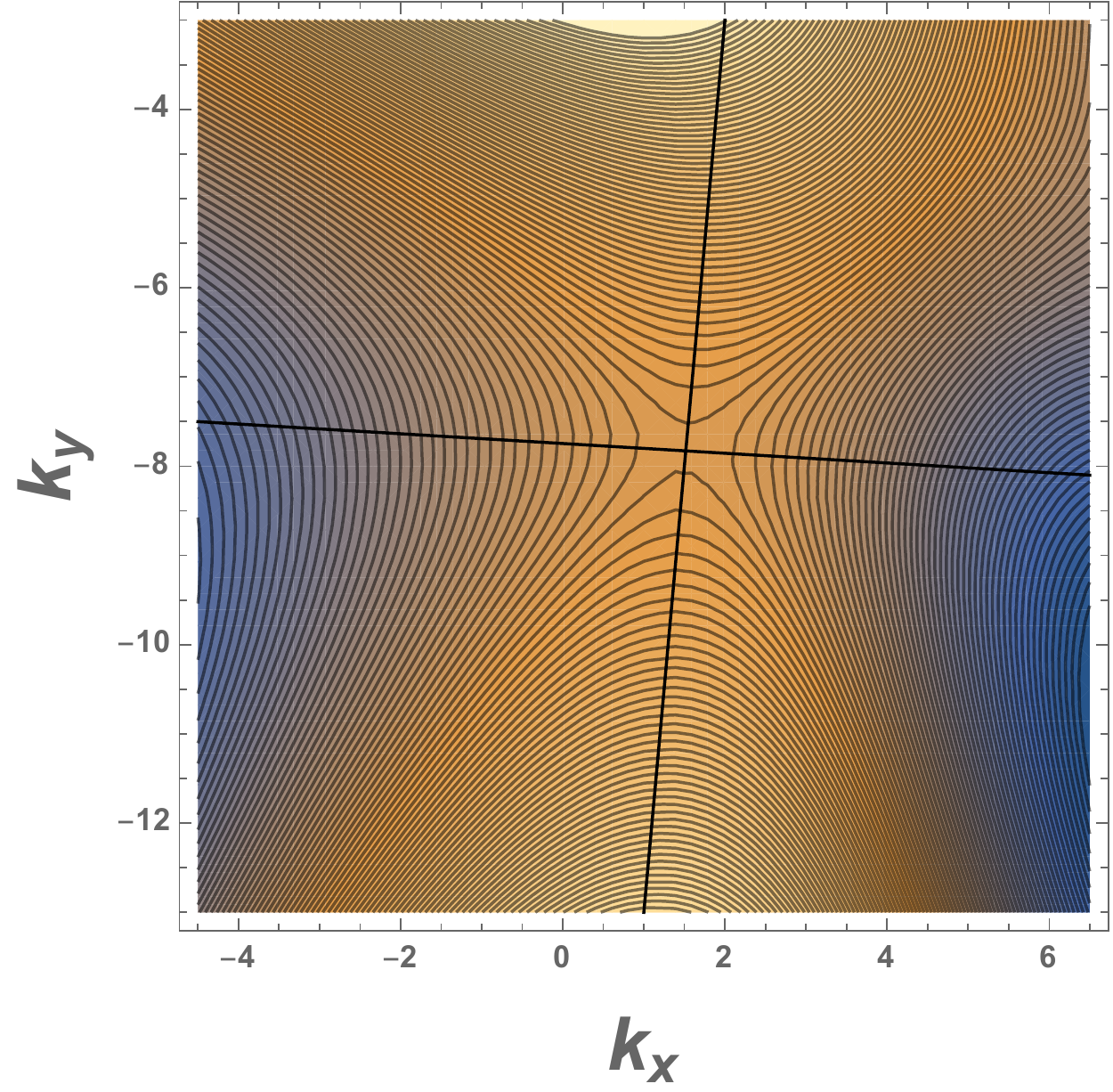}
\includegraphics[width=0.241\columnwidth]{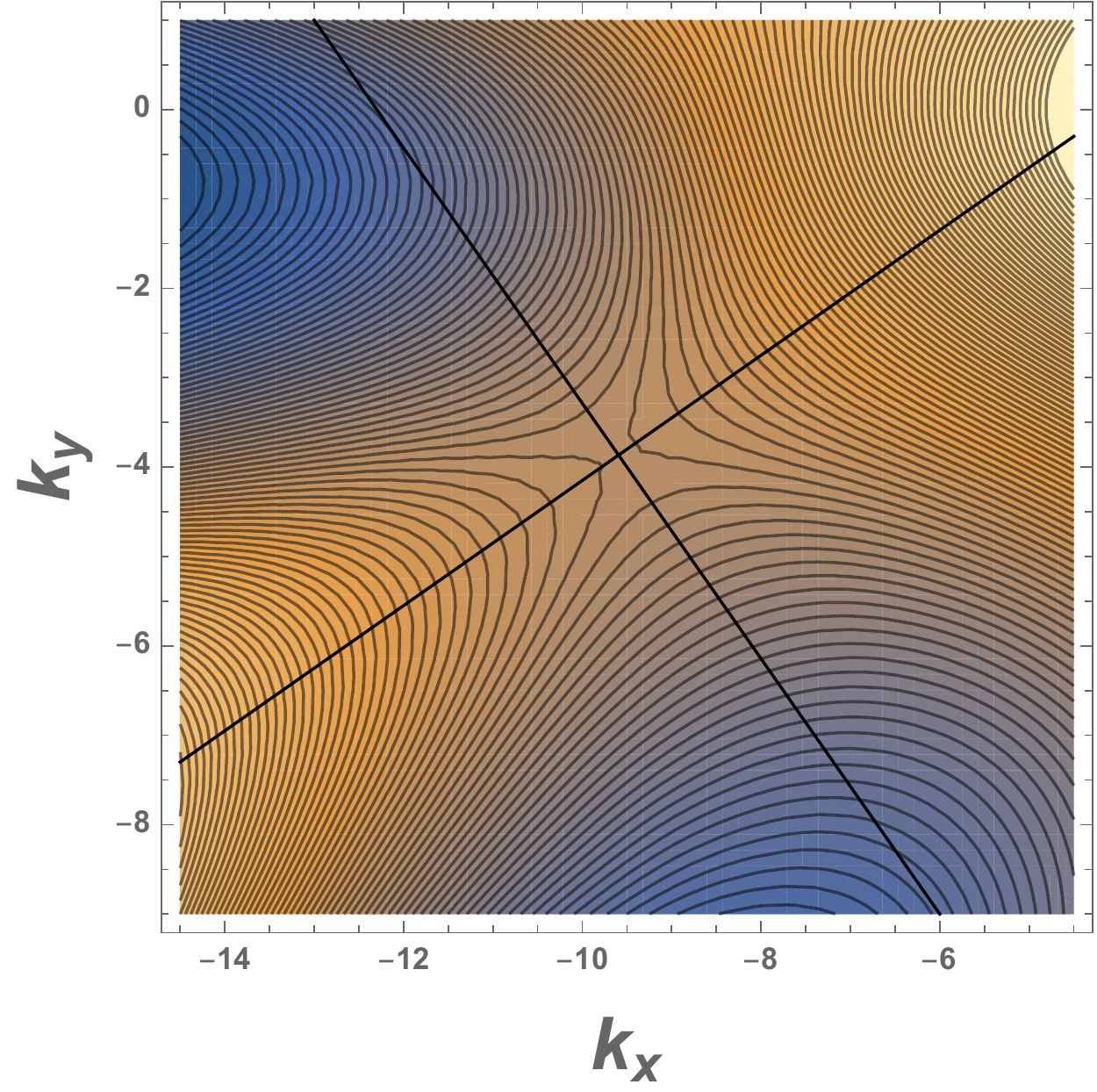}
\caption{Zoom-in of the van Hove singularities 1, 2, 3, 4 (from left to right). Also shown are the approximated principle axes along which the parameters of the saddle points are fitted. The wave numbers are in units of $\frac{1}{20}\frac{4\pi}{3a_M}$ with $a_M$ the moir\'e lattice constant.}
\label{PrincipleAxes}
\end{figure*}

\begin{figure*}
\includegraphics[width=0.32\columnwidth]{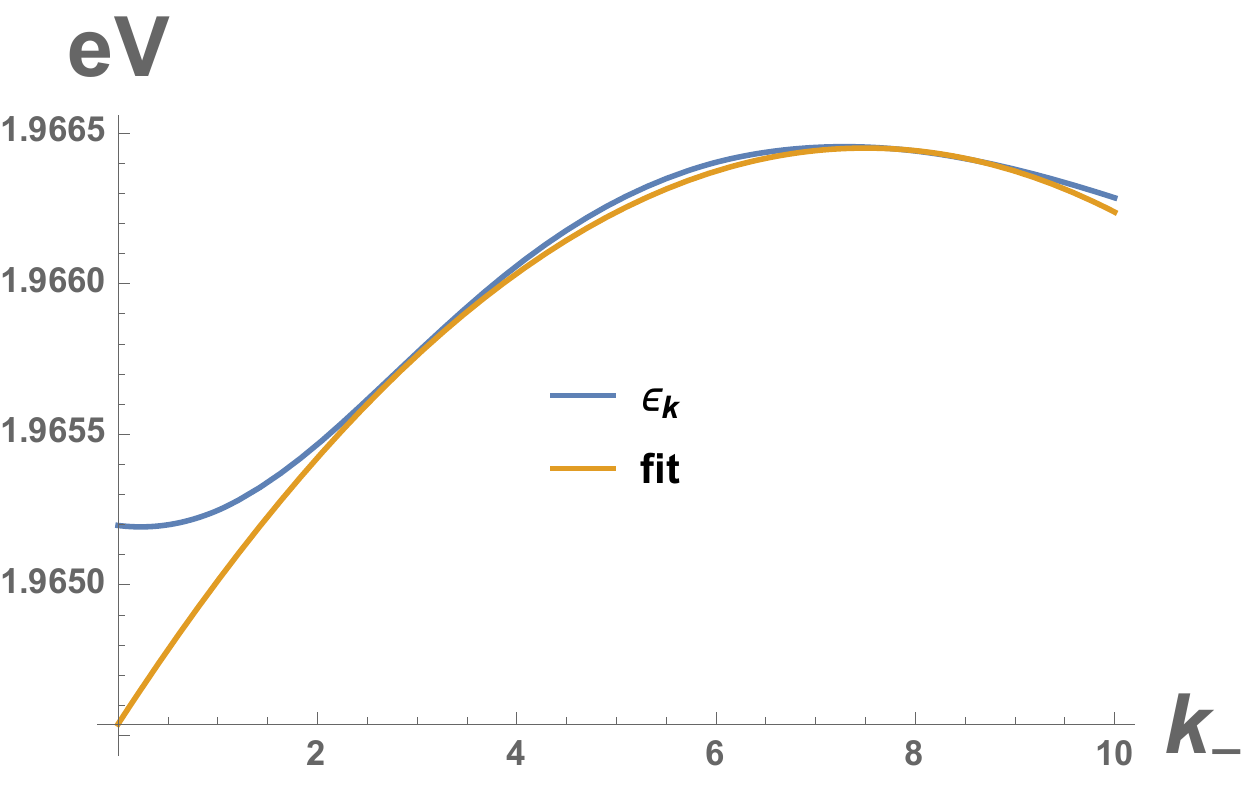}
\includegraphics[width=0.32\columnwidth]{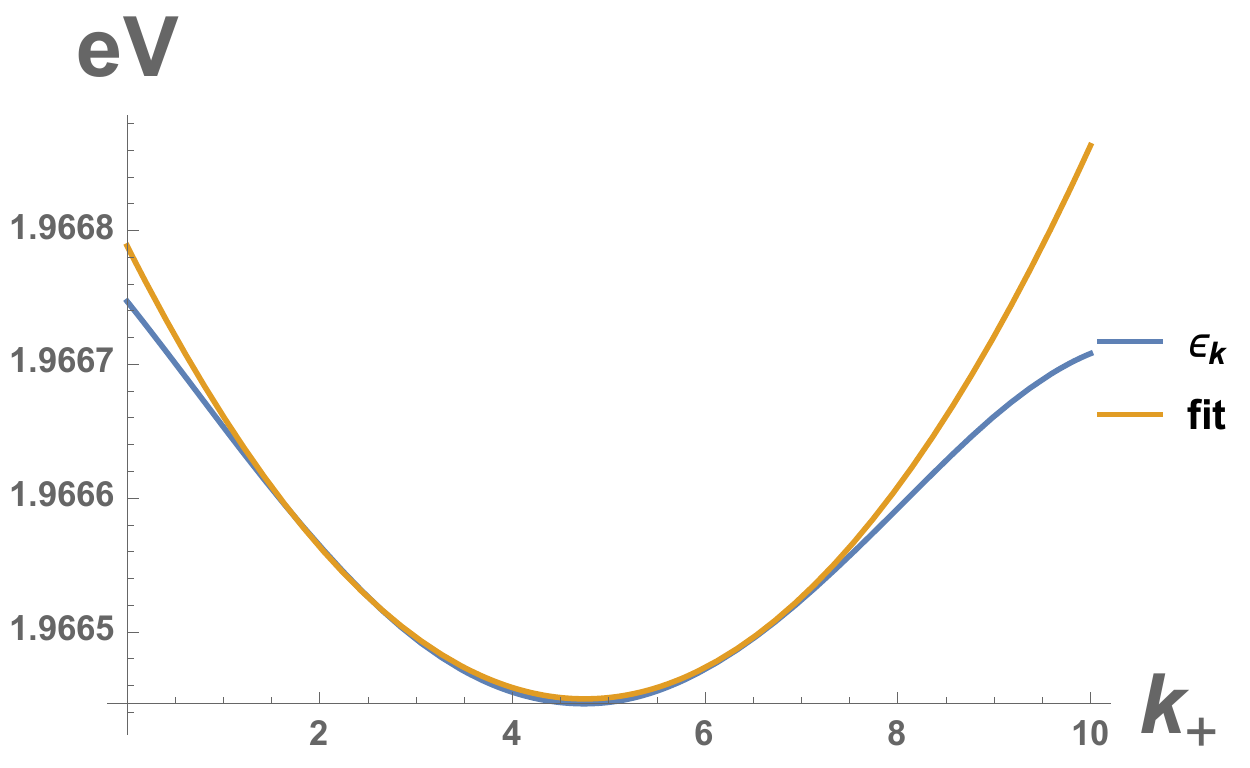}\\
\includegraphics[width=0.32\columnwidth]{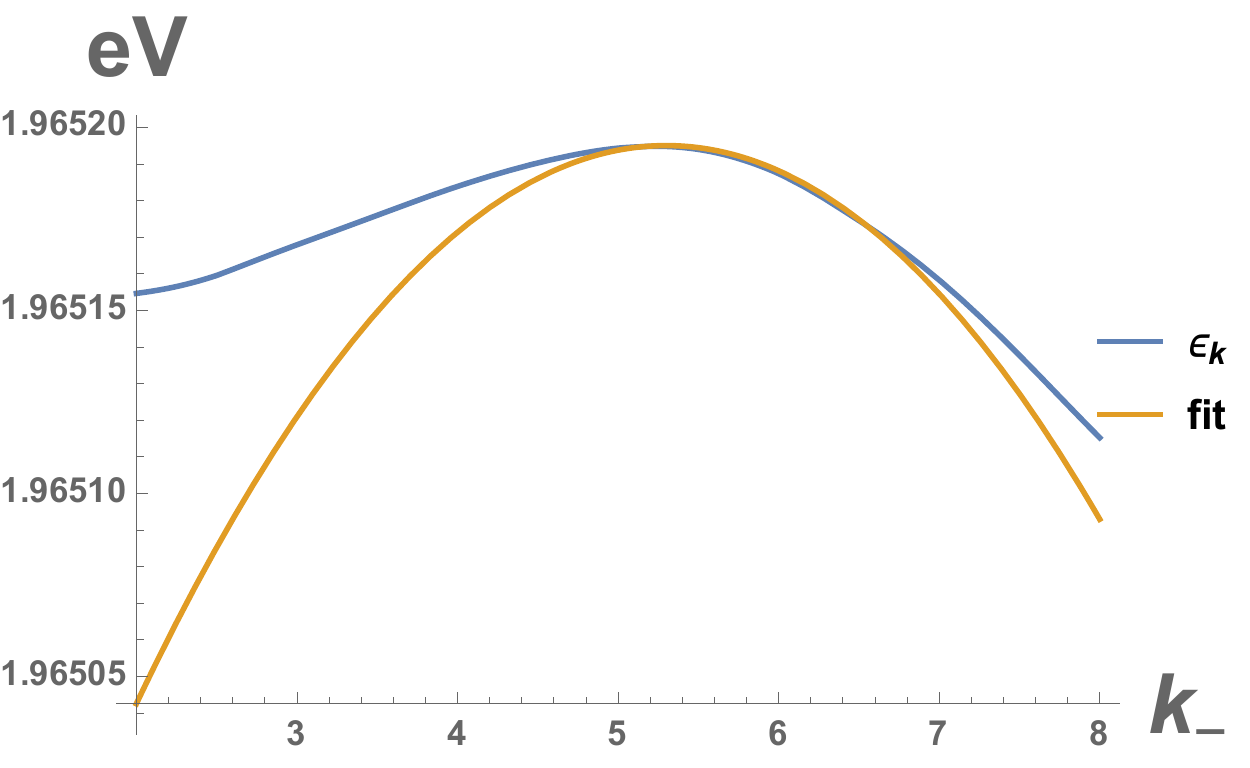}
\includegraphics[width=0.32\columnwidth]{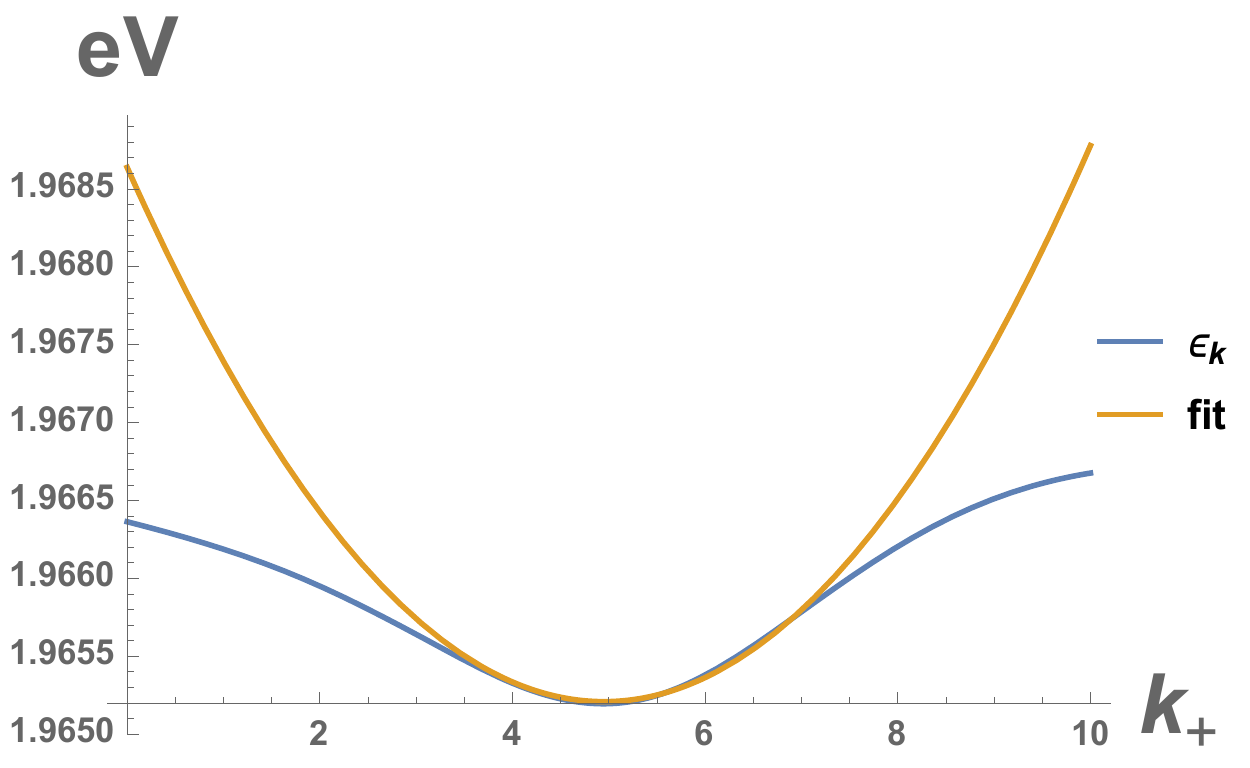}\\ 
\includegraphics[width=0.32\columnwidth]{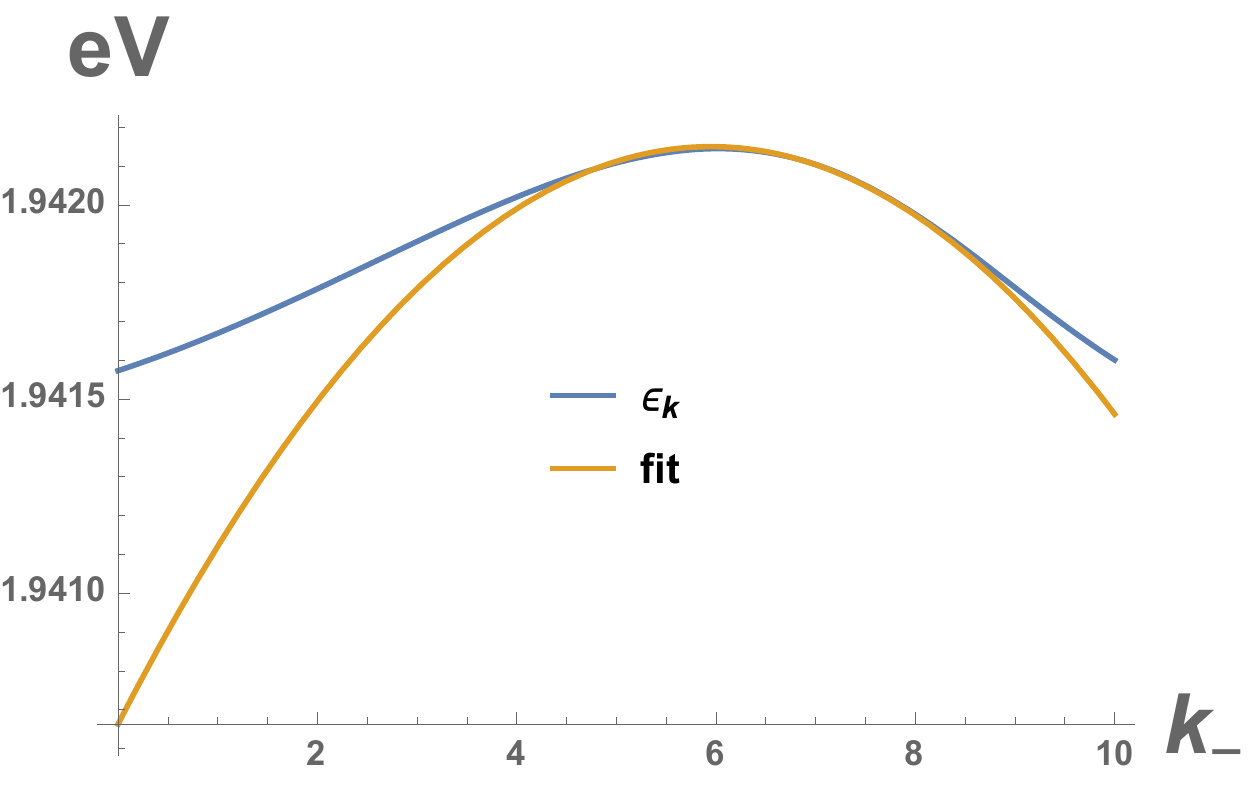}
\includegraphics[width=0.32\columnwidth]{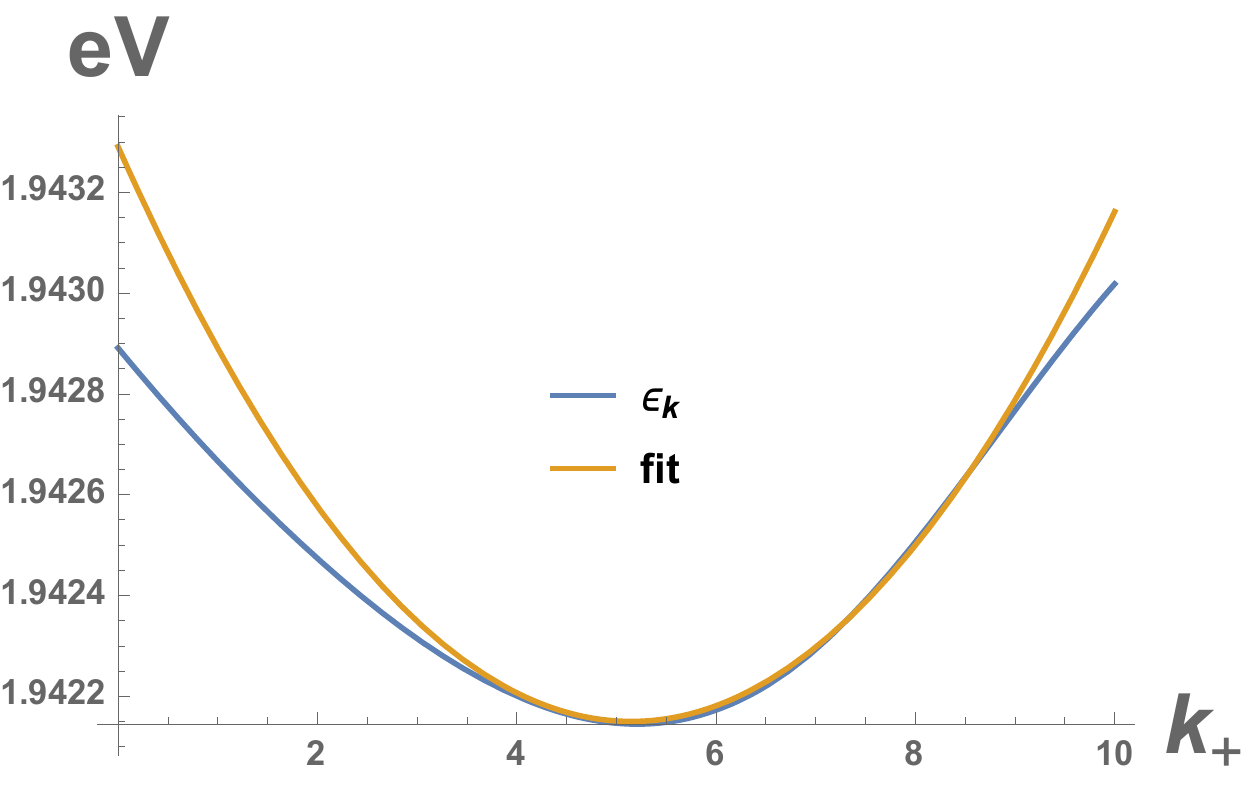}\\ 
\includegraphics[width=0.32\columnwidth]{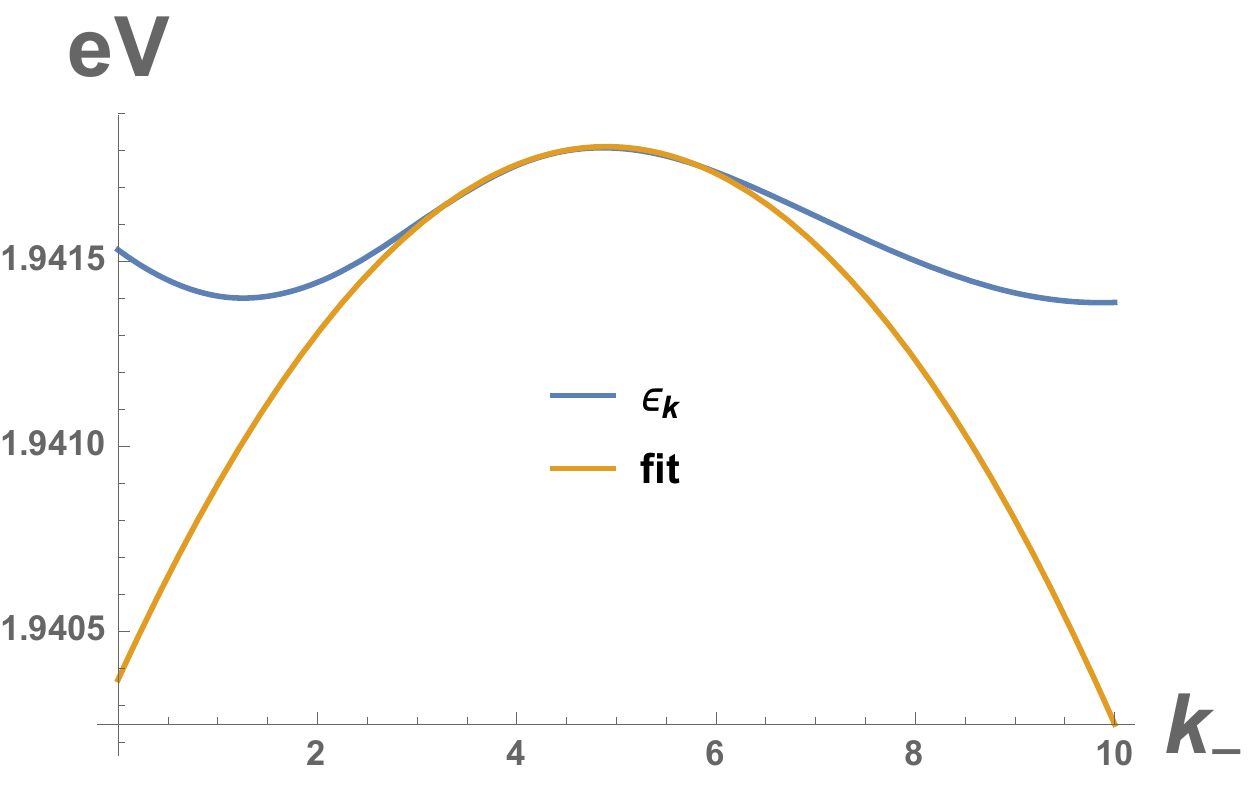}
\includegraphics[width=0.32\columnwidth]{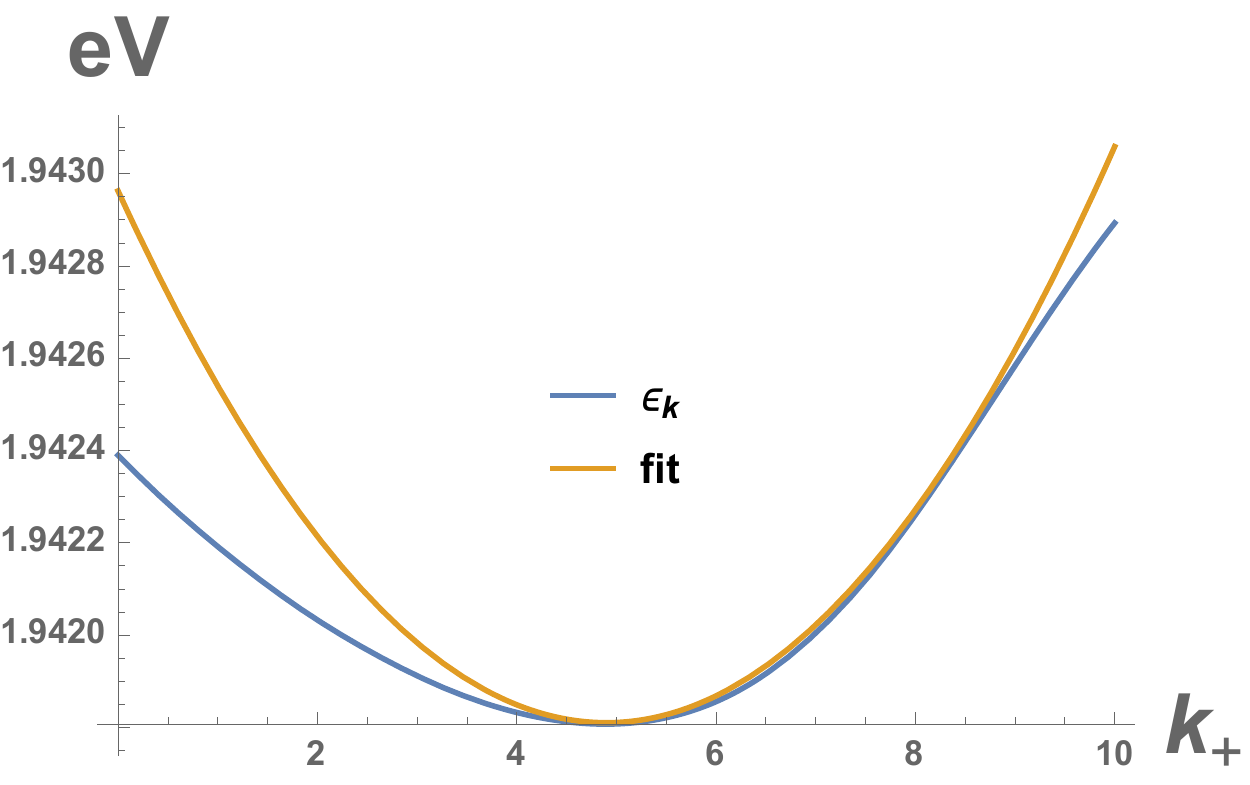} 
\caption{Fitting results of the van Hove singularities 1, 2, 3, 4 (from top to bottom) for the negative (left) and positive (right) inverse mass.}
\label{VanHoveFit}
\end{figure*}

We now fit the van Hove singularities along the principle axes as indicated in Fig. \ref{PrincipleAxes}. Obviously, this procedure could be improved by considering curved trajectories obeying the $C_3$-symmetry, however, we checked that the overall result hardly depends on it. By fitting the dispersion along the principle axes, there is also sometimes another scaling factor to be considered as we usually parametrize our curves by either $k_x$ or $k_y$. The explicit comparison between the dispersion and the fitting result as function of these scaled wave numbers $k_-$ and $k_+$ is shown in Fig. \ref{VanHoveFit}.
The results are listed in Table \ref{TableFits}.
\begin{table}[t]
\centering
\begin{tabular}{c|c|c|c|c|c|c|c|c}
$i$&$\epsilon_{vH}$&$n_{vH}$&$\alpha_-$&$\alpha_+$&$\alpha$&$\Lambda_{0\text{K}}$&$\Lambda_{70\text{mK}}$&$\Lambda_{1\text{K}}$\\
\hline
1&1.9552&-1.12&0.0334&0.0668&0.0447&4.83&5.15&8.25\\
\hline
2&1.9551&-1.51&0.0668&0.0334&0.0447&4.20&4.56&7.90\\
\hline
3&1.9422&-2.73&0.1998&0.2032&0.2015&3.88&3.97&4.99\\
\hline
4&1.9418&-3.21&0.1922&0.1538&0.1709&3.57&3.68&4.94\\
\end{tabular}
\caption{Energy of the van Hove singularities $\epsilon_{vH}$ (in eV), the corresponding filling number $n_{vH}$, and the fitting parameters $\alpha_\pm$ in units of meV $a_M^2$ defining the saddle point dispersion $\epsilon_\k=-\alpha_-k_-^2+\alpha_+k_+^2$. We also list the scale that enters the expression of the Hall density, $\alpha=\frac{2\alpha_+\alpha_-}{\alpha_++\alpha_-}$ (in meV $a_M^2$), and the band cutoff $\Lambda_T$ (in $a_M^{-1}$) for $T=0,$ 70 mK, 1 K.}
\label{TableFits}
\end{table}

What is left is the determination of the band cutoff $\Lambda$. This is done by demanding continuity with the universal regime $n_H=n$ at the cross-over chemical potential $\mu^*$ corresponding to the crossover density $n^*$:
\begin{align}
\Lambda^2=\frac{\mu^*}{\alpha}e^{\pi}
\end{align}
For the crossover density $|n^*-n_{vH}|=\Delta n$, we set $\Delta n=0.15$. For doping levels between two of the van Hove singularities, i.e., in the range $|n_{vH,1}-n_{vH,2}|-2\Delta n$, we choose a linear interpolation of the two logarithmic singularities. Note that by construction, there appears a discontinuity at $n_{vH}+\Delta n$ for the second van Hove singularity due to the change from hole to electron transport. This abrupt  change should be smeared out in more realistic models. 

Finite temperature effects as well as possible disorder effects are included by substituting $|\mu|\to|\mu|+k_BT$ that smears out the logarithmic singularity. This makes the cutoff parameter dependent of the temperature and/or disorder. In Table \ref{TableFits}, we report the results for $T=0,$ 70 mK, 1 K. Let us note that $T=70$ mK is the temperature used in the experiments of Ref. \onlinecite{Park21}, however, we obtain the best fit for $T=1$ K which suggests that there is considerable disorder in the sample without gate voltage.

Let us finally comment on the contribution of the Dirac cone that has been neglected in our analysis, so far. Due to mirror reflection symmetry, the flat bands and the Dirac cone decouple and can be treated separately. Dirac cones lead to circular trajectories due to their conical nature and thus lead again to universal behavior $n_H=n$ (assuming hole doping). However, the hole doping is only a fraction of the doping of the moir\'e supercell and can usually be neglected. Only, for $\nu\approx-4$, the contribution should be measurable and in fact, a small offset of $n_H$ at $\nu=-4$ is seen in the experiments of Ref. \onlinecite{Park21} which we attribute to the Dirac cone contribution. 
\iffalse
The results are shown in Fig. \ref{HallDensityExp}. Clearly seen are the two van Hove singularities in each sub-band. Also shown are the maximal values for each sub-band of the Hall density measured in \cite{Park21}. Obviously, if the van Hove singularities were located at different chemical potentials/filling factors . For $T=1$K, there is relatively good agreement with the experimental measurement and we attribute this to disorder effects that additional smear out the logarithmic divergences. 
\begin{figure*}
\includegraphics[width=0.88\columnwidth]{HallDensity.pdf}
\caption{Hall density as function of the filling factor for different temperatures $T=0,70$mK,1K. Also shown are the maximal value (on average) measured in Ref. \onlinecite{Park21} as well as the dashed purple lines indicating the universal behavior.}
\label{HallDensityExp}
\end{figure*}
\fi
\\
\section*{Supplementary Note VII.\\ Pairing instabilities} 
\begin{figure}[h!]
\includegraphics[width=0.8\columnwidth]{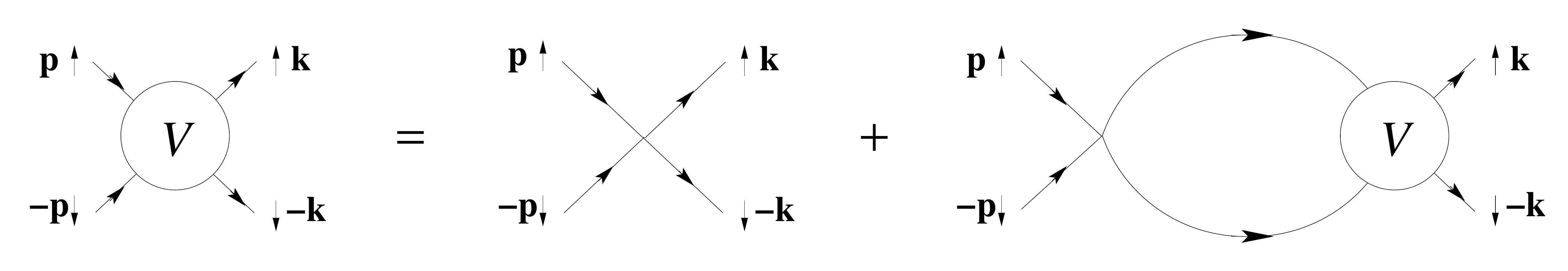}
\caption{Self-consistent diagrammatic equation for the BCS vertex $V$ encoding the iteration of Cooper-pair scattering.}
\label{bcsv}
\end{figure}

Pairing instabilities can be studied by looking for singularities in the so-called BCS vertex, when incoming and outgoing electrons have total momentum equal to zero. For this purpose, we may collect the most divergent contributions in this channel, which leads to the iteration of particle-particle diagrams encoded in the diagrammatic equation shown in Fig. \ref{bcsv}. The particle-particle loop at the right-hand-side of the equation involves an integration in momentum space, that can be parametrized in terms of the components $k_{\parallel }$ and $k_{\perp }$ which are parallel and normal, respectively, to the contour lines of constant energy. Alternatively, we can make a change of variables to the energy $\varepsilon $ of the contour lines and the angle $\theta $ along them. Then, the self-consistent equation for the BCS vertex $V$ becomes
\begin{equation}
V(\theta, \theta'; \omega) = V_0 (\theta, \theta') - 
\frac{1}{(2\pi )^2} \int_0^{\Lambda } d \varepsilon \int_0^{2\pi } d \theta'' 
\frac{\partial k_\perp }{\partial \varepsilon}  \frac{\partial k_\parallel }{\partial \theta''} 
V_0 (\theta, \theta'')   \frac{1}{\varepsilon  - \omega }  
V(\theta'', \theta'; \omega)
\label{selfbcs}
\end{equation}
where $\theta, \theta'$ are the angles of the respective momenta of the spin-up incoming and outgoing electrons and $\omega $ is the sum of the frequencies of the modes in the pair.

Eq. (\ref{selfbcs}) can be cast in a more compact form by making the change of variables
\begin{equation}
\widehat{V} (\theta, \theta'; \omega)  =
\sqrt{ \frac{1}{2\pi } \frac{\partial k_\perp (\theta)}{\partial \varepsilon}
	\frac{\partial k_\parallel (\theta)}{\partial \theta }  }
\sqrt{ \frac{1}{2\pi } \frac{\partial k_\perp (\theta')}{\partial \varepsilon}
	\frac{\partial k_\parallel (\theta')}{\partial \theta' }  }
V (\theta, \theta'; \omega)
\label{redef}
\end{equation}
After that, we can take the derivative with respect to the high-energy cutoff $\Lambda $ and apply the self-consistency at the right-hand-side of the equation, to end up in the scaling equation
\begin{equation}
\Lambda \frac{\partial \widehat{V}(\theta, \theta'; \omega)}{\partial \Lambda } 
=  - \frac{1}{2\pi }  \int_0^{2\pi } d \theta''  
\widehat{V} (\theta, \theta''; \omega)  \widehat{V}(\theta'', \theta'; \omega)
\label{scaling2}
\end{equation}
In Eq. (\ref{scaling2}) it is implicit that the BCS vertex must be actually a function of the ratio $\omega /\Lambda $. Then, the scaling equation can be also used to find the behavior of the vertex in the low-energy limit $\omega \rightarrow  0$. 
 
The analysis of Eq. (\ref{scaling2}) is facilitated by expanding the vertex in a set of orthogonal modes $\Psi_m^{(\gamma )} (\theta)$ corresponding to the different representations $\gamma $ of the point symmetry group, 
\begin{equation}
\widehat{V} (\theta, \theta'; \omega) = \sum_{\gamma, m, n} V_{m,n}^{(\gamma )}(\omega)
\Psi_m^{(\gamma )} (\theta)   \Psi_n^{(\gamma )} (\theta')
\label{dec}
\end{equation} 
We arrive then at the set of equations
\begin{equation}
\omega  \frac{\partial V_{m,n}^{(\gamma )}}{\partial \omega } =
  \sum_{s}  V_{m,s}^{(\gamma )}  V_{s,n}^{(\gamma )}
\label{rg}
\end{equation}
where we have assumed that the vertex must depend on the combination $\omega /\Lambda $.

In this framework, a pairing instability arises when any of the eigenvalues in the expansion (\ref{dec}) has a negative value $V^{(\gamma )} (\Lambda_0 ) < 0$ at the high-energy cutoff. Then, the solution of (\ref{rg}) leads to a divergent flow given by
\begin{equation}
V^{(\gamma )} (\omega ) = \frac{V^{(\gamma )} (\Lambda_0 )}{1 +  V^{(\gamma )} (\Lambda_0 ) \log \left( \frac{\Lambda_0 }{\omega } \right)   }
\end{equation}
In the flow towards the low-energy limit $\omega \rightarrow  0$, a singularity is reached at a critical frequency 
\begin{equation}
\omega_c  \approx  \Lambda_0  \exp \{ -1/|V^{(\gamma )} (\Lambda_0)| \}
\label{ctemp}
\end{equation} 
which sets the scale of the superconducting instability.

\begin{figure}[h!]
\includegraphics[width=0.35\columnwidth]{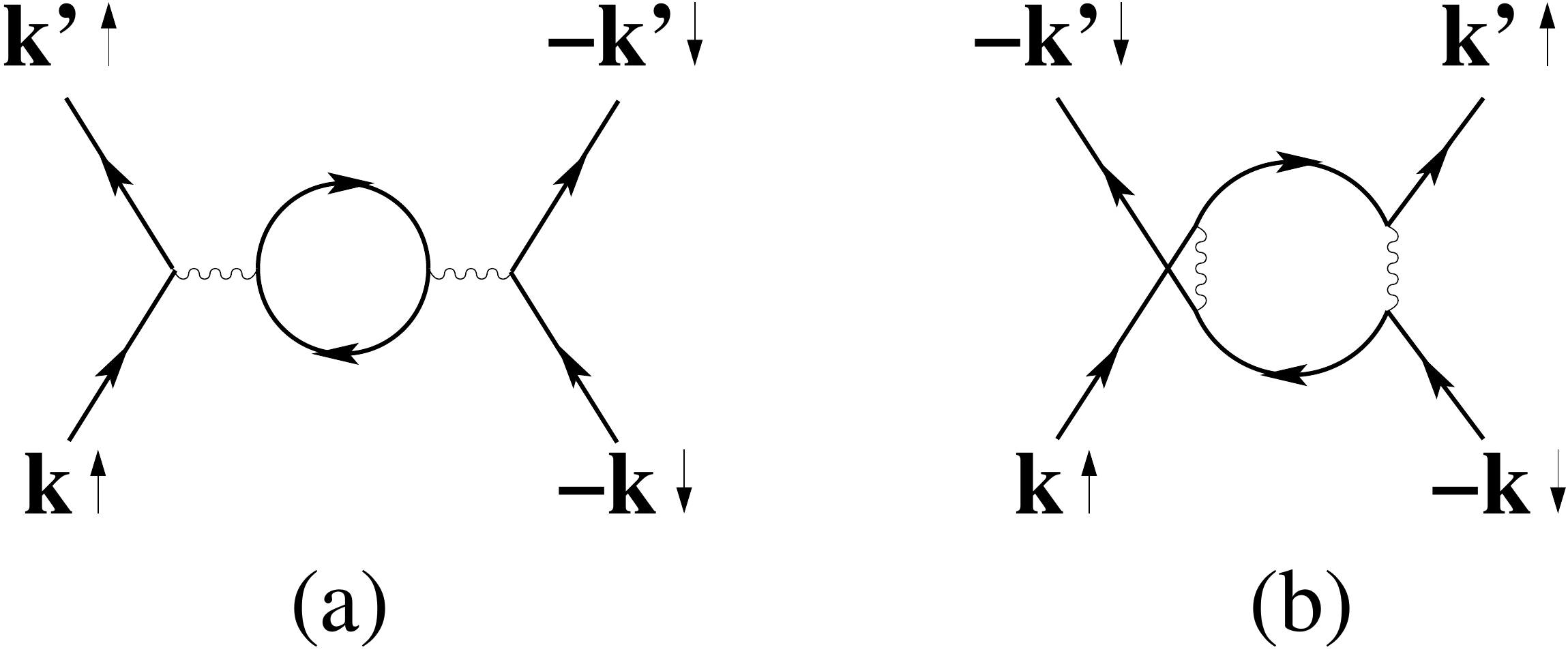}
\caption{Second-order diagrams contributing to the unrenormalized BCS vertex.}
\label{kohn}
\end{figure}

In practice, one has to start with a sensible representation of the vertex $V(\theta, \theta')$ at the high-energy cutoff $\Lambda_0 $. This can be obtained by performing a sum of particle-hole contributions, building on the original diagrams which were considered in the seminal work by Kohn and Luttinger. Usually, one resorts to iterate in the particle-hole scattering shown in Figs. \ref{kohn}(a)-(b) \cite{Scalapino87}. In our case, an important difference with respect to the discussion carried out for the Hubbard model is that the interaction is mediated by the extended Coulomb potential $v(\r)$, as we are dealing with all the atoms in the unit cell of the moir\'e superlattice. Then, the sum of RPA and ladder contributions leads to an expression for the vertex
\begin{equation}
V (\theta, \theta'; \Lambda_0) = 
  \frac{v(\mathbf{k}-\mathbf{k}^\prime)}
       {1 +  v(\mathbf{k}-\mathbf{k}^\prime)  \chi_{\rm ph}(\mathbf{k}-\mathbf{k}^\prime)}
  +  \frac{v^2(\mathbf{Q}) \widetilde{\chi}_{\rm ph}(\mathbf{k}+\mathbf{k}^\prime)}
         {1 - v(\mathbf{Q}) \widetilde{\chi}_{\rm ph}(\mathbf{k}+\mathbf{k}^\prime)}
\label{init2}
\end{equation}
where $\chi_{\rm ph}$ ($\widetilde{\chi}_{\rm ph}$) stands for the susceptibility in the series of bubble (ladder) diagrams. The interaction $v(\mathbf{Q})$ is a function of the momentum transfer $\mathbf{Q}$ which depends on the sum of the momenta $\mathbf{k},\mathbf{k}^\prime$ of incoming and outgoing electrons as well as on the momentum of the internal loop. In Eq. (\ref{init2}), the sum of RPA diagrams leads to screening of the interaction, making its contribution less relevant, while it is the sum of ladder diagrams encoded in the second term what may enhance potential pairing instabilities.

Once we compute the BCS vertex according to Eq. (\ref{init2}), the last stage of the analysis is the evaluation of the different coefficients in the expansion (\ref{dec}) at the high-energy cutoff. This can be easily made using the orthogonality of the modes, so that 
\begin{equation}
V_{m,n}^{(\gamma )}(\Lambda_0 )  =  \int_0^{2\pi } d \theta  \int_0^{2\pi } d \theta' \; \widehat{V} (\theta, \theta'; \Lambda_0) \Psi_m^{(\gamma )} (\theta)   \Psi_n^{(\gamma )} (\theta')
\label{conv}
\end{equation}
This is the approach we have followed to determine the different eigenvalues for the BCS vertex, applying in particular the convolution (\ref{conv}) with a large set of harmonics to capture the modulations along the energy contour lines of the second valence band.

\begin{figure*}
\hspace{1.5cm}
\includegraphics[width=0.3\columnwidth]{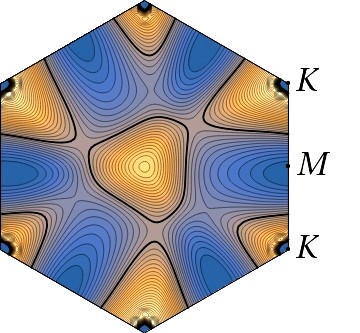}
\hspace{1.5cm}
\includegraphics[width=0.3\columnwidth]{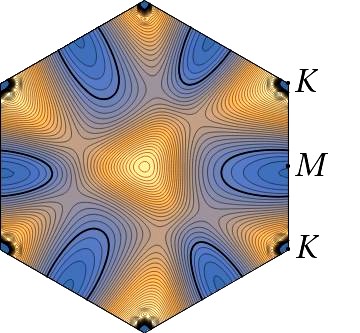}
\\
 \hspace*{1.0cm} (a) \hspace{7.7cm} (b)
\caption{Energy contour maps of the second valence band (for spin-up projection) in the Brillouin zone of twisted trilayer graphene at twist angle $\theta \approx 1.61^\circ$, computed in a self-consistent Hartree-Fock approximation with dielectric constant $\epsilon = 48$ and filling fraction of 2.8 holes (a) and 3.6 holes (b) per moir\'e unit cell. The thick contours stand for the Fermi lines. Contiguous contour lines differ by a constant step of 0.1 meV, from lower energies in blue to higher energies in light color.}
\label{flines}
\end{figure*}

The results of this decomposition of the BCS vertex in harmonics may differ significantly, depending on whether the Fermi line displays the triangular patches found above the van Hove singularity at $\nu \approx -2.8$ or it has evolved into elliptical shape below that filling fraction. These two different possibilities are illustrated in Fig. \ref{flines}. The first regime has been discussed at $\nu = -2.4$ in the main text, and we analyze here the two instances shown in Figs. \ref{flines}(a)-(b) for filling fraction $\nu = -2.8$ and $\nu = -3.6$.

\begin{table}[h!]
	\begin{tabular}{c|c|c}
		Eigenvalue $\lambda$  &   harmonics  &   Irr. Rep.\\
		\hline \hline
		2.31  &     $     1     $         &                  \\
		\hline
		1.25   &    \multirow{2}{*}{ $\{\cos (\phi),\sin (\phi)\}$ }    & \multirow{2}{*}{E}  \\
		1.24   &                                  &                                      \\
		\hline
		0.46  &     $     \cos (3\phi)     $         &       $A_1$            \\
		\hline
		$-0.30$  &     \multirow{2}{*}{ $\{\cos (4\phi),\sin (4\phi)\}$ }    &  \multirow{2}{*}{E}  \\
		$-0.29$   &                                 &                                         \\
		\hline
		$-0.29$   &     $\sin (3\phi)  $       &     $A_2$               \\                
		\hline
		0.27   &    \multirow{2}{*}{  $\{\cos (4\phi),\sin (4\phi)\}$  }     &  \multirow{2}{*}{E}   \\
		0.26   &                             &                                               \\
		\hline
		0.15   &     $\cos (6\phi)  $       &     $A_1$                     \\
		\hline
		0.15   &    \multirow{2}{*}{  $\{\cos (5\phi),\sin (5\phi)\}$  }     &  \multirow{2}{*}{E}   \\
		0.13   &                                  &                        
		
		%\hline
		
		%0.09   &     $\{\cos (8\phi),\sin (8\phi)\}$     &E\\
		%-0.05   &     $     \sin (3\phi)     $         &$A_2$

	\end{tabular}
	\caption{Eigenvalues of the Cooper-pair vertex with largest magnitude and dominant harmonics, grouped according to the irreducible representations of the approximate $C_{3v}$ symmetry, for the Fermi line shown in Fig. \ref{flines}(a). The modes $\{\cos (4\phi),\sin (4\phi)\}$ appear twice in the list, as they only denote the dominant harmonic, but they actually represent different eigenvectors.} 
	\label{table1}
\end{table}

The Fermi line shown in Fig. \ref{flines}(a) has an approximate $C_{3v}$ symmetry, so that the eigenmodes in the expansion of the BCS vertex can be assorted in irreducible representations of that group. The first terms in the series of eigenvalues an respective eigenvectors can be seen in Table \ref{table1}. We find that there are two irreps with relatively prominent negative eigenvalues, although slightly smaller in absolute value than those obtained in the expansion at $\nu = -2.4$. We can introduce these values into Eq. (\ref{ctemp}) to estimate $T_c$. Taking $\Lambda_0 \approx 1.5$ meV, we obtain an estimate of $T_c \sim 1$ K. Although the order of magnitude is similar to that found for $T_c$ at $\nu = -2.4$, the absolute value of the dominant negative coupling is smaller at $\nu = -2.8$, implying that the critical temperature has to be necessarily smaller at such a larger hole doping.

Turning to the Fermi line in Fig. \ref{flines}(b), the elliptical shape has an approximate $C_{2v}$ symmetry, which means that the different eigenvalues of the BCS vertex correspond to one-dimensional representations. The different couplings for the particular case shown in Fig. \ref{flines}(b) are listed in Table \ref{table2}. We observe that there are several negative eigenvalues, which imply that the elliptic Fermi line still may support a pairing instability. The critical energy scale has to be obtained according to Eq. (\ref{ctemp}), bearing in mind that $\Lambda_0$ must be a symmetric cutoff dictated by the effective bandwidth, here constrained by the proximity of the Fermi line to the bottom of the band. We estimate $\Lambda_0 \approx  0.4$ meV which leads, for the dominant negative coupling in Table \ref{table2}, to a critical temperature $T_c \sim 0.1$ K.

\begin{table}[h!]
\begin{tabular}{c|c|c}
Eigenvalue $\lambda$  &   harmonics  &   Irr. Rep.\\
\hline \hline
0.85  &     $     1     $         &                  \\
\hline
$-0.25$  &     $     \sin (\phi)     $         &       $B_2$            \\
\hline
$-0.11$   &     $\cos (\phi)  $       &     $B_1$               \\                
\hline
0.08   &     $\cos (3\phi)  $       &     $B_1$                     \\
\hline
$-0.06$   &     $\sin (2\phi)  $       &     $A_2$                     \\
\hline
$-0.04$   &     $\cos (2\phi)  $       &     $A_1$

\end{tabular}
\caption{Eigenvalues of the Cooper-pair vertex with largest magnitude and dominant harmonics along the elliptic Fermi lines shown in Fig. \ref{flines}(b) for filling fraction of 3.6 holes per moir\'e unit cell.} 
  \label{table2}
\end{table}

We arrive at the general conclusion that the Kohn-Luttinger instability is stronger in the regime where twisted trilayer graphene develops the rather regular triangular Fermi lines observed in the second valence band, in the range within filling fractions $\nu \approx -2$ and $\nu \approx -2.8$. The pairing instability then looses strength for larger hole doping, as a consequence of having smaller pairing modulation as well as much smaller energy range for the scattering of Cooper pairs, which produces a substantial decrease in the critical temperature when approaching the bottom of the band.      
\\
\\
\\
\section*{Supplementary Note VIII.\\ Effective spin-orbit coupling, Ising superconductivity and violation of the Pauli-limit} 
The spin-selective valley symmetry breaking is driven by the emerging flux that is generated by the imaginary part of the next-nearest neighbour hopping $t_X^{(\ell)}$, with the two sublattices $X=A,B$ and the layer $\ell=1,2,3$. This flux has opposite sign for the two sublattices and thus valley symmetry breaking in each spin channel is the dominate order parameter related to $(t_A^{(\ell)}-t_B^{(\ell)})/2$. Nevertheless, there is also a net Haldane flux that leads to a time-reversal symmetry broken gap related to $(t_A^{(\ell)}+t_B^{(\ell)})/2$.  

\begin{figure*}
\includegraphics[width=0.29\columnwidth]{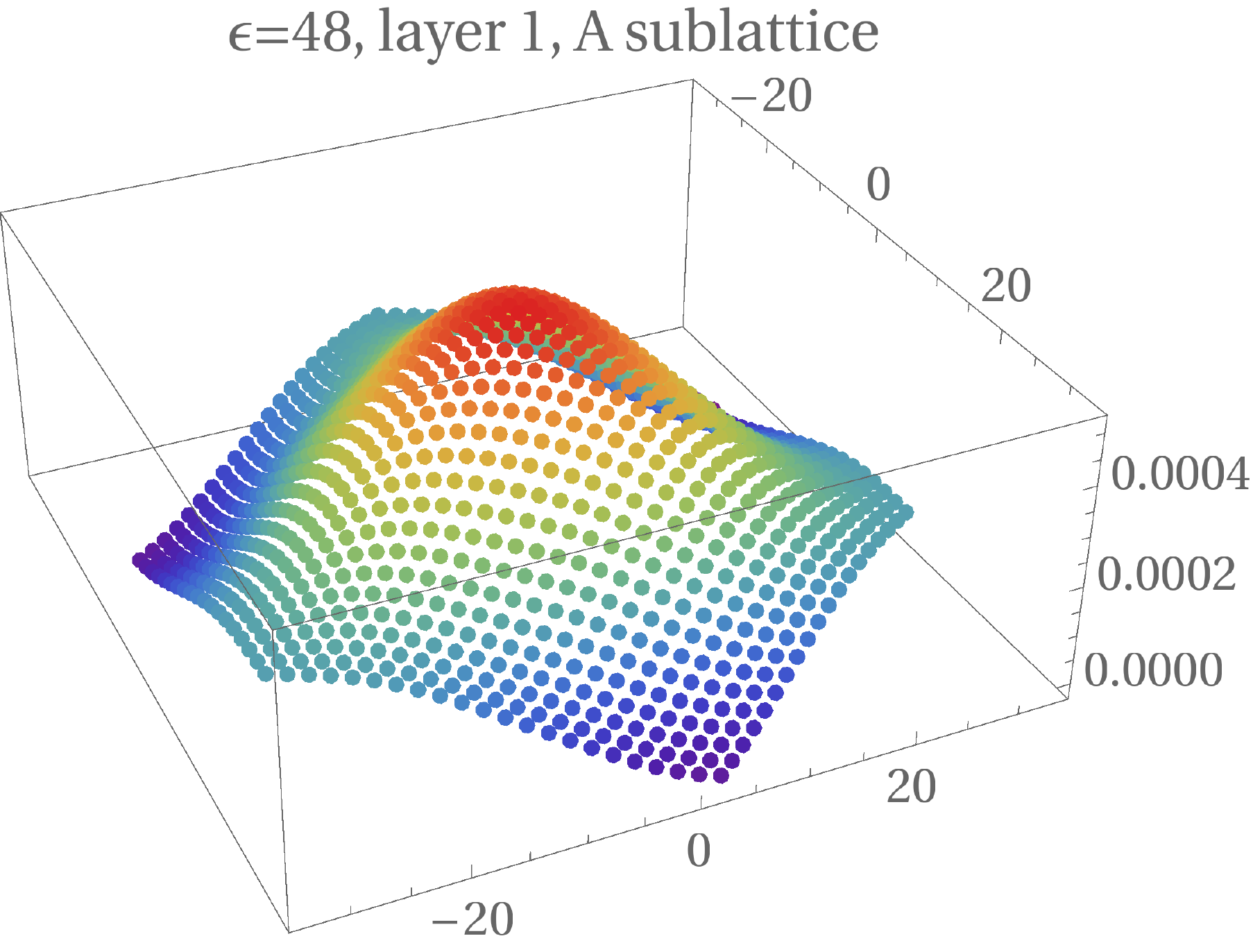}
\includegraphics[width=0.32\columnwidth]{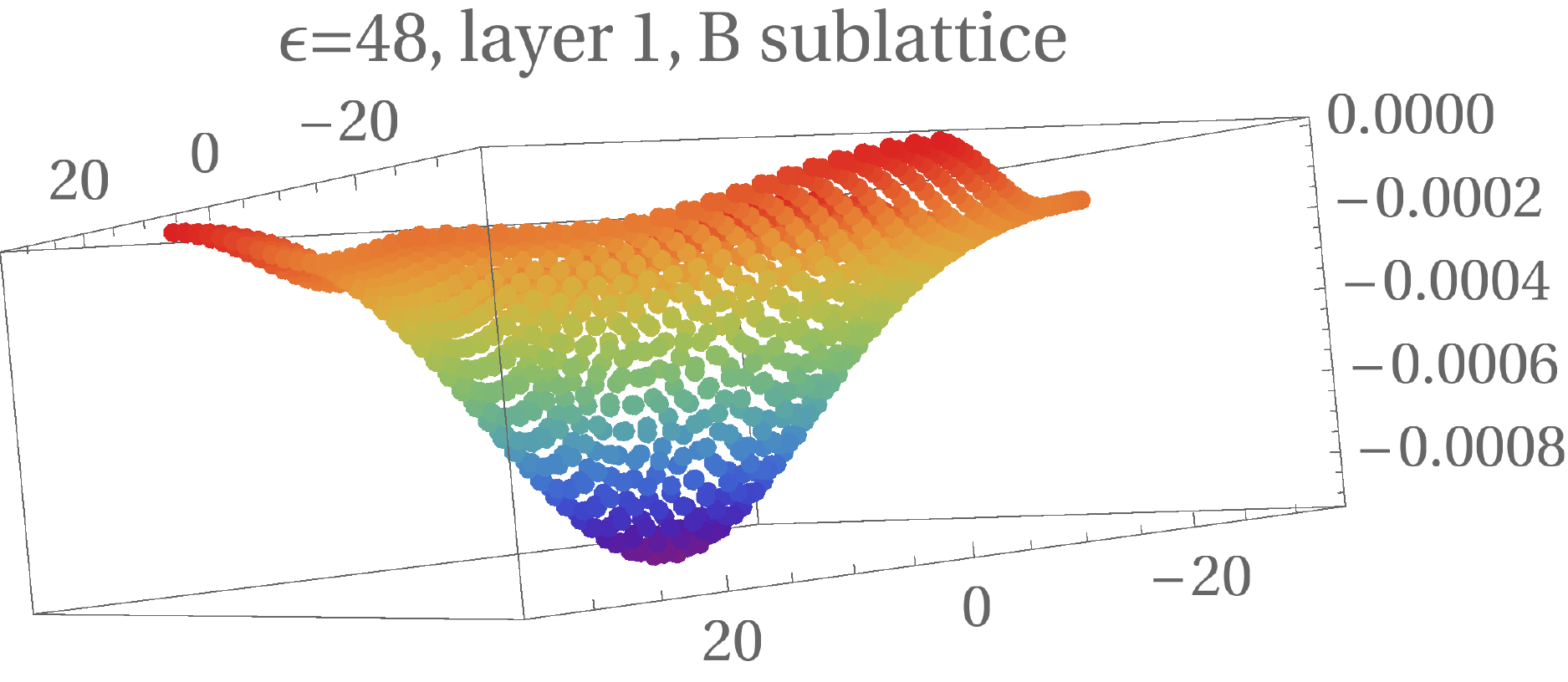}\\
\vspace{0.5cm}
\includegraphics[width=0.32\columnwidth]{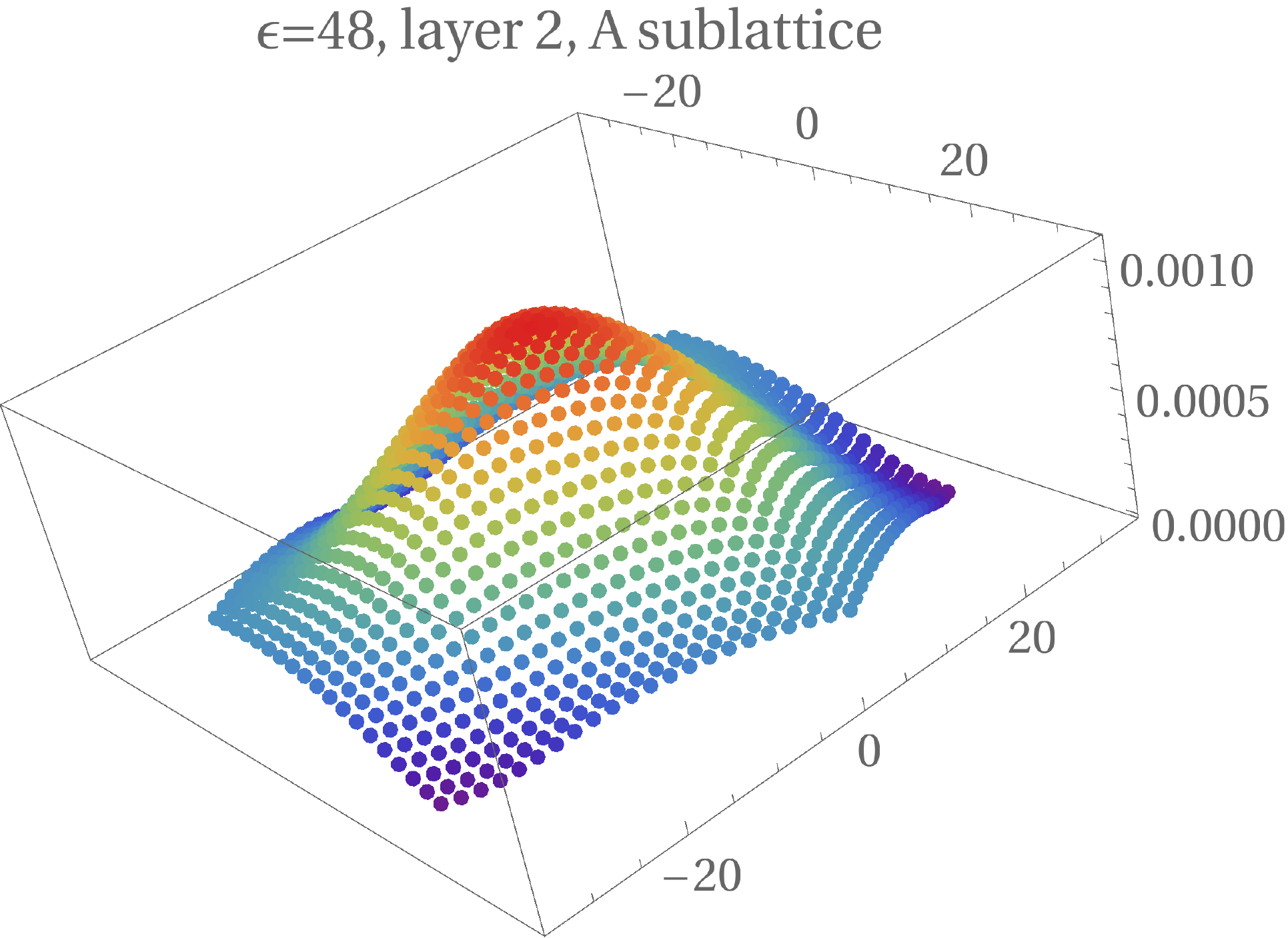}
\includegraphics[width=0.34\columnwidth]{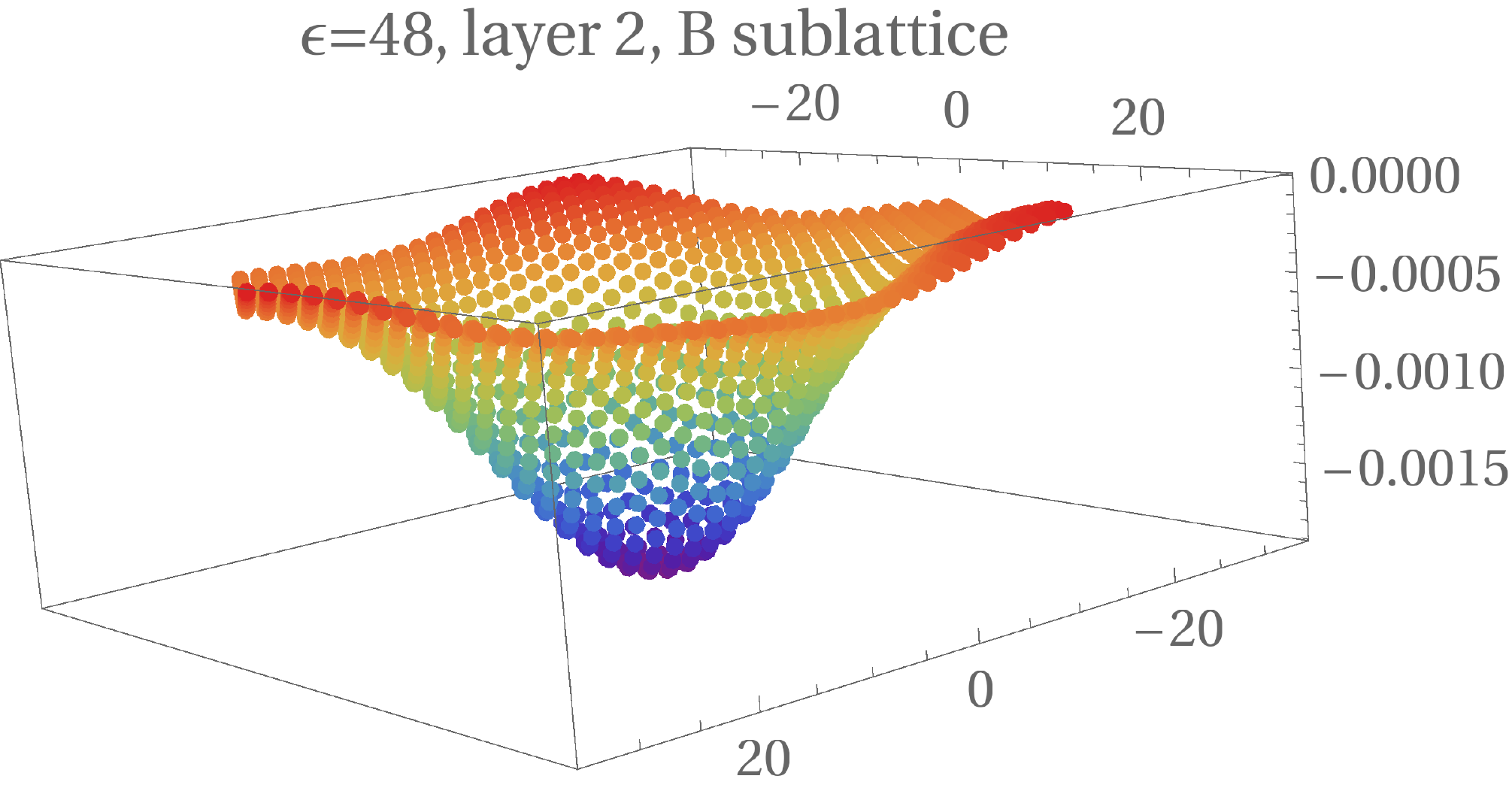}\\ 
\vspace{0.5cm}
\includegraphics[width=0.29\columnwidth]{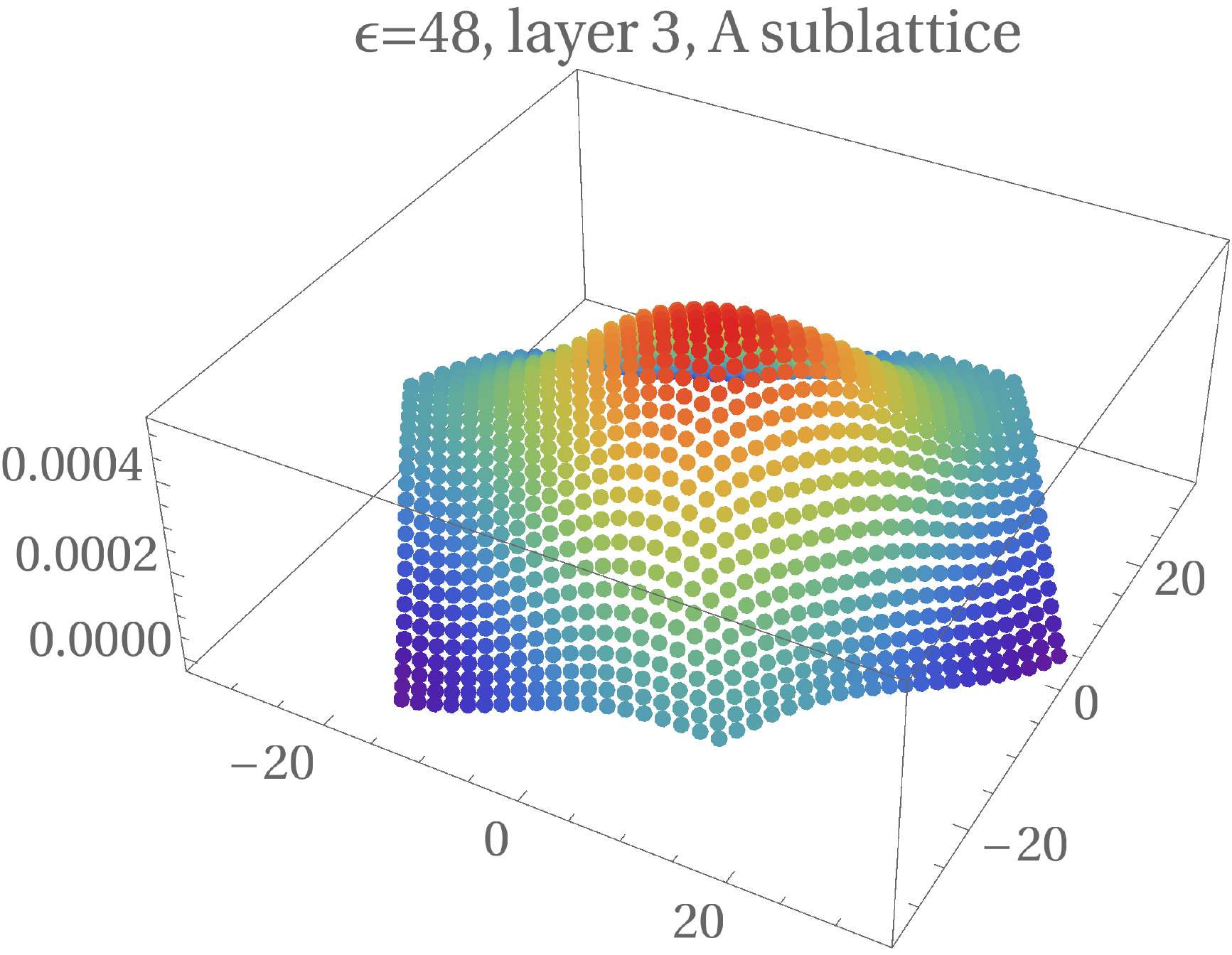}
\includegraphics[width=0.32\columnwidth]{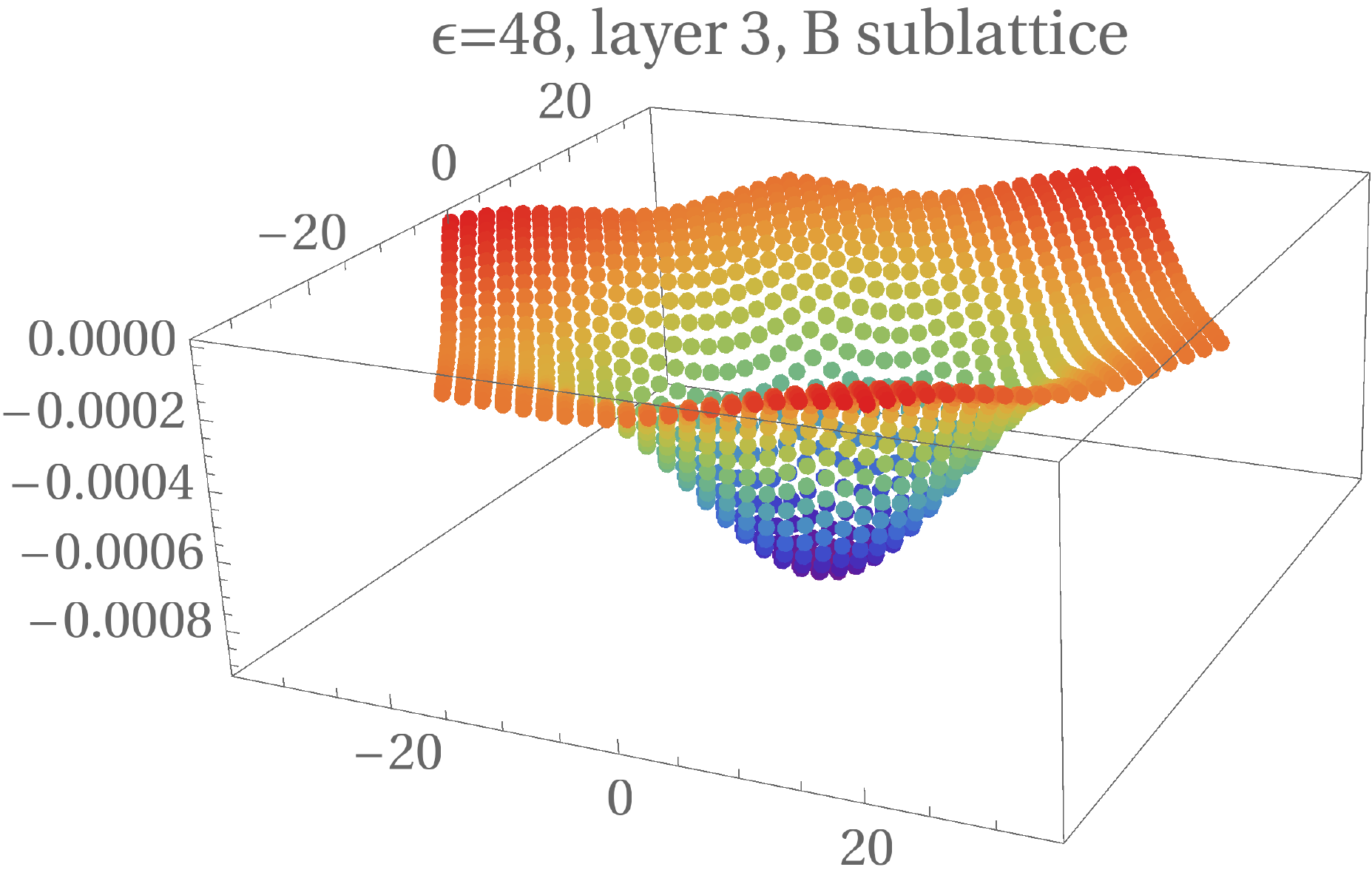}
\caption{Plot on the moir\'e cell of the imaginary part of the next-nearest neighbour hopping $t_X^{(\ell)}$ with sublattice index $X=A,B$ and layer index $\ell=1,2,3$ for one spin-projection. The panels show 3$t_X^{(\ell)}$ and the absolute value is concentrated around the $AA$-stacked region. The vertical scale is in units of eV.}
\label{Flux}
\end{figure*}

The graphs shown in Fig. \ref{Flux} display the results for one spin projection. The values are reversed for the other spin projection such that time-reversal symmetry is only broken for each spin sector individually. Combining the two spin-channels, time-reversal symmetry is restored just as it is the case in the Kane-Mele model,\cite{Kane05} only with an effective intrinsic spin-orbit coupling. This leads to a pinning of the spin polarization perpendicular to the layer. 

To make the discussion quantitative, let us set the maximal imaginary tunnel-matrix element of layer 2 and of sublattice A/B as $3t_A^{(2)}=0.001$ eV and $3t_B^{(2)}=-0.0015$ eV, respectively, as shown in Fig. \ref{Flux}. The energy scale for the spin gap $\Delta=2\sqrt{3}3t$ is thus given by $\Delta=2\sqrt{3} \times 0.25$meV $\sim 1$ meV. 

So far, the initial Hamiltonian had no spin-orbit coupling such that the spin-polarization of the Cooper pairs would be arbitrary. However, due to the bare intrinsic spin-orbit coupling of single-layer graphene, the spin-degeneracy is broken and leads to an out-of-plane spin-polarization. The effective, renormalized intrinsic spin-orbit coupling thus also leads to out-of-plane polarized spin-singlet Cooper-pairs as was already discussed the context of graphene by Kane and Mele.\cite{Kane05} 

Because of the out-of-plane polarization, these spin-states are unaffected by an in-plane magnetic field unless the magnetic field energy surpasses the pinning energy. In this case, the singlet of the Cooper-pair is first rotated parallel to the field and then broken up due to the energy gained from the magnetic susceptibility, characterized by the Pauli-limit which corresponds to an energy less than the pinning energy. This leads to a violation of the Pauli-limit by a factor of 2-3 as argued in the main text.

\end{document}